%% file: main.tex
\documentclass{sig-alternate}

\newcommand{\ignore}[1]{}
\usepackage{fancyhdr}
\usepackage[normalem]{ulem}
\usepackage[hyphens]{url}
\usepackage{hyperref}

\usepackage{subfig}
\usepackage{multirow}
\usepackage{array}
\usepackage{xcolor,colortbl}
\usepackage[justification=centering]{caption}
\usepackage{siunitx}
\usepackage{enumitem}
\usepackage{gensymb}

\newcommand{\upperRomannumeral}[1]{\uppercase\expandafter{\romannumeral#1}}
\newcommand{\lowerromannumeral}[1]{\romannumeral#1\relax}

\fancypagestyle{firstpage}{
  \fancyhf{}

  \pagenumbering{arabic}
}  

\title{DReAM: Dynamic Re-arrangement of Address Mapping\\to Improve the Performance of DRAMs} 

\author{Mohsen Ghasempour$^\dagger$,  Jim Garside$^{\dagger}$, Aamer Jaleel$^{\star}$ and Mikel Luj{\'a}n$^{\dagger}$\\
School of Computer Science, University of Manchester$^{\dagger}$ \\
  NVidia Research$^{\star}$}

\begin{document}
\maketitle
\thispagestyle{firstpage}
\pagestyle{plain}


\input{Abstract}

\input{Introduction}

\input{Background}

\input{DReAM}

\input{EvaluationMethodology}

\input{ResultsAndDiscussion}

\input{RelatedWork}

\input{Conclusion}


\section{Acknowledgements}
The research leading to these results has received funding from the European Union's Seventh Framework Programme (FP7/2007-2013) under grant agreement n$^{\circ}$ 318633; AXLE project http://axleproject.eu/. Mikel Luj{\'a}n is funded by a Royal Society University Research Fellowship and further supported by UK EPSRC grants DOME EP/J016330/1 and PAMELA EP/K008730/1.

\bibliographystyle{ieeetr}
\bibliography{ref}


\end{document}

%% file: Abstract.tex
\begin{abstract}

The initial location of data in DRAMs is determined and controlled by the `address-mapping' and even modern memory controllers use a fixed and run-time-agnostic address mapping. On the other hand, the memory access pattern seen at the memory interface level will dynamically change at run-time. This dynamic nature of memory access pattern and the fixed behavior of address mapping process in DRAM controllers, implied by using a fixed address mapping scheme, means that DRAM performance cannot be exploited efficiently. 

DReAM is a novel hardware technique that can detect a workload-specific address mapping at run-time based on the application access pattern which improves the performance of DRAMs. The experimental results show that DReAM outperforms the best evaluated address mapping on average by 9\%, for mapping-sensitive workloads, by 2\% for mapping-insensitive workloads, and up to 28\% across all the workloads. DReAM can be seen as an insurance policy capable of detecting which scenarios are not well served by the predefined address mapping.

\end{abstract}



%% file: Introduction.tex
\section{Introduction}

Increasing the number of general purpose cores and accelerator cores (e.g.\ GPU cores) integrated into a single chip and competing for access to DRAM, demands better performance from the main memory. In this situation, exploiting the maximum performance obtainable from the memory system is crucial. However, due to the internal structure and organization of DRAMs, described in Section \ref{sec:background}, there is always some memory bandwidth (Performance) wasted due to internal conflicts. One of the most serious conflicts in a DRAM memory system is referred to as `page conflict'. This happens when two consecutive memory requests go to different rows within the same bank. In this situation, these memory requests must be serviced one after another which causes a high access latency for the second request. Dealing with page conflicts becomes even more challenging considering the fact that they are completely dependent on the memory access pattern. This means that the rate of page conflicts and the time of their occurrence change dynamically according to the application behavior. To mitigate the vulnerability of DRAMs performance to page conflicts, state-of-the-art memory controllers have evolved into complex hardware components employing subsystems such as schedulers. These schedulers take advantage of workload run-time information (the sequence of memory requests) to reduce page conflicts. An important role of the scheduler is to minimize DRAM page conflicts by reordering the memory commands that are available to issue to the DRAM. However, the main limitation for schedulers is the number of options (memory requests) that they have to choose from at the time of scheduling. In general, the number of available memory requests at the time of scheduling is limited by data dependencies between memory requests, the number of running threads, the number of cores etc. 
Therefore, there are conflicts that schedulers cannot eliminate. These page conflicts result from the address-mapping and data placement in DRAMs. As discussed in the next section, the address mapping is a process that maps the physical address bits provided by processors to the internal structure of DRAMs. This process controls the initial data placement in memory. Thus, it is important to understand how to select a good address-mapping scheme to place and distribute data in DRAM devices to mitigate page conflicts. This is possible using a software-only approach; e.g.\ with OS support and intelligent memory allocators. However, this option faces complex problems when considering multiple independent applications executing concurrently, or with virtualized scenarios (both hypervisors and containers) and relies on software being compiled for specific memory hardware.

This paper presents DReAM, a novel hardware technique based on approximating the entropy of each memory address bit for a set of memory requests, to generate workload specific address-mappings at run-time. To rearrange the address mapping at run-time DReAM needs to support the online-data migration imposed by changing the address-mapping. DReAM investigates different scenarios for data migration with different levels of complication. The proposed solutions were evaluated over a wide range of mapping-sensitive and mapping-insensitive workload mixes. Three different address mapping schemes were evaluated over all the workloads and the best one was chosen to compare against DReAM. Overall, DReAM is the first on-the-fly mechanism capable of generating workload specific address-mappings without requiring to stop the running applications.

%% file: Background.tex
\section{Background}
\label{sec:background}

\begin{figure}[!htb]
\centering
\includegraphics[scale=0.35]{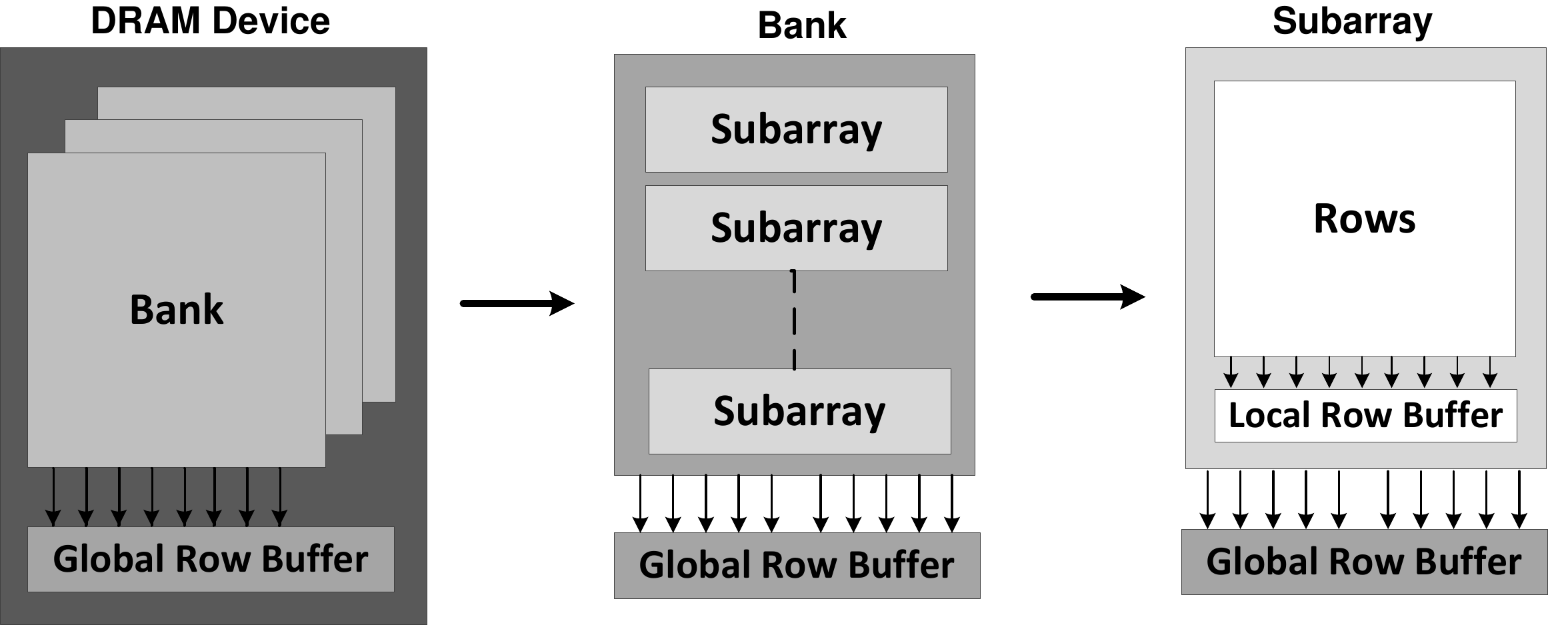}
\caption{DRAM device organisation. }
\label{fig:DRAM_organisation}
\end{figure}

Figure \ref{fig:DRAM_organisation} presents the basic organization of a DRAM device. Each DRAM device consists of multiple banks each of which has a data array and one row buffer. In practice, the data array within a bank consists of multiple subarrays, each of which has its own local row buffer. The local row buffers within a bank are connected to other local row buffers as well as the global row buffer. There are some interesting works by Chang \textit{et al}.\ \cite{changimproving}, Kim \textit{et al}.\ \cite{kim2012case} and Seshadri \textit{et al}.\ \cite{seshadri2013rowclone} to exploit these subarrays to improve the DRAM performance and bulk data copy in DRAMs. 

\begin{figure}[!htb]
\centering
\includegraphics[scale=0.3]{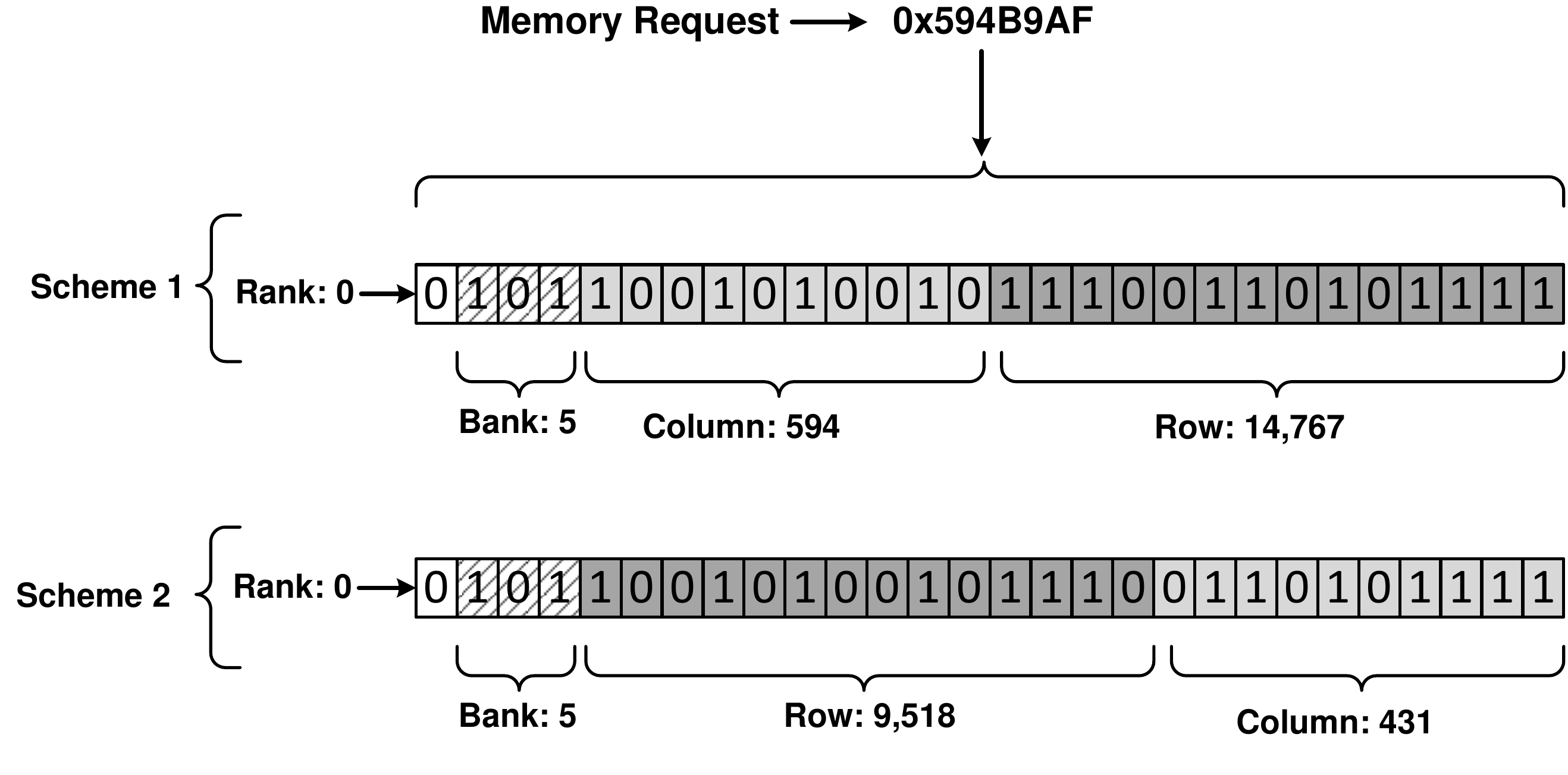}
\caption{Two different address mapping schemes.}
\label{fig:address_mapping}
\end{figure}

The address mapping mechanism for DRAMs transforms the flat 1D of physical addresses into the internal 2D structure of DRAMs devices (row \& column). Figure~\ref{fig:address_mapping} illustrates how one physical address can be interpreted with two different mapping schemes. Most memory systems contain DIMMs and a DIMM can have multiple \textit{ranks} of DRAMs. Multiple DIMMs can be placed on a \textit{channel}; i.e.\ the physical connection between a memory controller and DRAMs \cite{jacob2010memory}. The reason for these many hierarchical levels is to maximize the parallelism that can be exploited when servicing multiple memory requests.

In general an address-mapping scheme extracts the corresponding address for Channel, Rank, Bank, Row and Column from the physical address. Due to the internal structure and electronic circuit characterization of the DRAMs, consecutive access to different memory locations can have a different memory cost depending on the previous state of the memory. For instance, if there are two consecutive accesses to the same row in the same bank of a DRAM, the second access can have significantly smaller latency than the first access since the target row has been `opened' by the first memory request. On the other hand, if there are two consecutive accesses to different rows within the same bank, the second access has significantly higher latency in comparison with the first access. The reason is that, in this case, the previous row must be `closed' before the new row is `activated'. These scenarios describe a \textit{page conflict} and degrades the overall performance of DRAMs. Page conflicts are sensitive to the data placement in DRAMs and data placement is determined by the address-mapping schemes in the first place. Therefore, choosing an address mapping scheme carefully can reduce the page conflicts and improve the performance of DRAMs.

\subsection{Motivation -  Address Mapping Analysis} \label{subsec:DReAMMotivaiton}

\begin{figure}[!htb]
	\centering
	\subfloat[Mapping 1: Maximise row-buffer locality (Baseline)]{\includegraphics[scale=0.35]{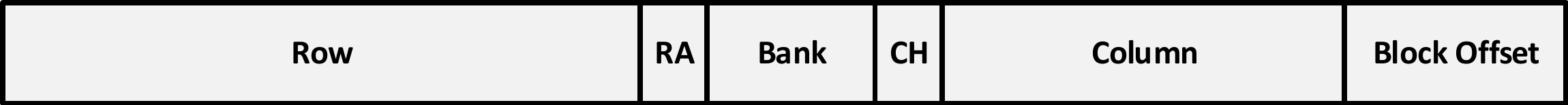}\label{fig:dream_mapping_1}} \\
	\subfloat[Mapping 2: Permutation-based Page Interleaving \cite{zhang2000permutation}]{\includegraphics[scale=0.35]{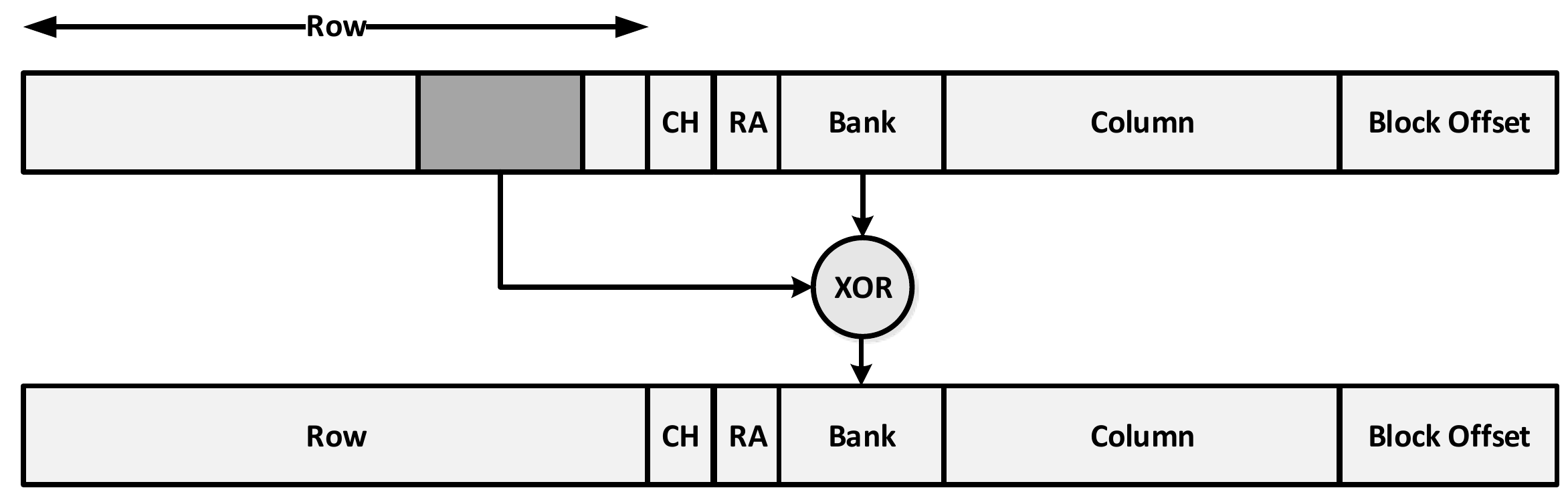}\label{fig:dream_mapping_3}} \\
	\subfloat[Mapping 3:Minimalist Open-Page Scheme \cite{kaseridis2011minimalist}]{\includegraphics[scale=0.35]{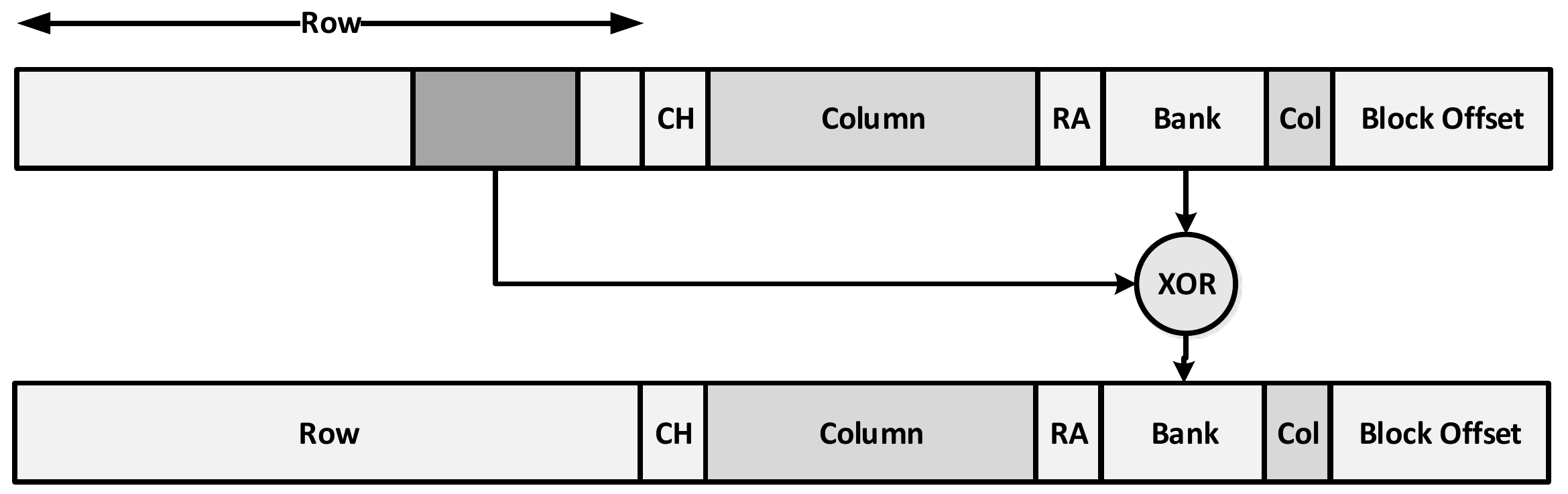}\label{fig:dream_mapping_4}} 		
	\caption{Different address mapping schemes.}	
	\label{fig:DReAM_Address_Mapping_fig}
\end{figure}

Figure \ref{fig:DReAM_Address_Mapping_fig} presents three different well-known address-mapping schemes currently employed by modern DRAM controllers. The first mapping (Figure \ref{fig:dream_mapping_1}) is a standard mapping intended to exploit the spacial locality by placing column address at bottom. The next two address interleaving policies are schemes proposed by Kaseridis~\textit{et al}.\ \cite{kaseridis2011minimalist} and Zhang~\textit{et al}.\ \cite{zhang2000permutation}. The proposed mapping by Zhang~\textit{et al}.\ XORs some of the row address bits with the bank's address bits to produce a new bank index  (Figure~\ref{fig:dream_mapping_3}). This tries to change the bank ID whenever the Row ID is changed to reduce the page conflict. Kaseridis~\textit{et al}.\ \cite{kaseridis2011minimalist} extend this technique by producing the column index using a different section of the physical address (Figure~\ref{fig:dream_mapping_4}). Both techniques aim to reduce page conflicts in DRAMs. 

\begin{figure*}
\centering
\includegraphics[scale=0.253]{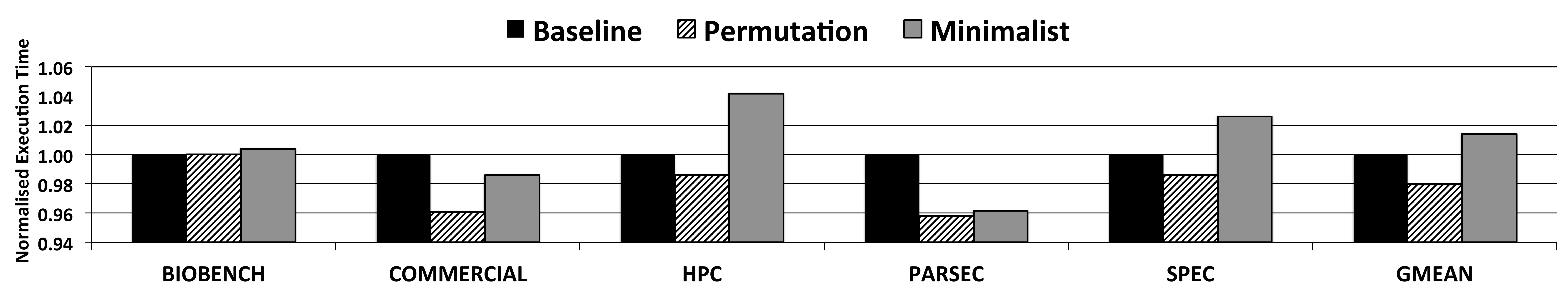}
\caption{Performance comparison among different address-mapping schemes.}
\label{fig:motivation_address_mapping}
\end{figure*}

There might be other variation of address-mapping schemes, than those presented in this figure, that can be used to perform the required translation phase to service a memory request. However, the important point to consider is that the current memory controllers can only use one of such address-mapping schemes to translate the physical address to the internal structure of DRAMs. Moreover, modern DRAM controllers are limited to perform read/write operations in bursts (typically bursts of 4 or 8 items). This implies some bits are used as a block offset, presented in Figure \ref{fig:DReAM_Address_Mapping_fig}. To motivate the technique presented in this paper, Figures \ref{fig:motivation_address_mapping} presents the performance comparison of different address-mapping schemes for all the benchmark suites evaluated in this work. Each bar in these graphs represents the normalized execution time to the baseline address-mapping scheme (address mapping 1 in Figure \ref{fig:DReAM_Address_Mapping_fig}). Our experimental results (considering the results of an individual workload) suggest that a predefined address mapping schemes is not efficient in all situations and thus employing a fixed address mapping scheme cannot deliver the best execution time across all workloads. As Figure \ref{fig:motivation_address_mapping} suggests, the permutation-based address mapping almost always (except for the BIOBENCH benchmark) delivers a better geometric average (GMEAN) execution time compared with other two address mapping schemes. This address mapping is chosen as the best baseline of those presented in this paper to be compared against DReAM mapping.

%% file: DReAM.tex
\section{DReAM: Dynamic Rearrangement of Address Mapping}
\label{cha:DReAM}

DReAM is a novel technique to analyze the memory access pattern (produced either by single or multithreaded applications) at run-time and estimate an efficient address-mapping scheme, that reduces page conflicts and improves page hits. DReAM consists of two main phases: `online prediction of address mapping' and `on-the-fly data migration'.

\subsection{Online Prediction of Address Mapping} \label{subsec:OnlinePrediction}

\begin{figure}[!htb]
\centering
\includegraphics[scale=0.3]{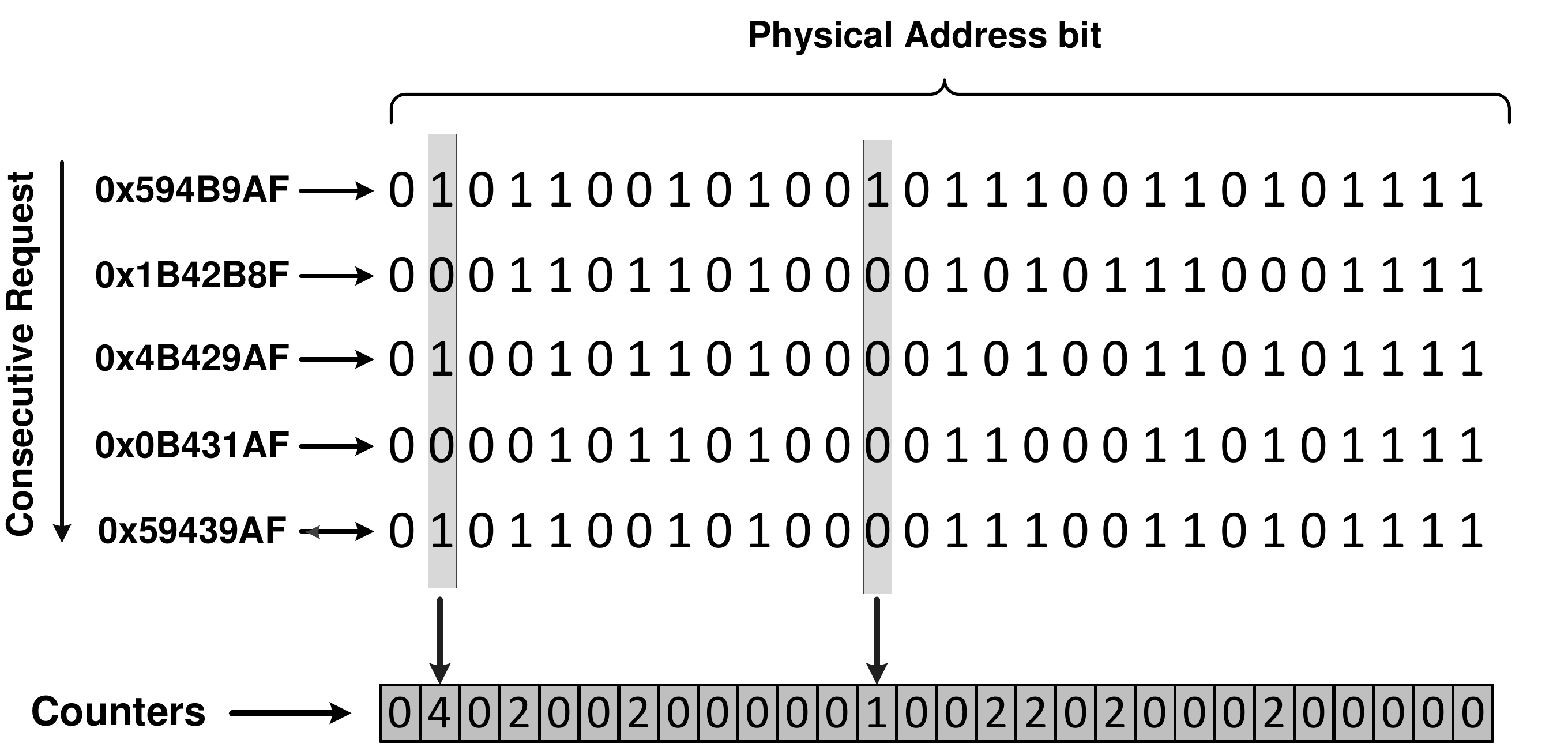}
\caption{Bit-counters mechanism. }
\label{fig:bit_counters}
\end{figure}

The first step is to discover whether the current workload, a set of executing applications, is a good match with the baseline address mapping scheme. A baseline address-mapping scheme decides which physical address bits should be used to address which specific part of a DRAM device (e.g.\ rank, bank, row, etc.). Therefore, a physical address is divided in to different sets of bits each set pointing to a specific part of internal hierarchy of the DRAM system. Considering consecutive requests to a DRAM module, the changing rate of each physical address bit (as a result of the changing rate of each bit within different sets) in comparison with the previous access has a strong correlation with the changing rate of a specific DRAM location that has been accessed. On the other hand, accessing different rows within the same bank causes page conflicts and imposes a power and performance overhead. Therefore, ideally, it is desired to keep the change rate of the physical address bits that are used to address the row, as low as possible to reduce the row switches within a bank. DReAM estimates how much each physical address bit changes by observing memory requests over a period of time as a means of generating improved memory mappings. The estimations of change per bit require minimum extra hardware; one counter per physical address bit per memory controller. Those bits changing the most have higher entropy and those bits changing the least have smaller entropy. For a given period, these counters (or frequency change estimators) keep track of the number of changes of each bit of the physical address in comparison with the previous memory address request. The given period creates time windows and can be based on number of clock cycles or number of memory requests. Figure~\ref{fig:bit_counters} shows an example of five consecutive accesses to demonstrate the function of these counters. 


\renewcommand{\thesubfigure}{.\arabic{subfigure}}

\begin{figure*}
	\centering
	\subfloat[Comm1]{\includegraphics[scale=0.2]{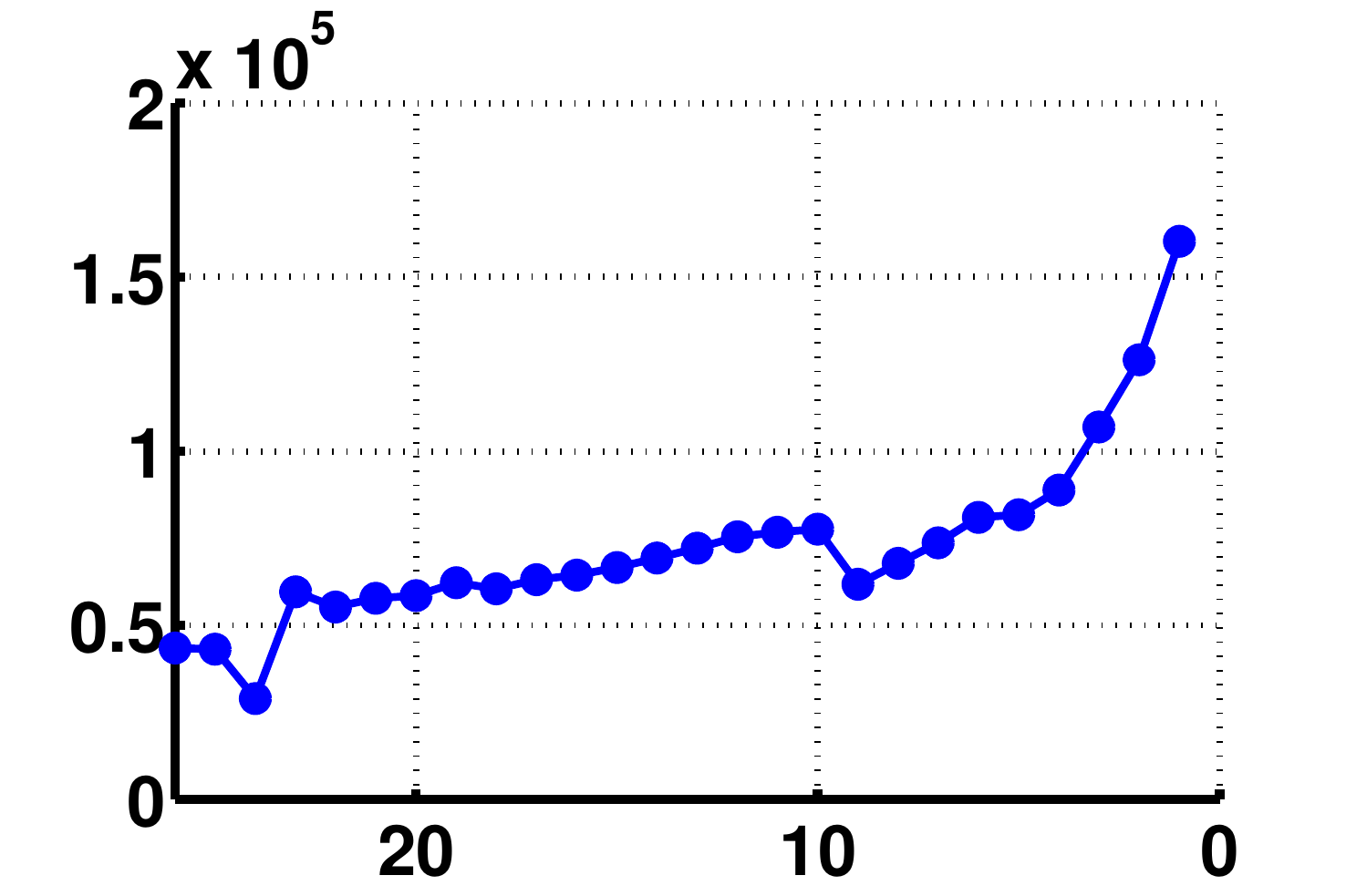}\label{fig:dream_comm1_pat}} 
	\subfloat[Comm2]{\includegraphics[scale=0.2]{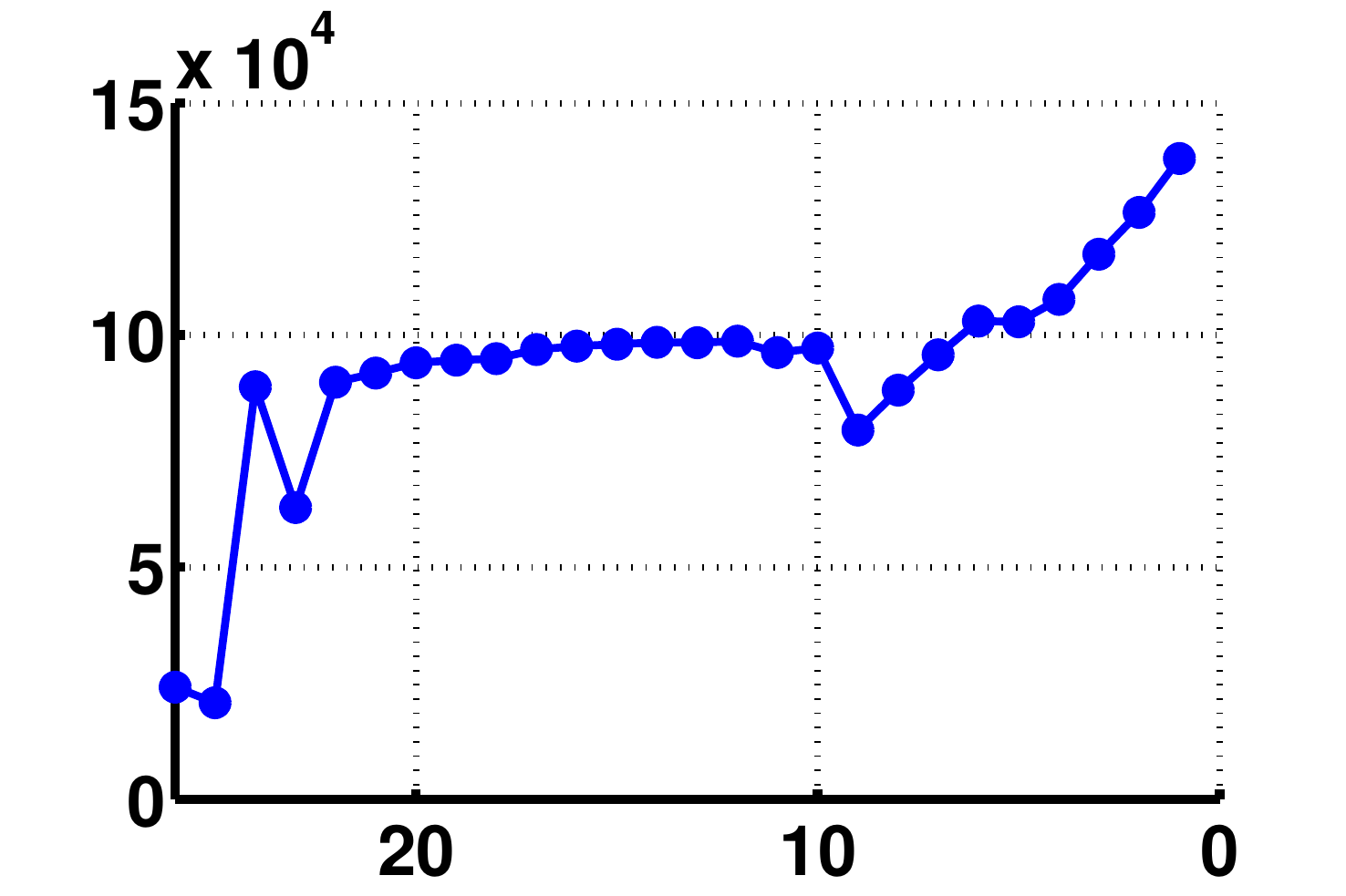}\label{fig:dream_comm2_pat}} 
	\subfloat[Comm3]{\includegraphics[scale=0.2]{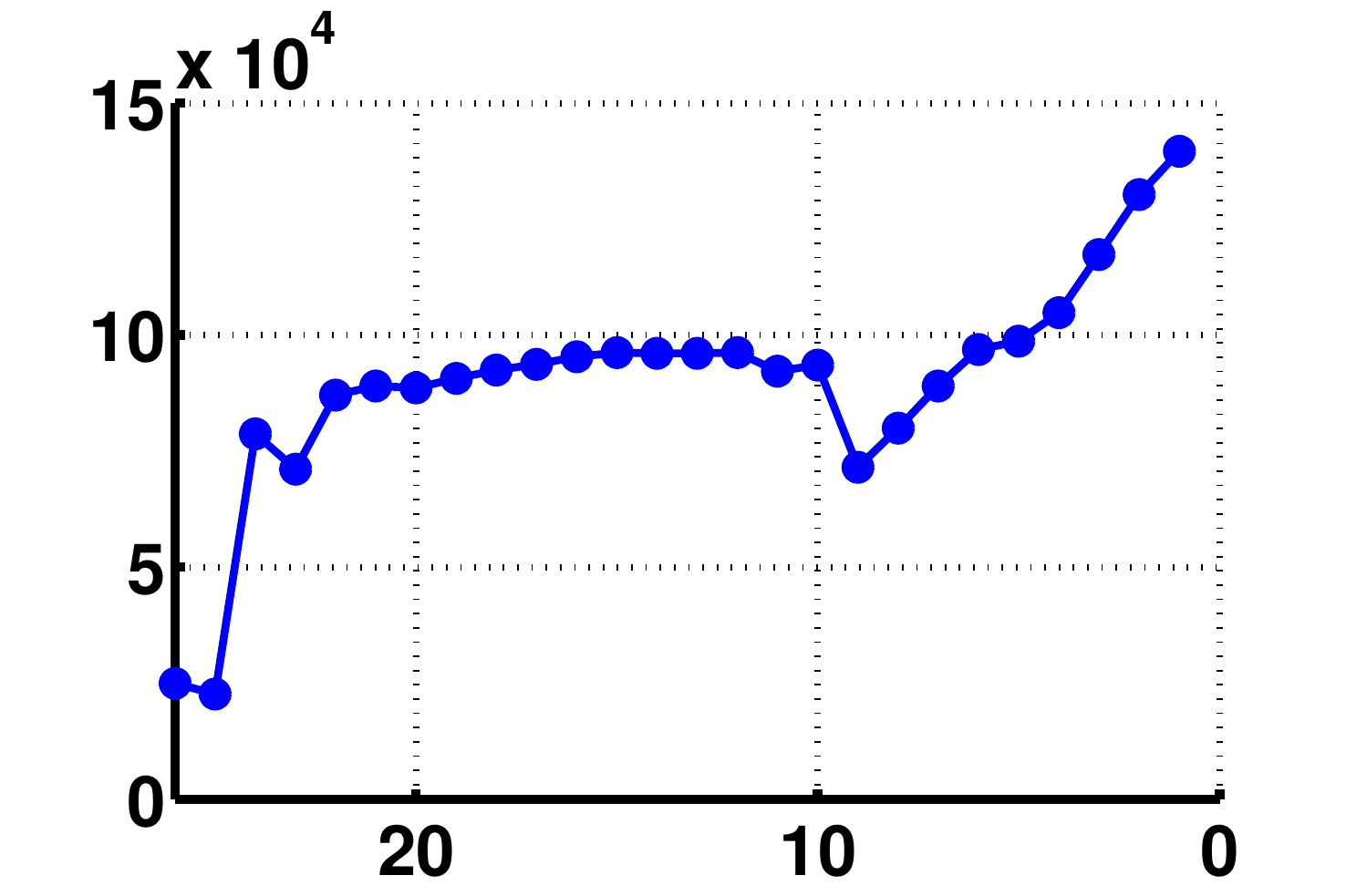}\label{fig:dream_comm3_pat}} 
	\subfloat[Comm4]{\includegraphics[scale=0.2]{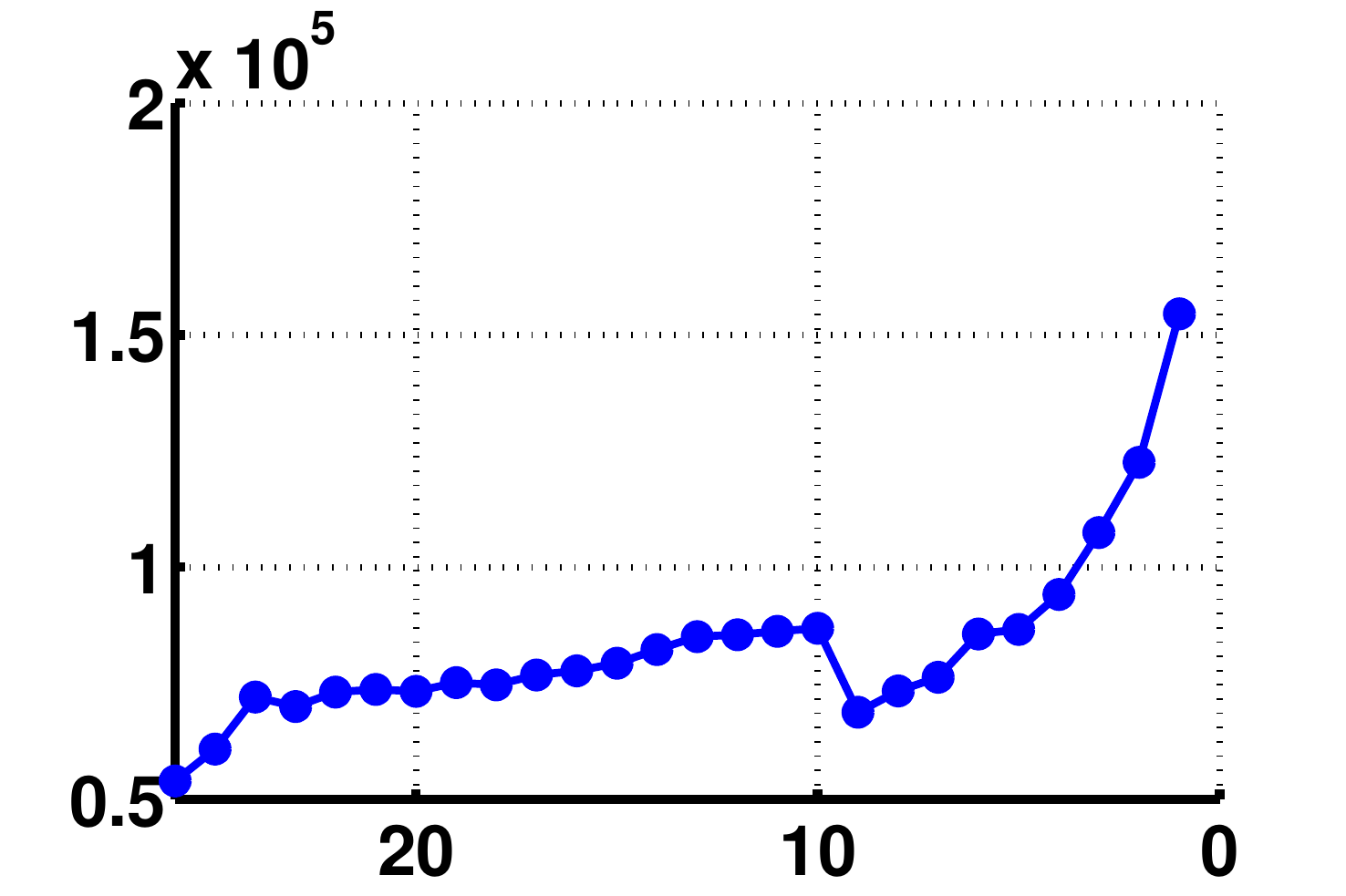}\label{fig:dream_comm4_pat}} 
	\subfloat[Comm5]{\includegraphics[scale=0.2]{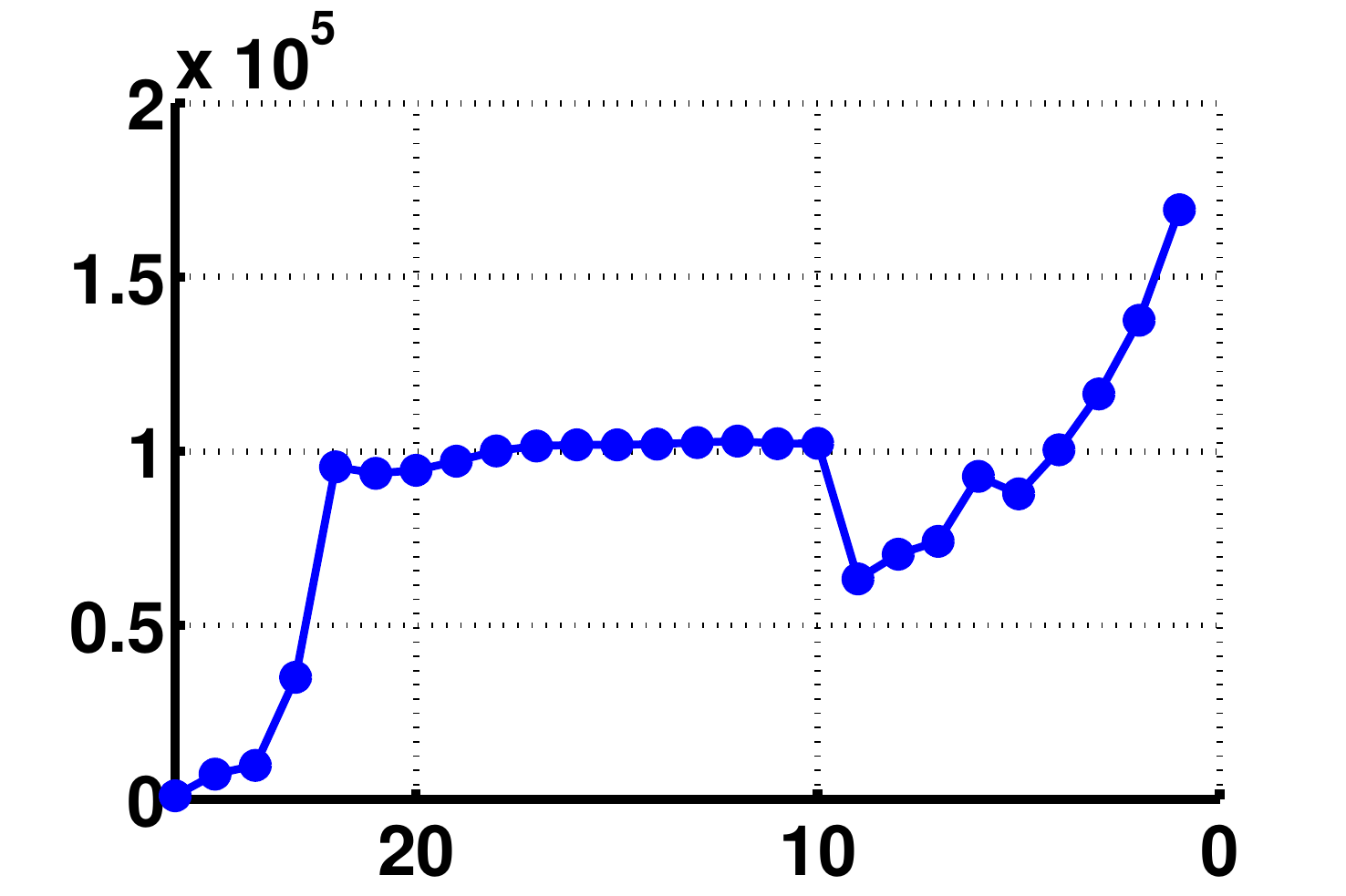}\label{fig:dream_comm5_pat}} 
	\subfloat[hpc1]{\includegraphics[scale=0.2]{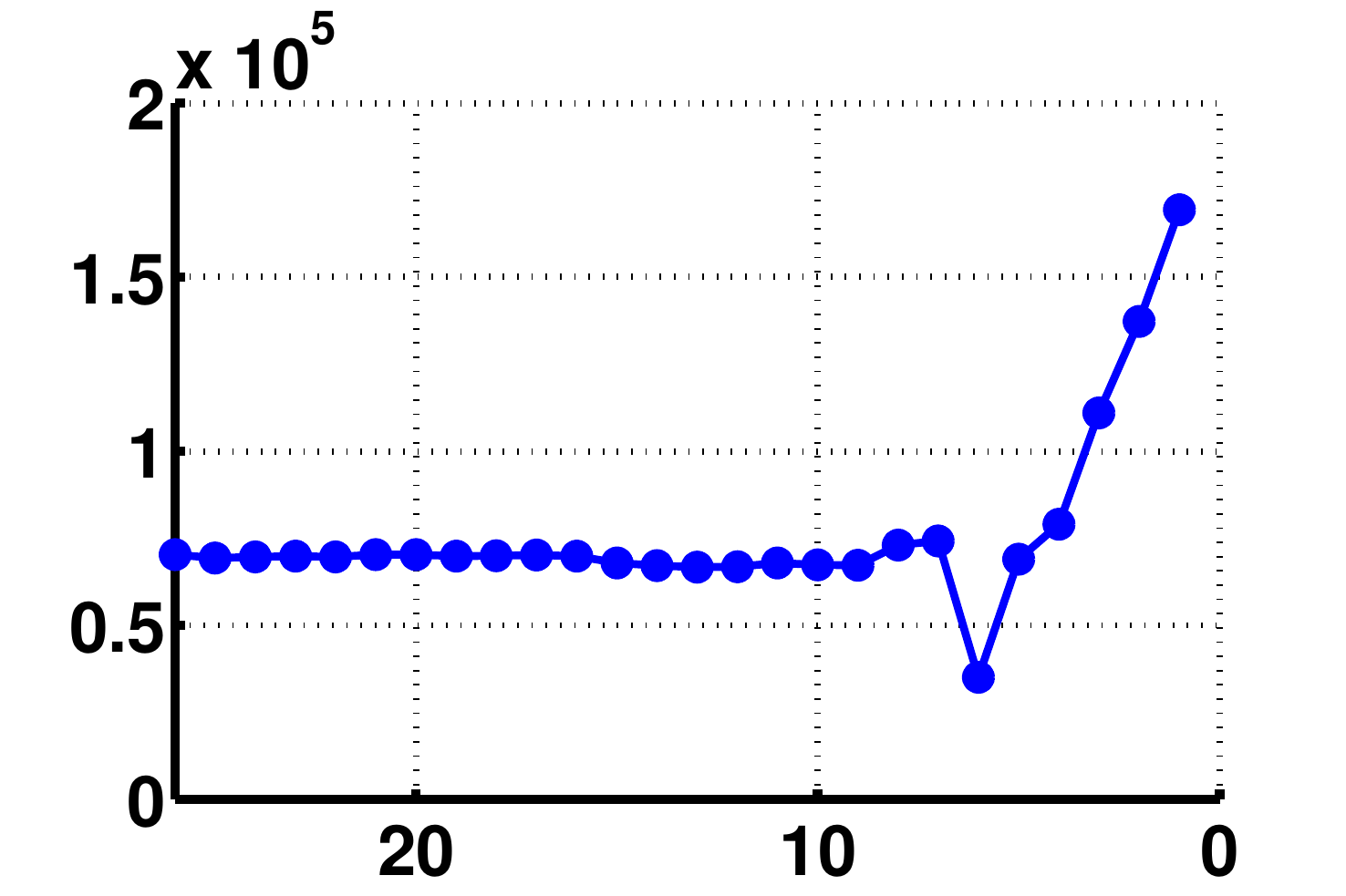}\label{fig:dream_hpc1_pat}}\\ 
	\subfloat[hpc2]{\includegraphics[scale=0.2]{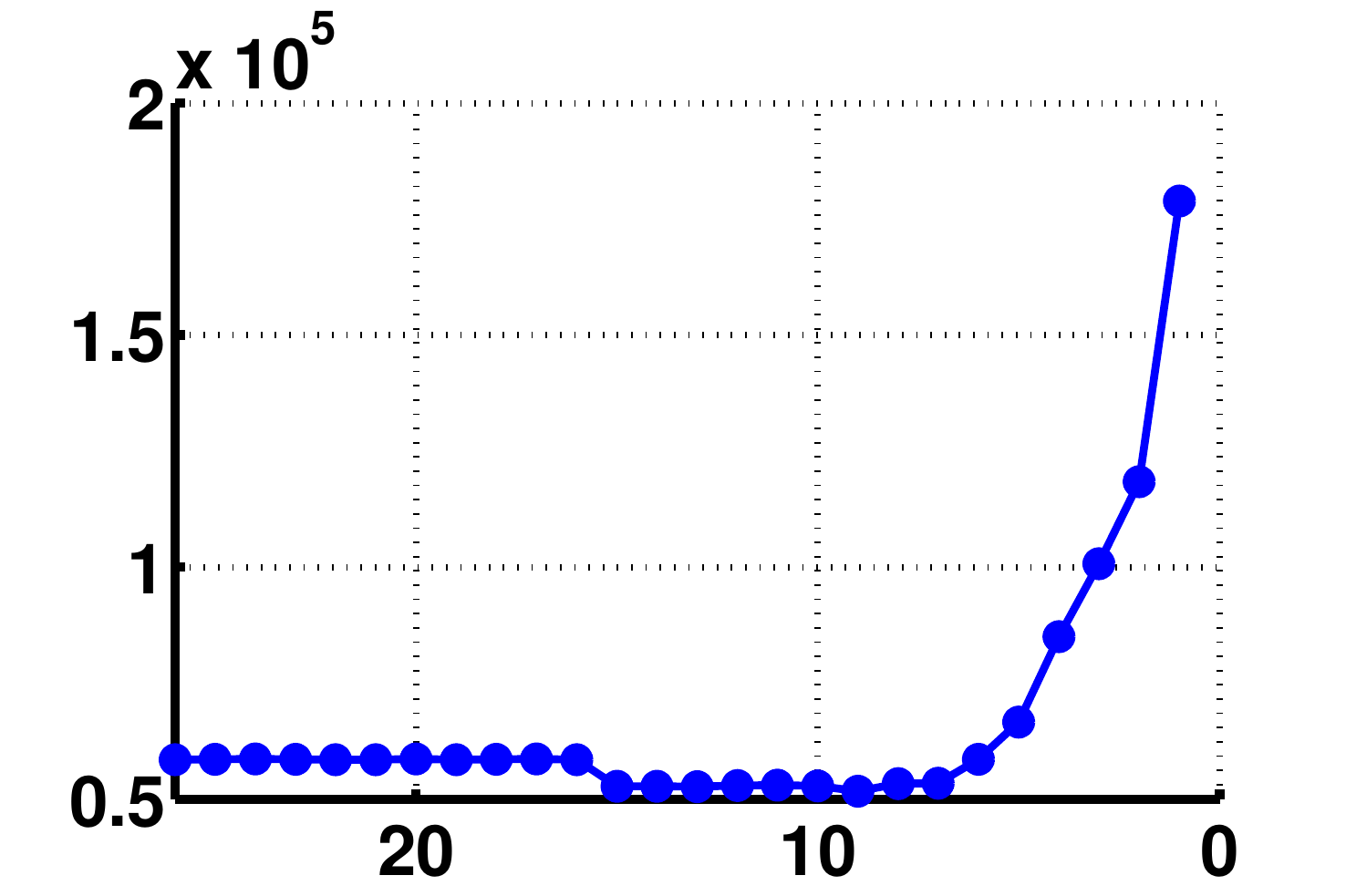}\label{fig:dream_hpc2_pat}} 
	\subfloat[hpc3]{\includegraphics[scale=0.2]{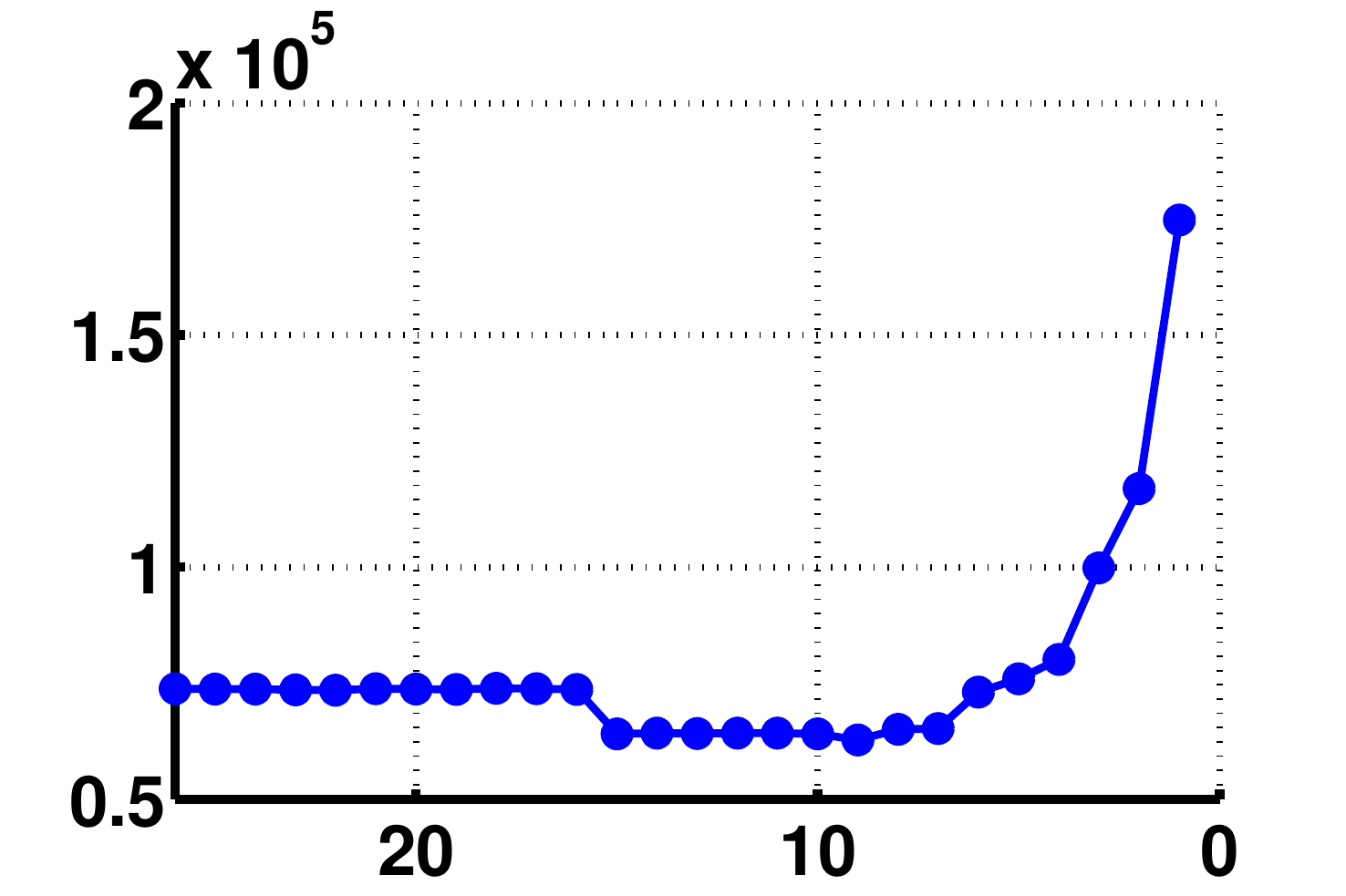}\label{fig:dream_hpc3_pat}} 
	\subfloat[hpc4]{\includegraphics[scale=0.2]{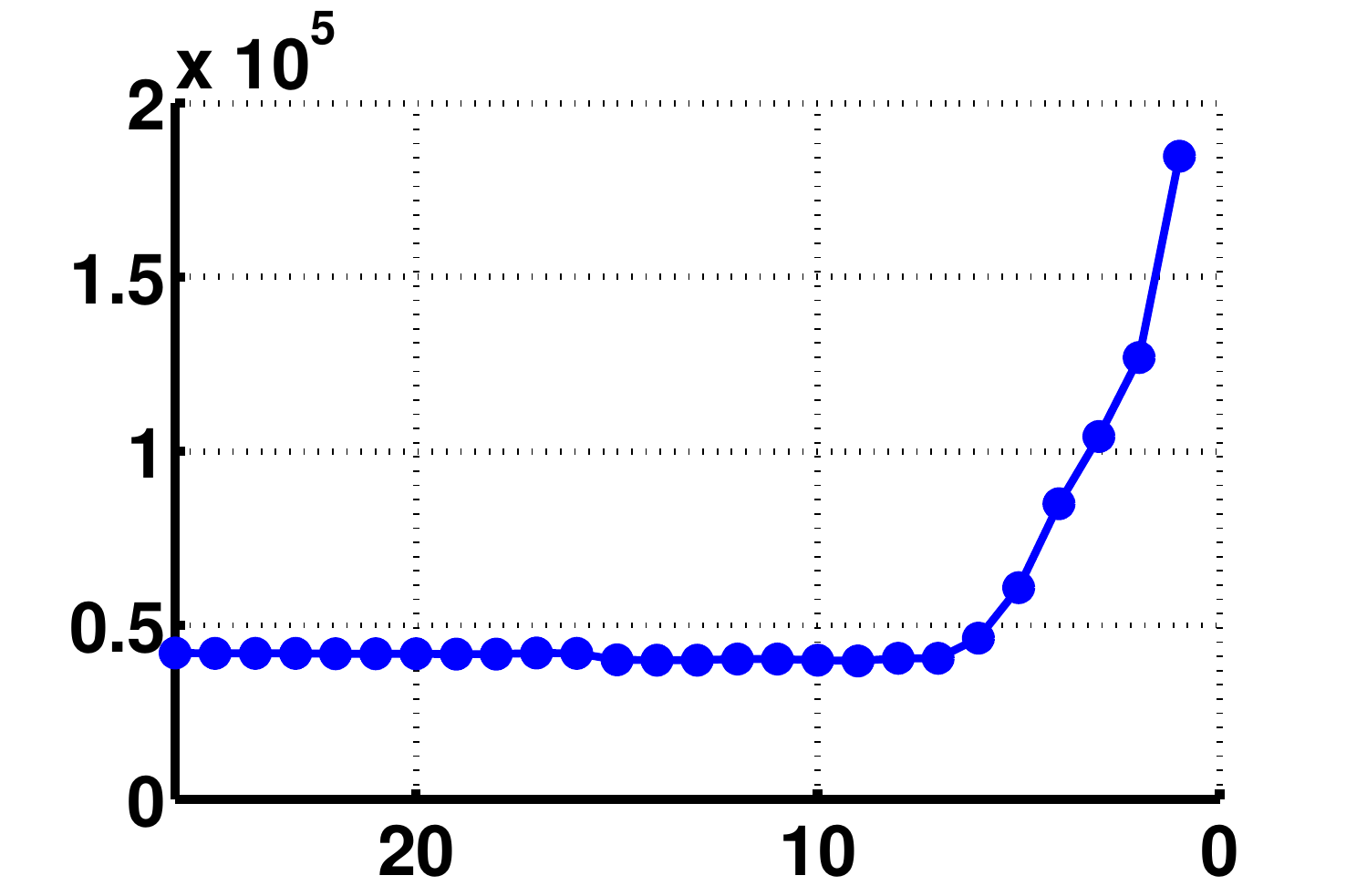}\label{fig:dream_hpc4_pat}} 	
	\subfloat[hpc5]{\includegraphics[scale=0.2]{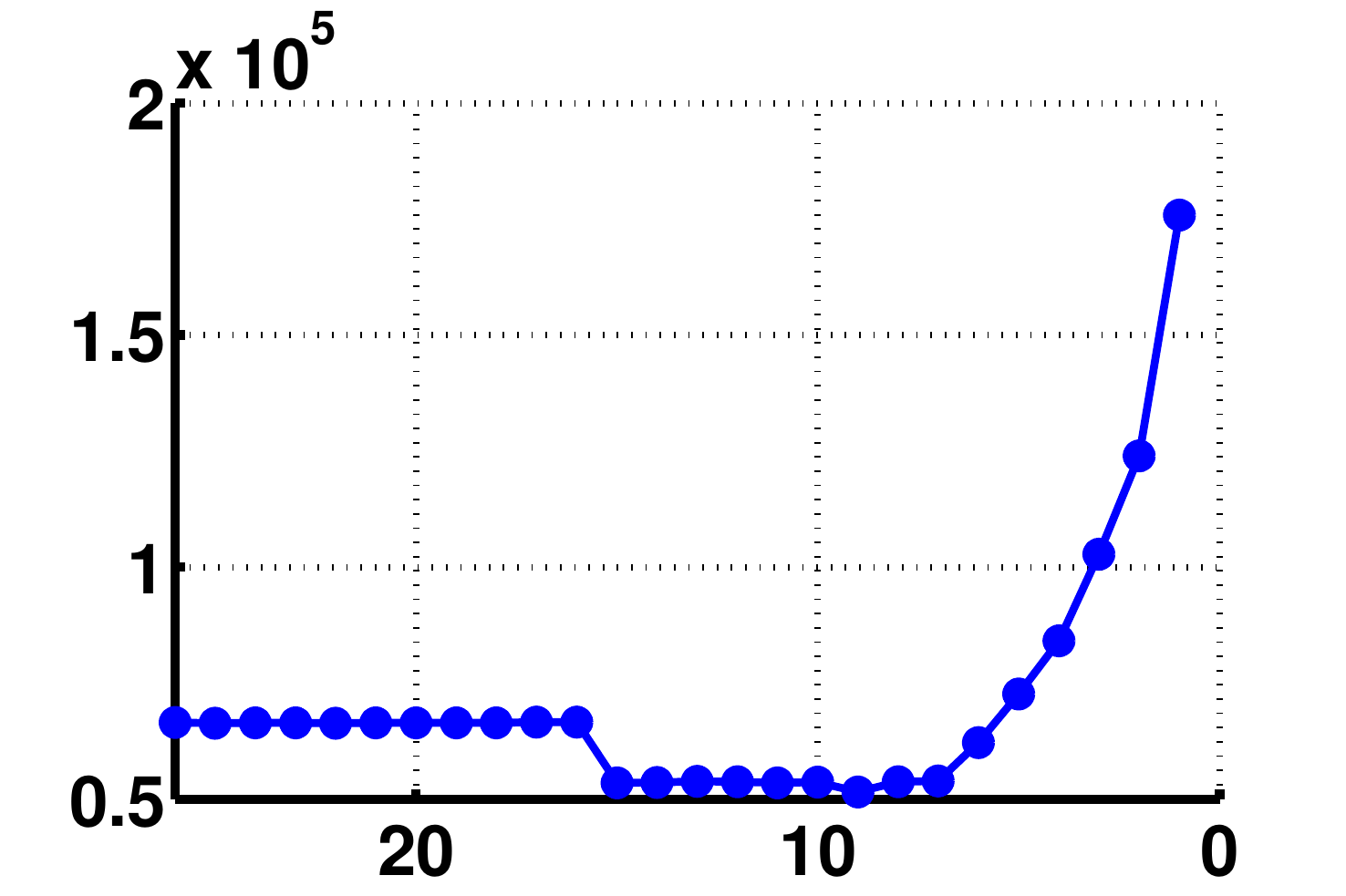}\label{fig:dream_hpc5_pat}} 	
	\subfloat[hpc6]{\includegraphics[scale=0.2]{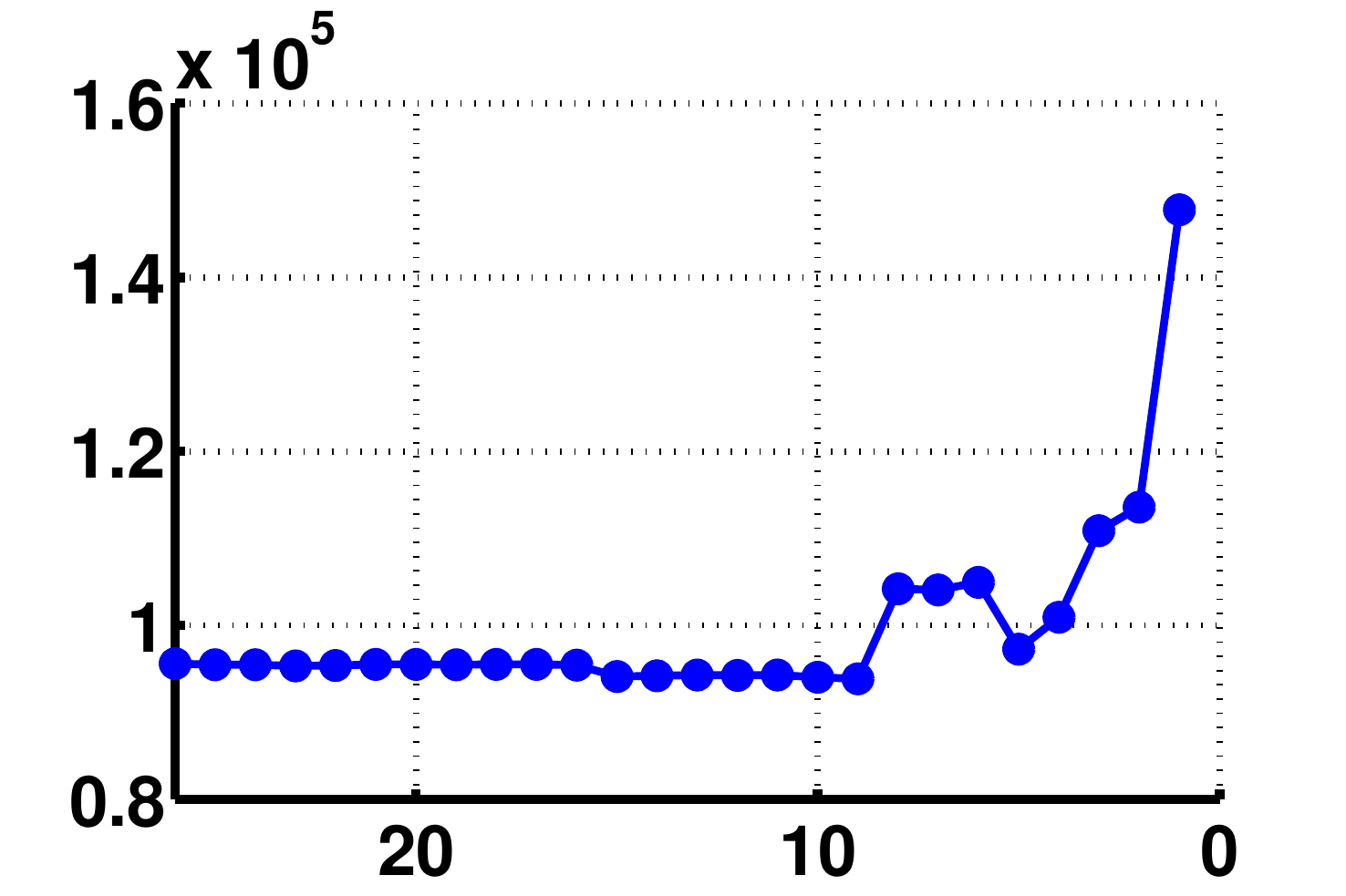}\label{fig:dream_hpc6_pat}} 
	\subfloat[hpc7]{\includegraphics[scale=0.2]{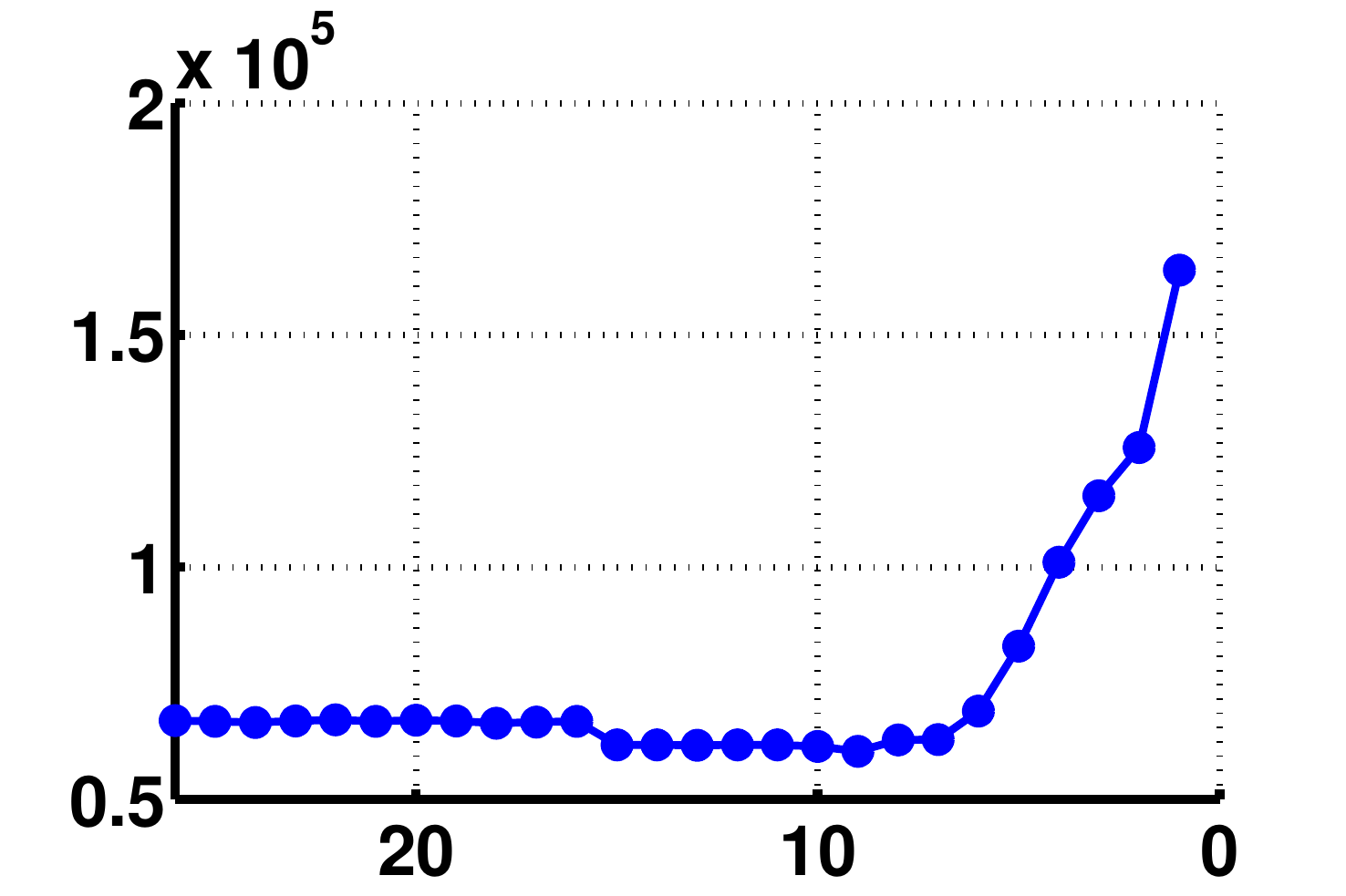}\label{fig:dream_hpc7_pat}}\\ 
	\subfloat[hpc8]{\includegraphics[scale=0.2]{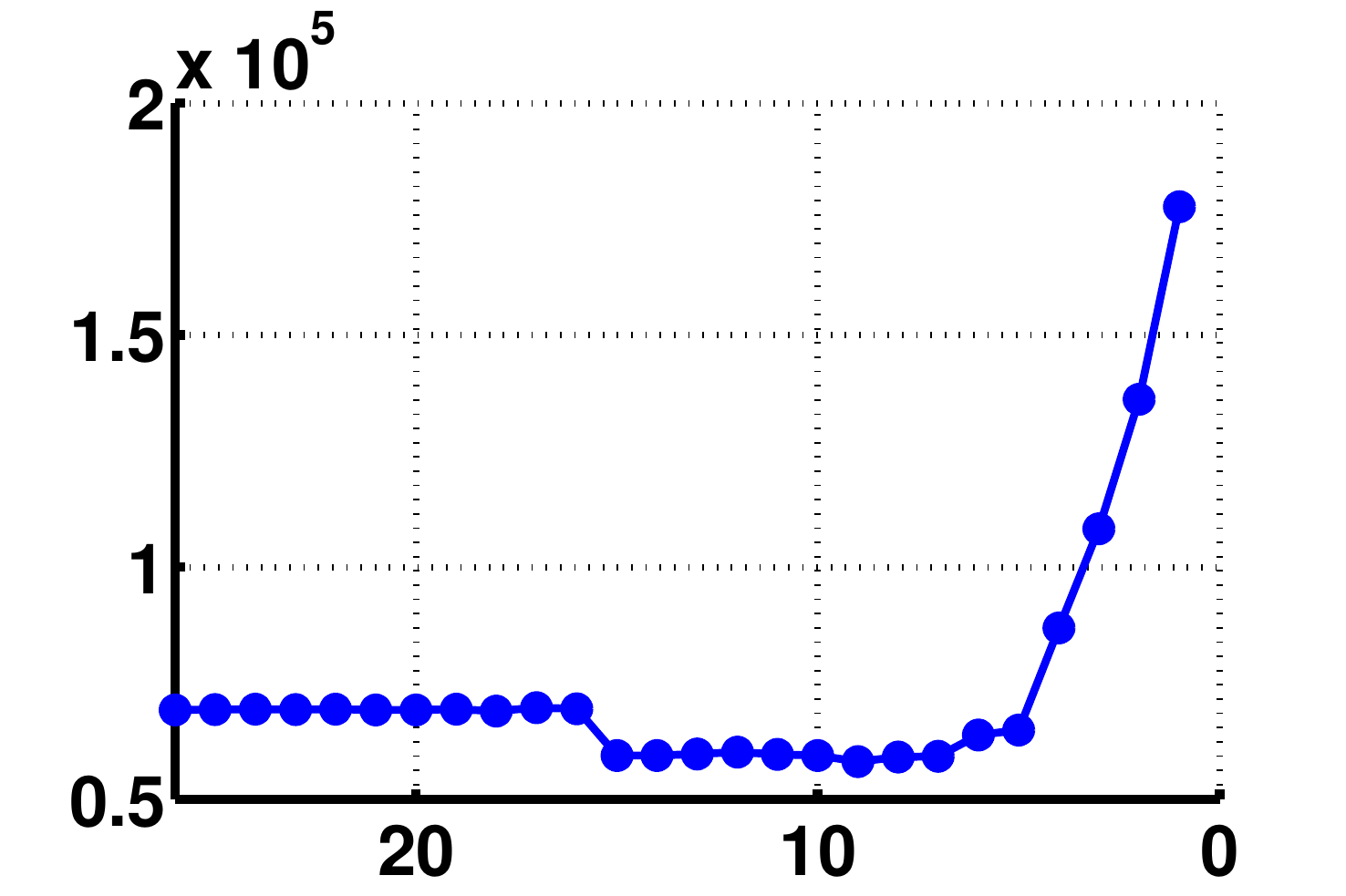}\label{fig:dream_hpc8_pat}} 	
	\subfloat[hpc9]{\includegraphics[scale=0.2]{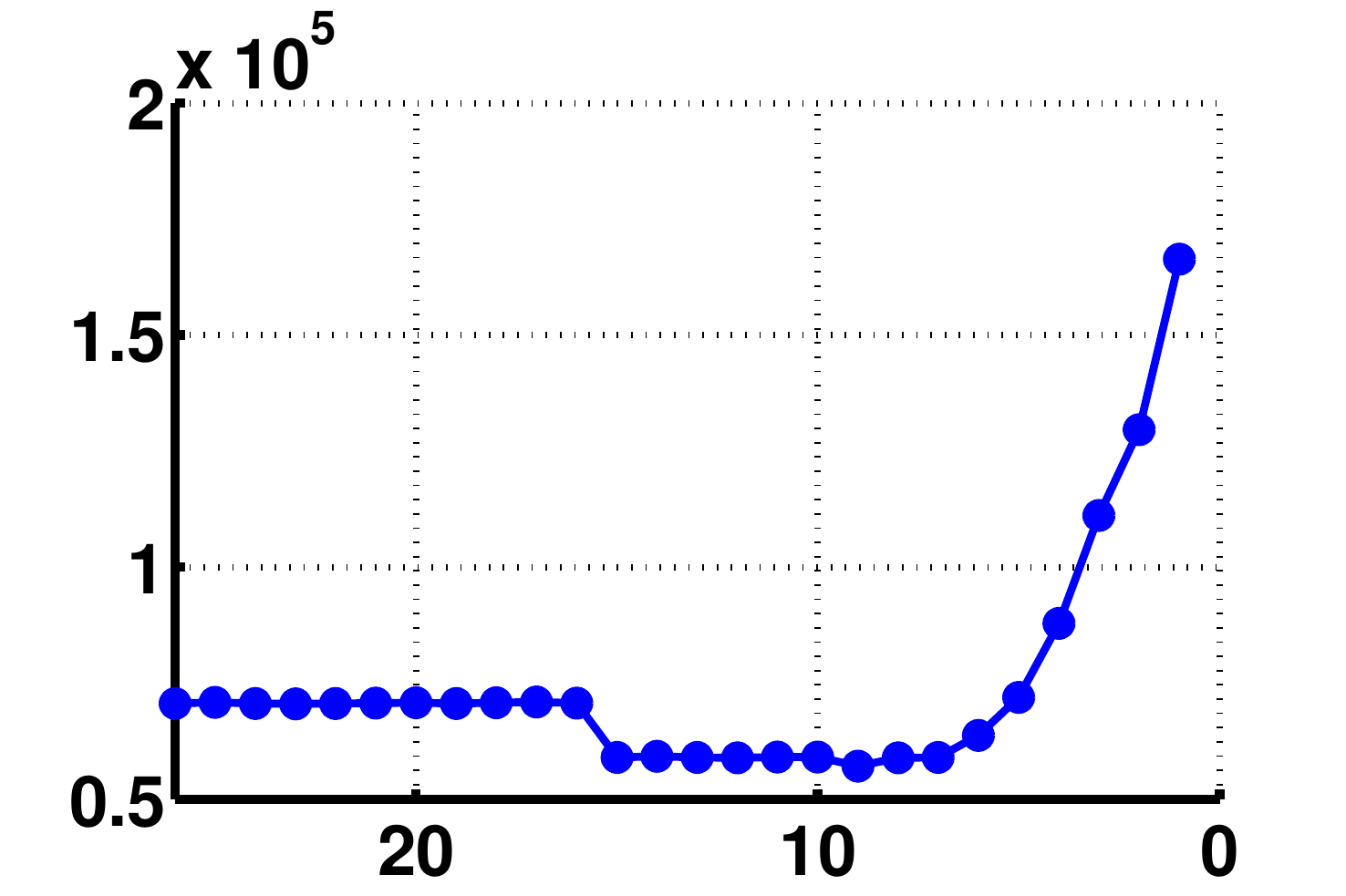}\label{fig:dream_hpc9_pat}} 
	\subfloat[hpc10]{\includegraphics[scale=0.2]{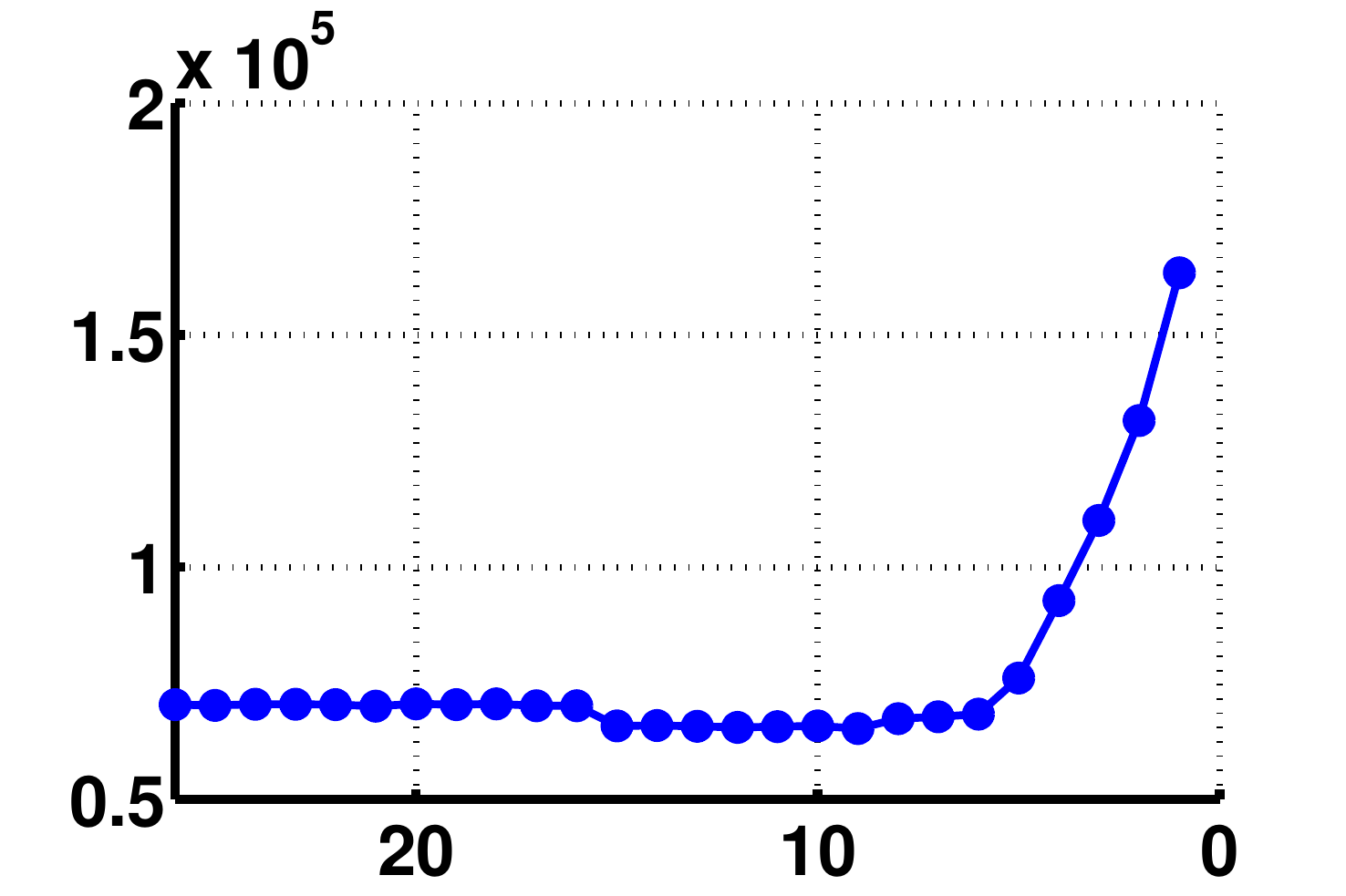}\label{fig:dream_hpc10_pat}} 
	\subfloat[hpc11]{\includegraphics[scale=0.2]{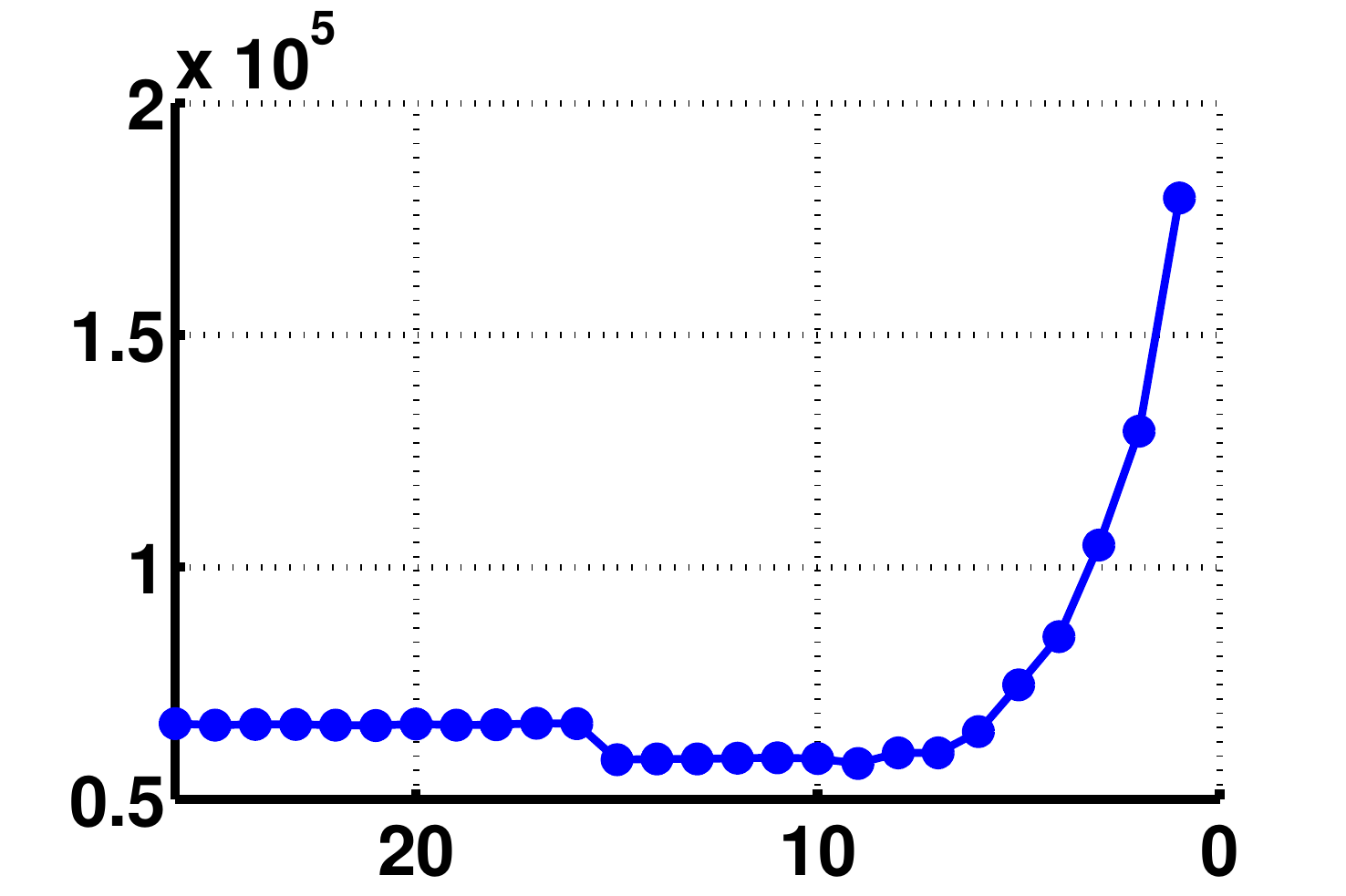}\label{fig:dream_hpc11_pat}} 
	\subfloat[hpc12]{\includegraphics[scale=0.2]{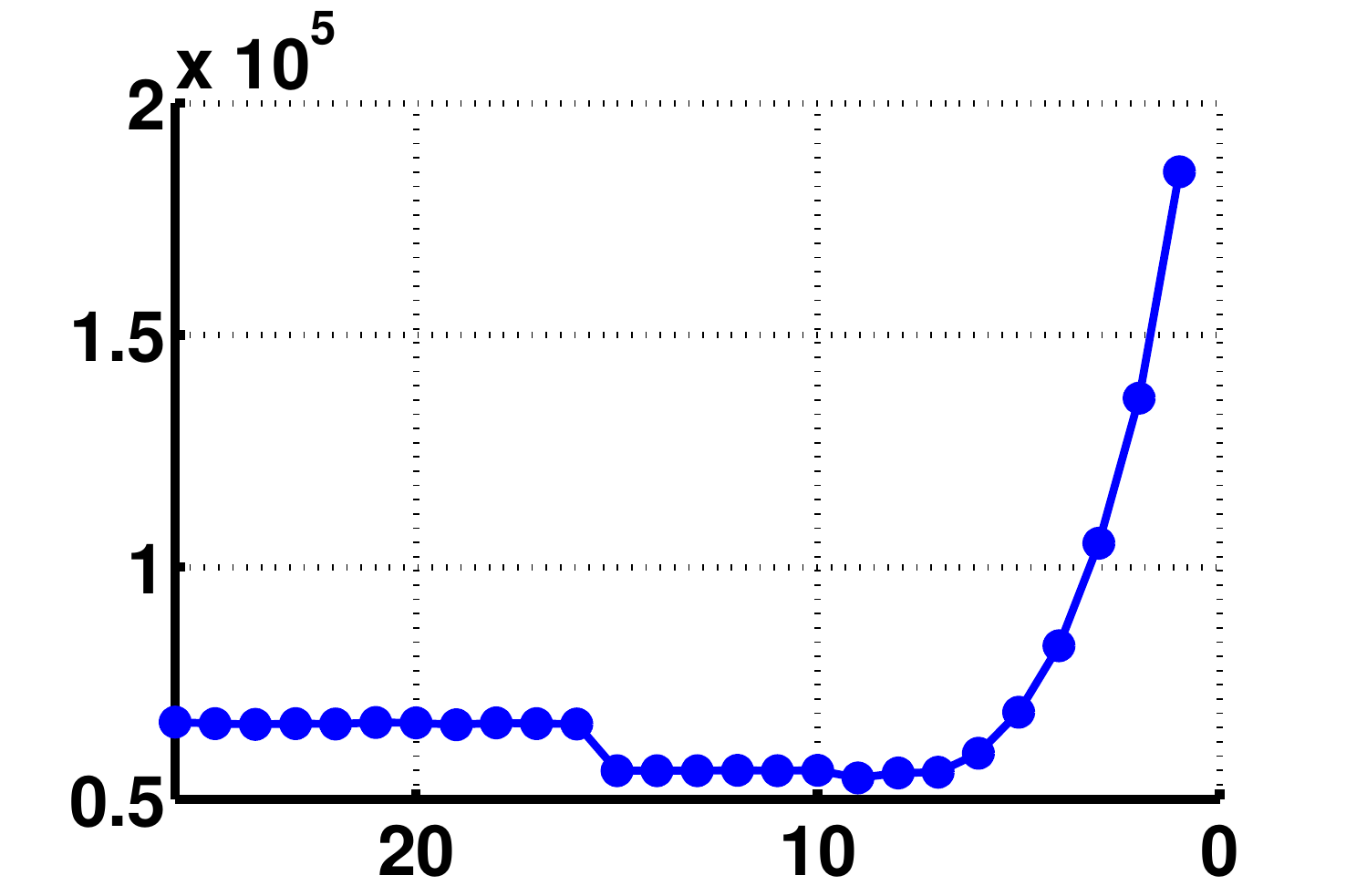}\label{fig:dream_hpc12_pat}} 	
	\subfloat[hpc13]{\includegraphics[scale=0.2]{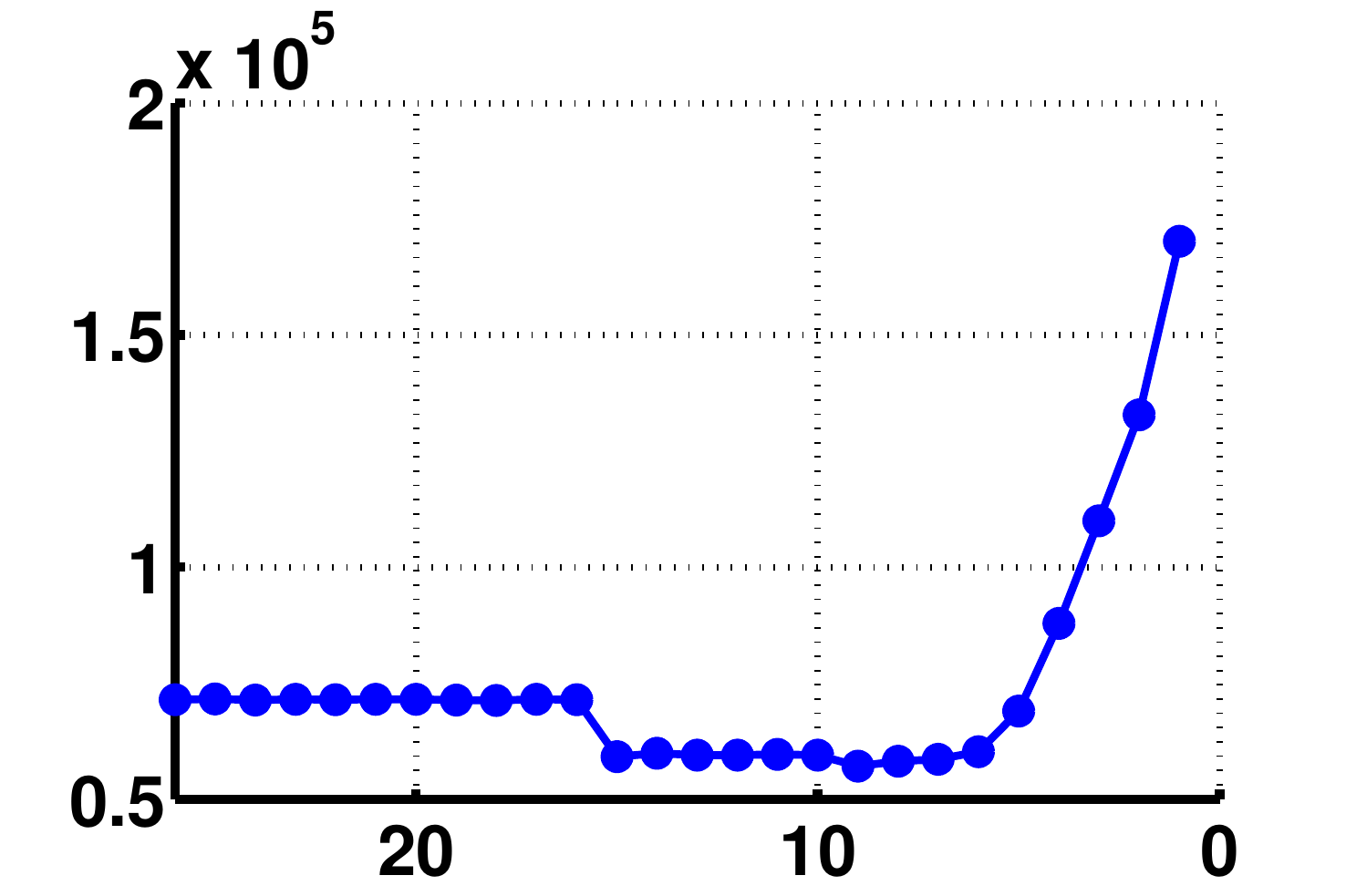}\label{fig:dream_hpc13_pat}}\\
	\subfloat[black]{\includegraphics[scale=0.2]{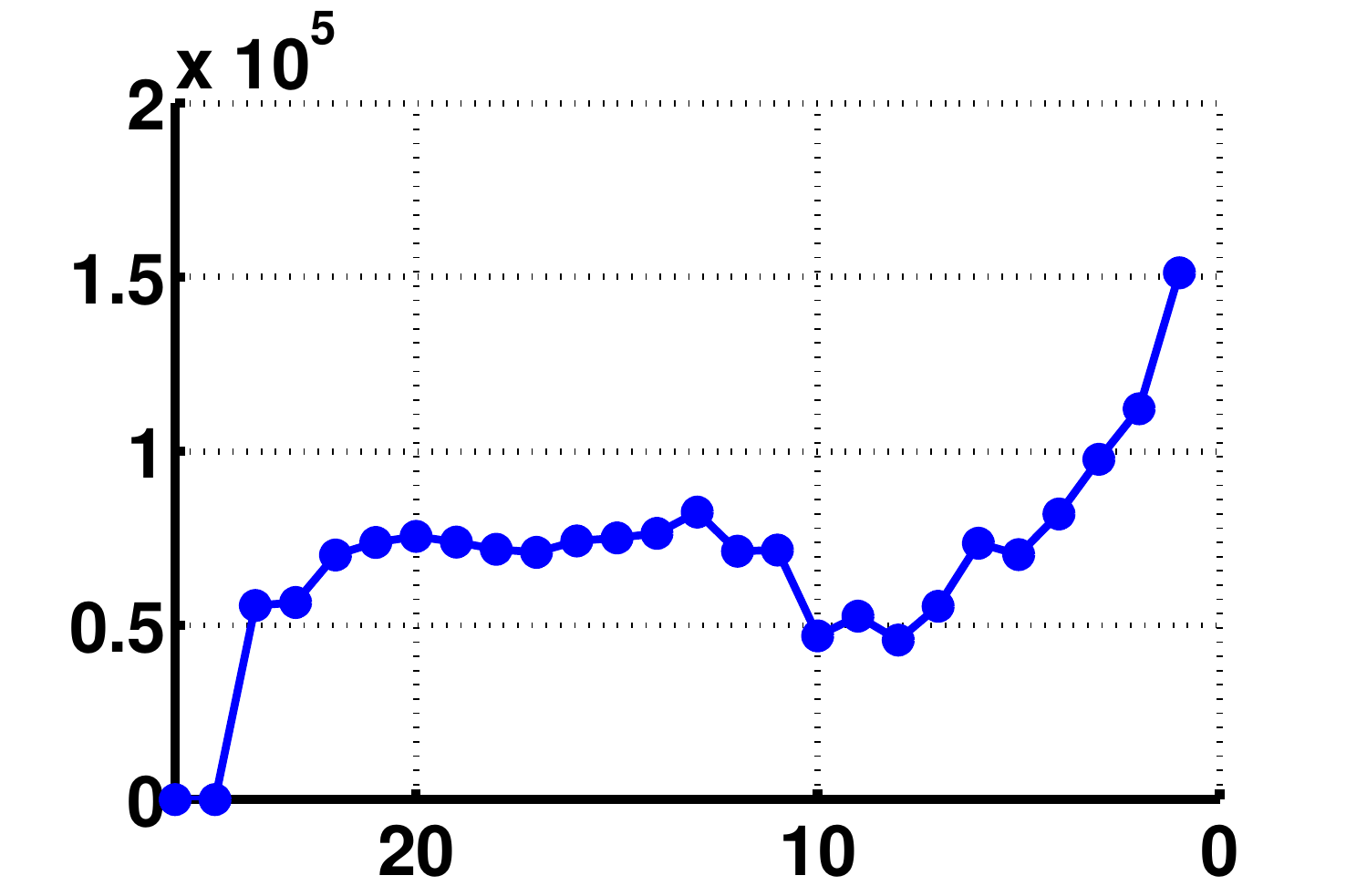}\label{fig:dream_black_pat}} 
	\subfloat[caneal]{\includegraphics[scale=0.2]{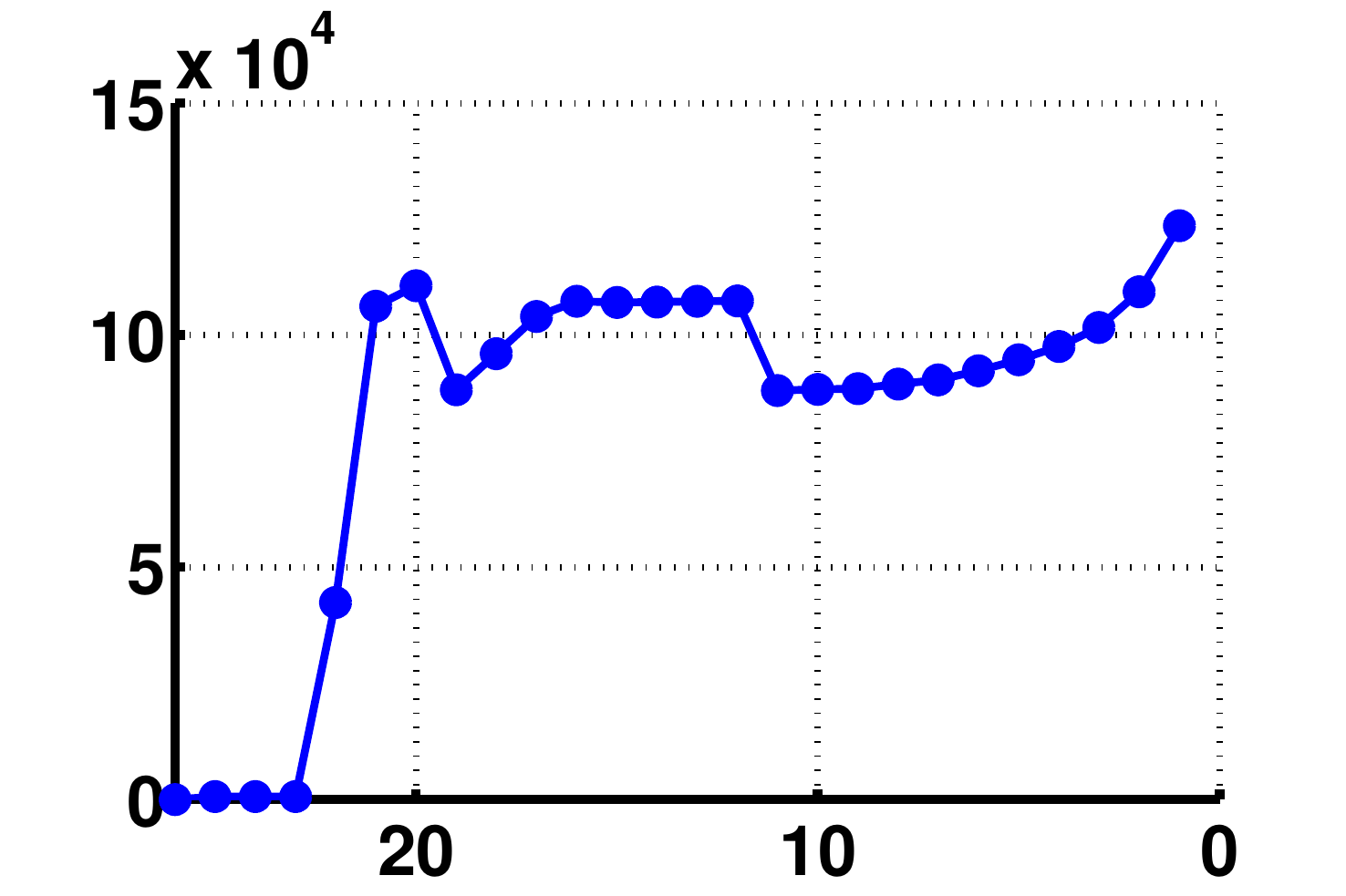}\label{fig:dream_caneal_pat}} 
	\subfloat[face]{\includegraphics[scale=0.2]{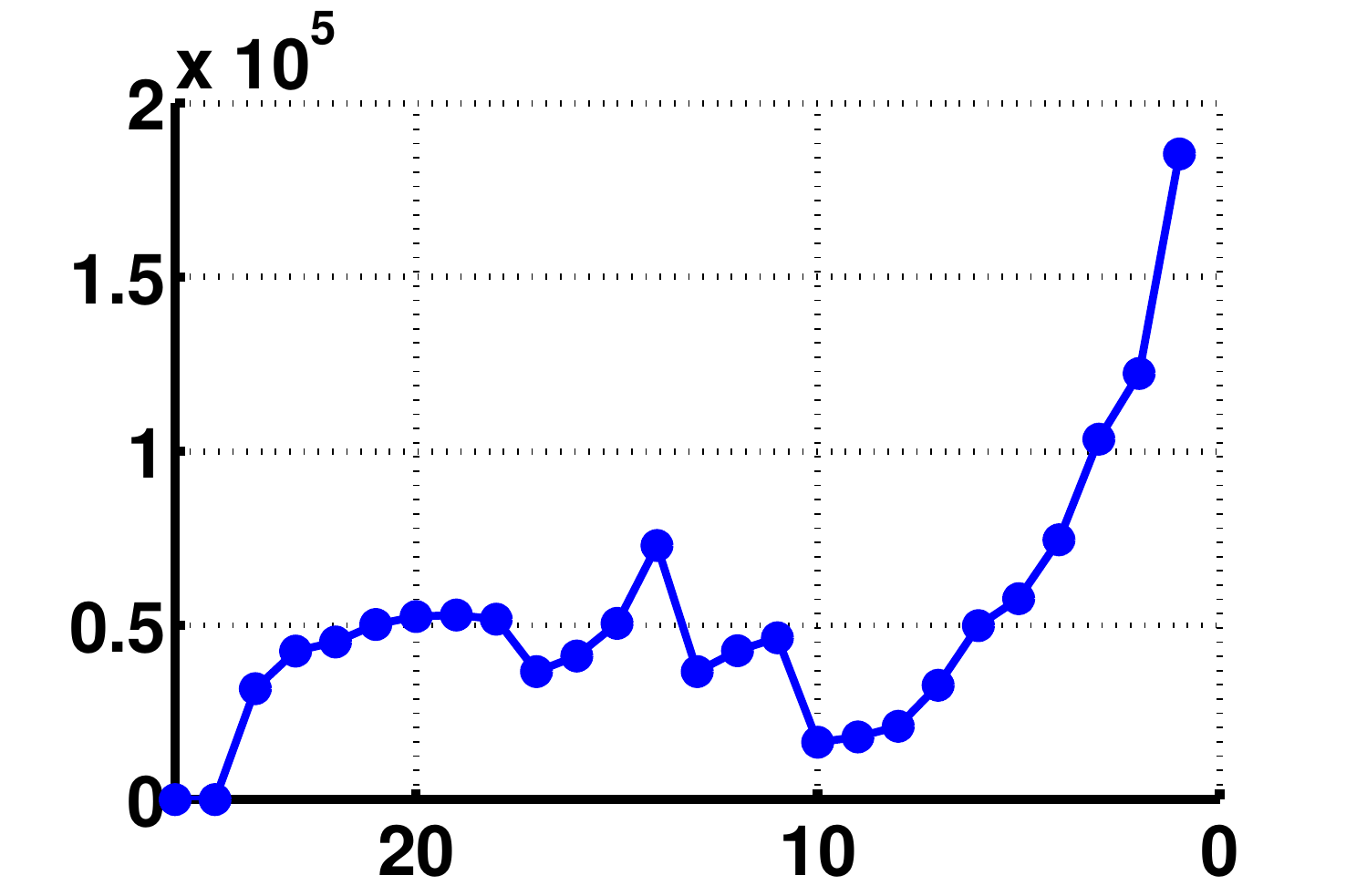}\label{fig:dream_face_pat}} 
	\subfloat[ferret]{\includegraphics[scale=0.2]{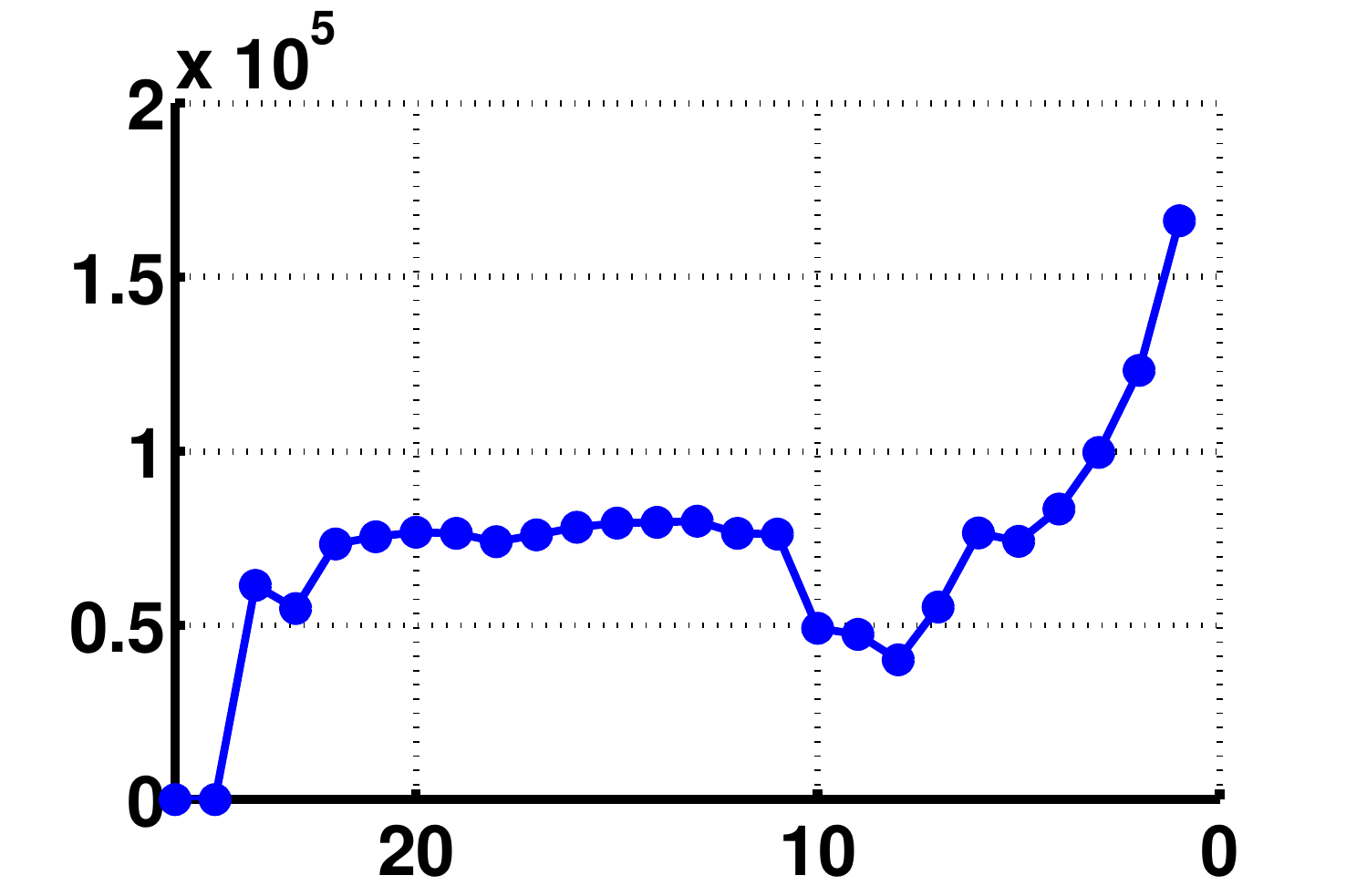}\label{fig:dream_ferret_pat}}	
	\subfloat[fluid]{\includegraphics[scale=0.2]{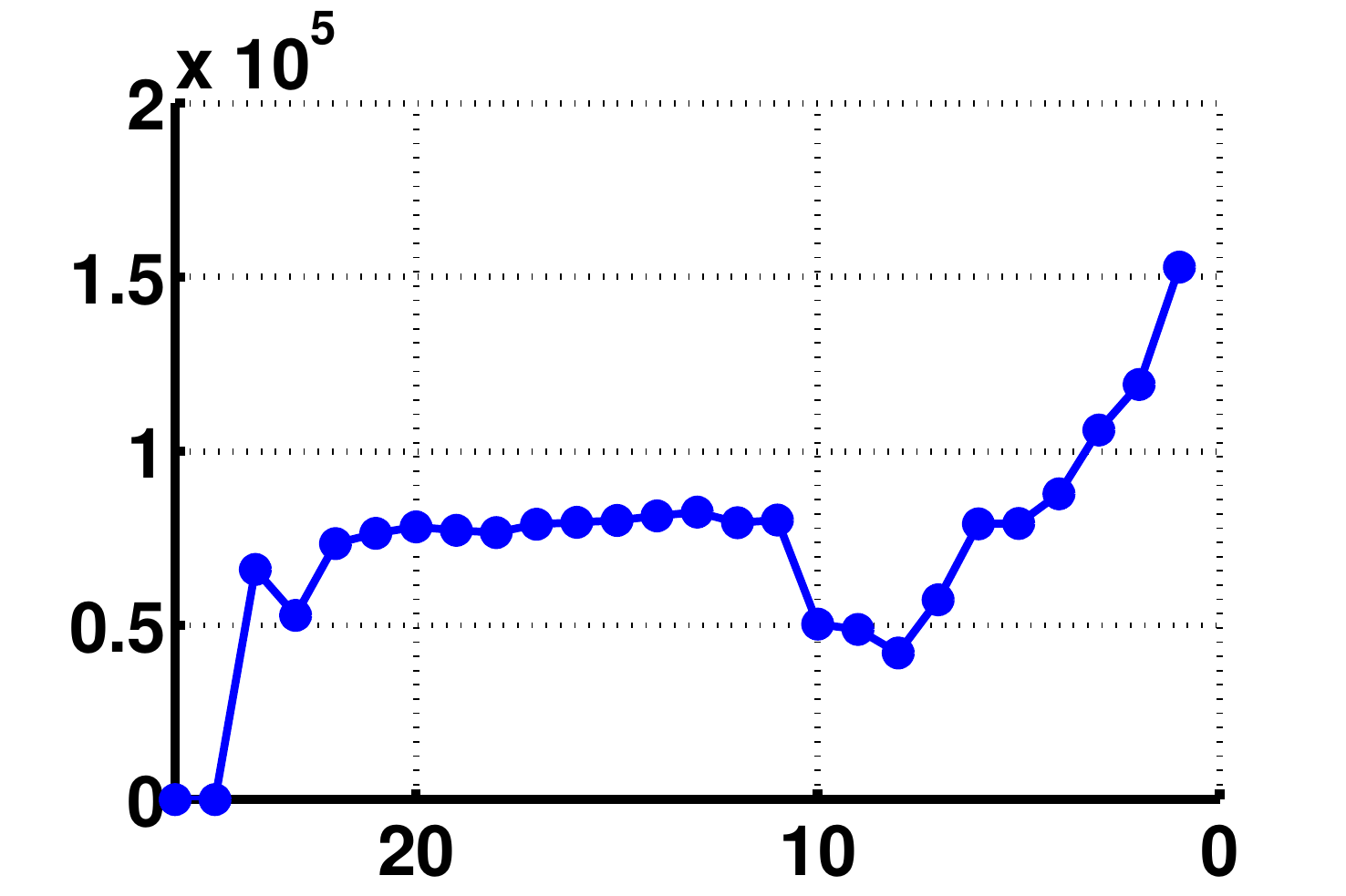}\label{fig:dream_fluid_pat}} 	
	\subfloat[freq]{\includegraphics[scale=0.2]{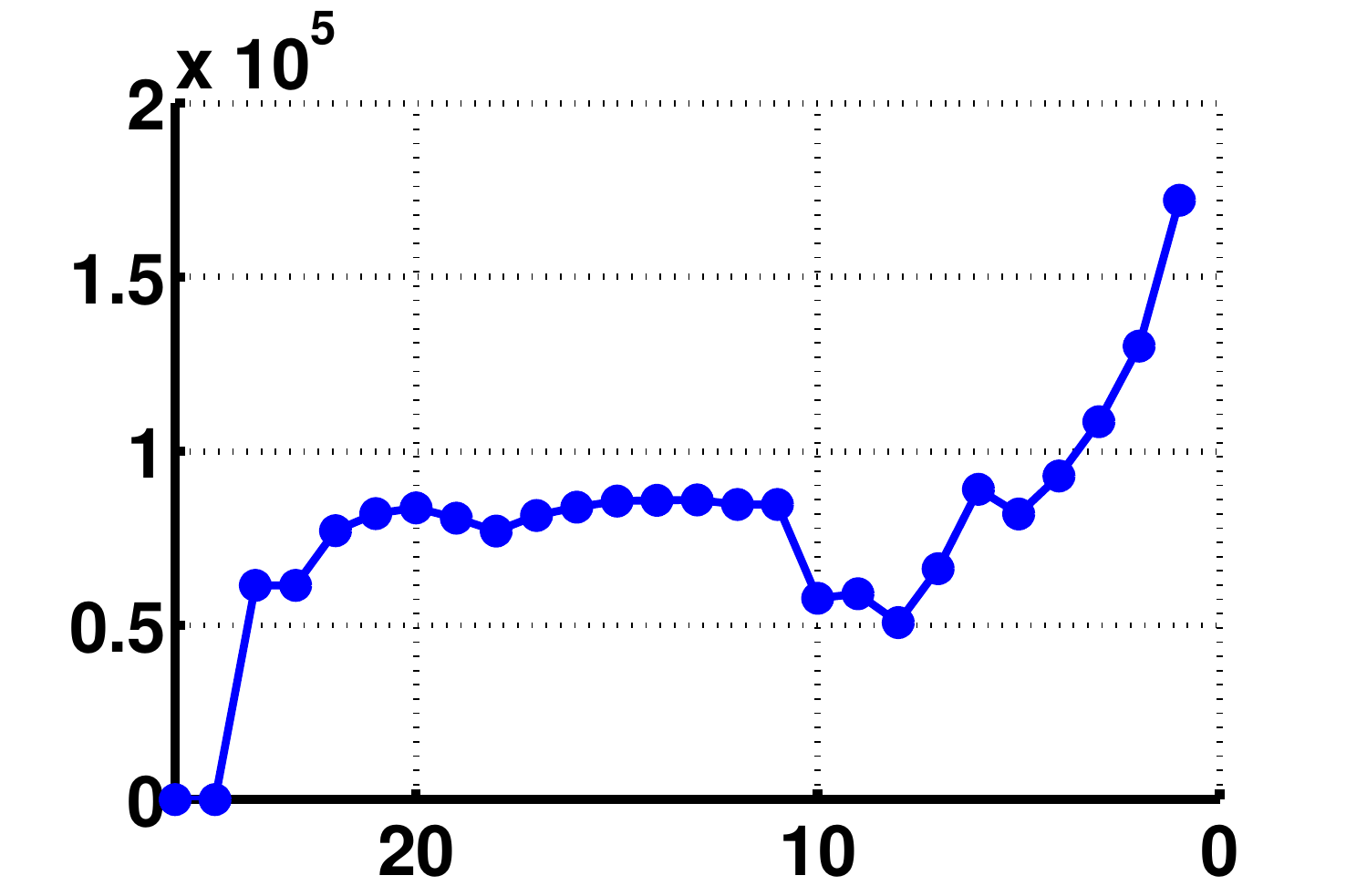}\label{fig:dream_freq_pat}} \\
	\subfloat[stream]{\includegraphics[scale=0.2]{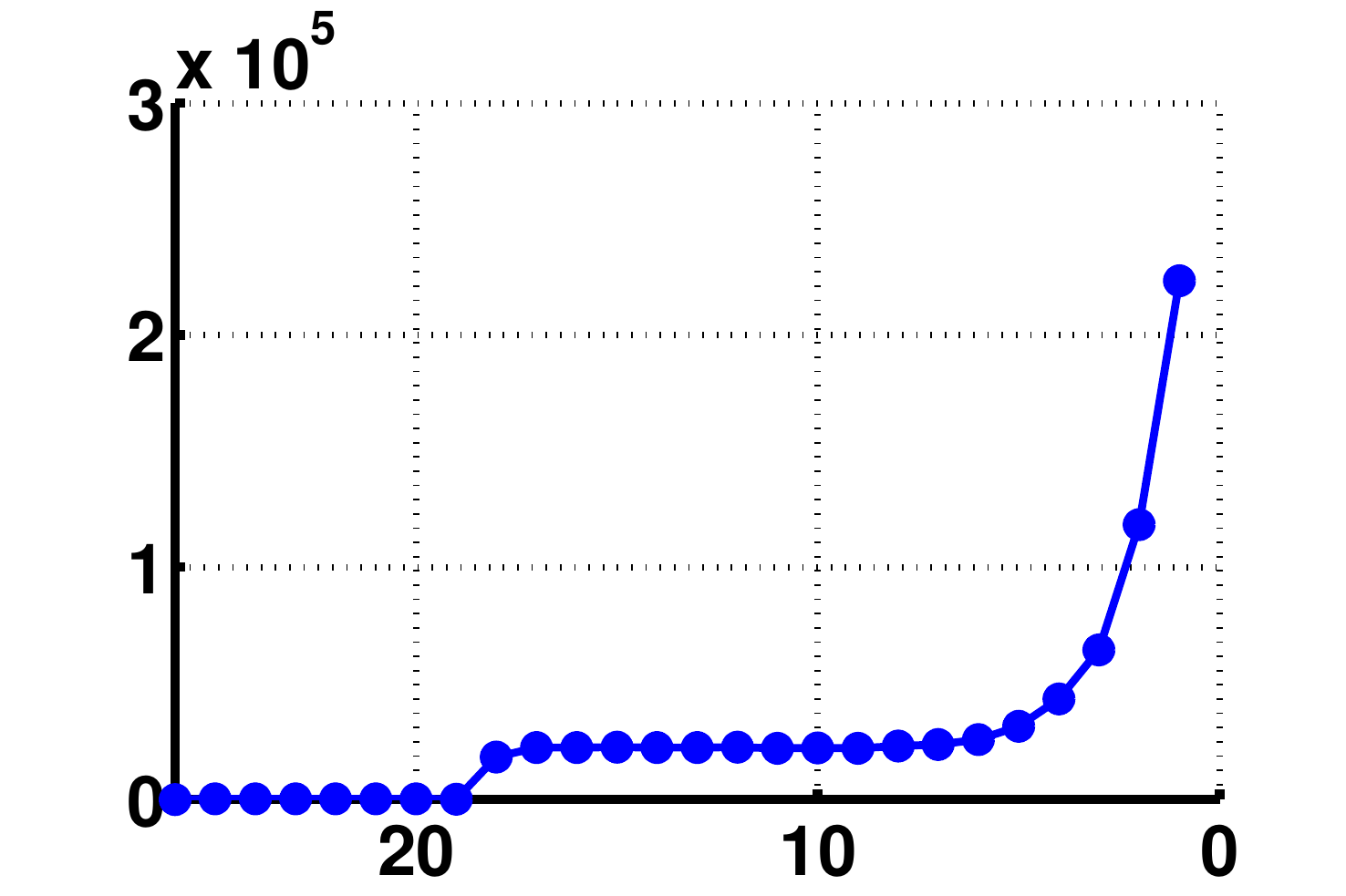}\label{fig:dream_stream_pat}} 
	\subfloat[swapt]{\includegraphics[scale=0.2]{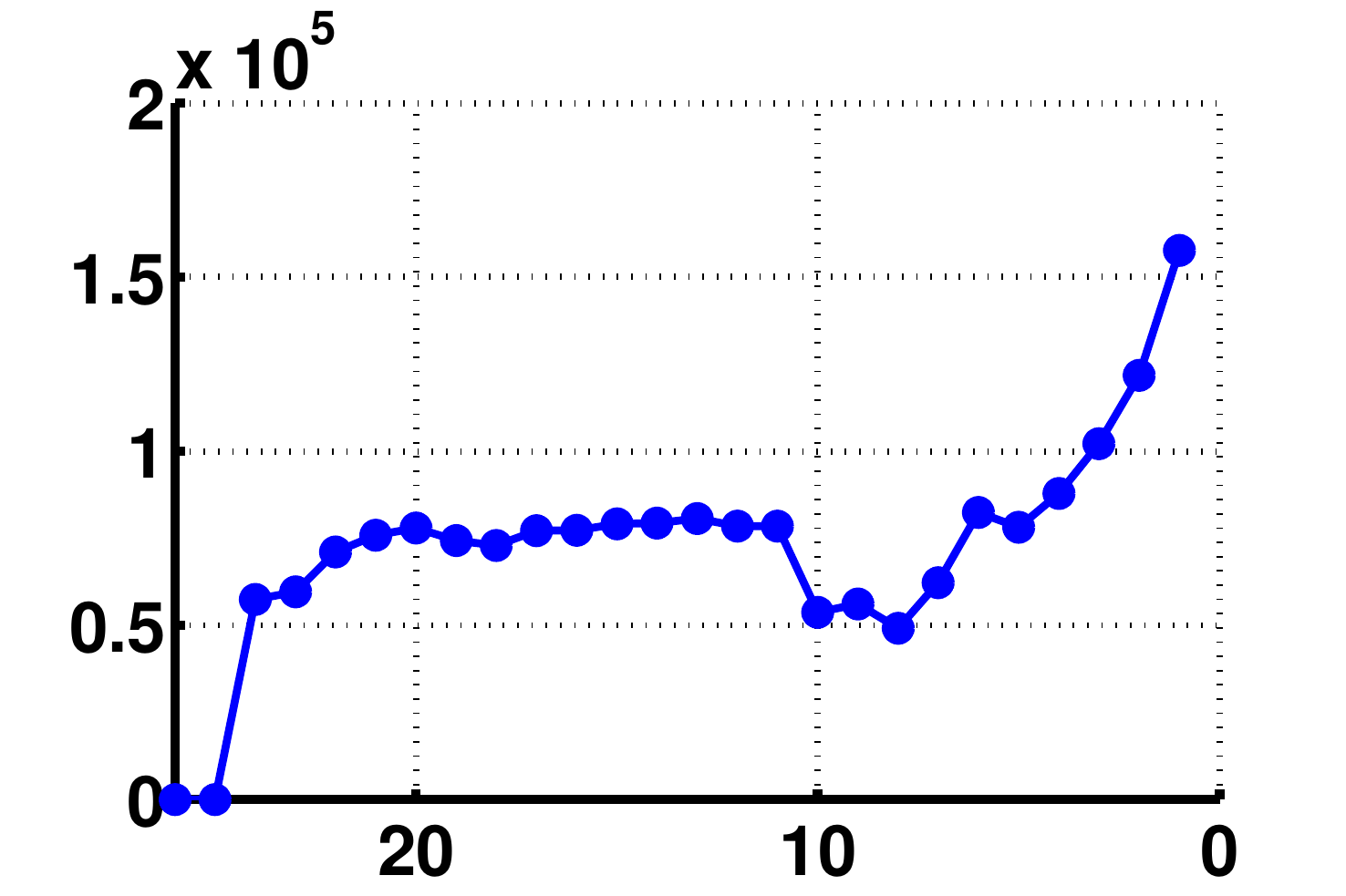}\label{fig:dream_swapt_pat}}
	\subfloat[astar-B]{\includegraphics[scale=0.2]{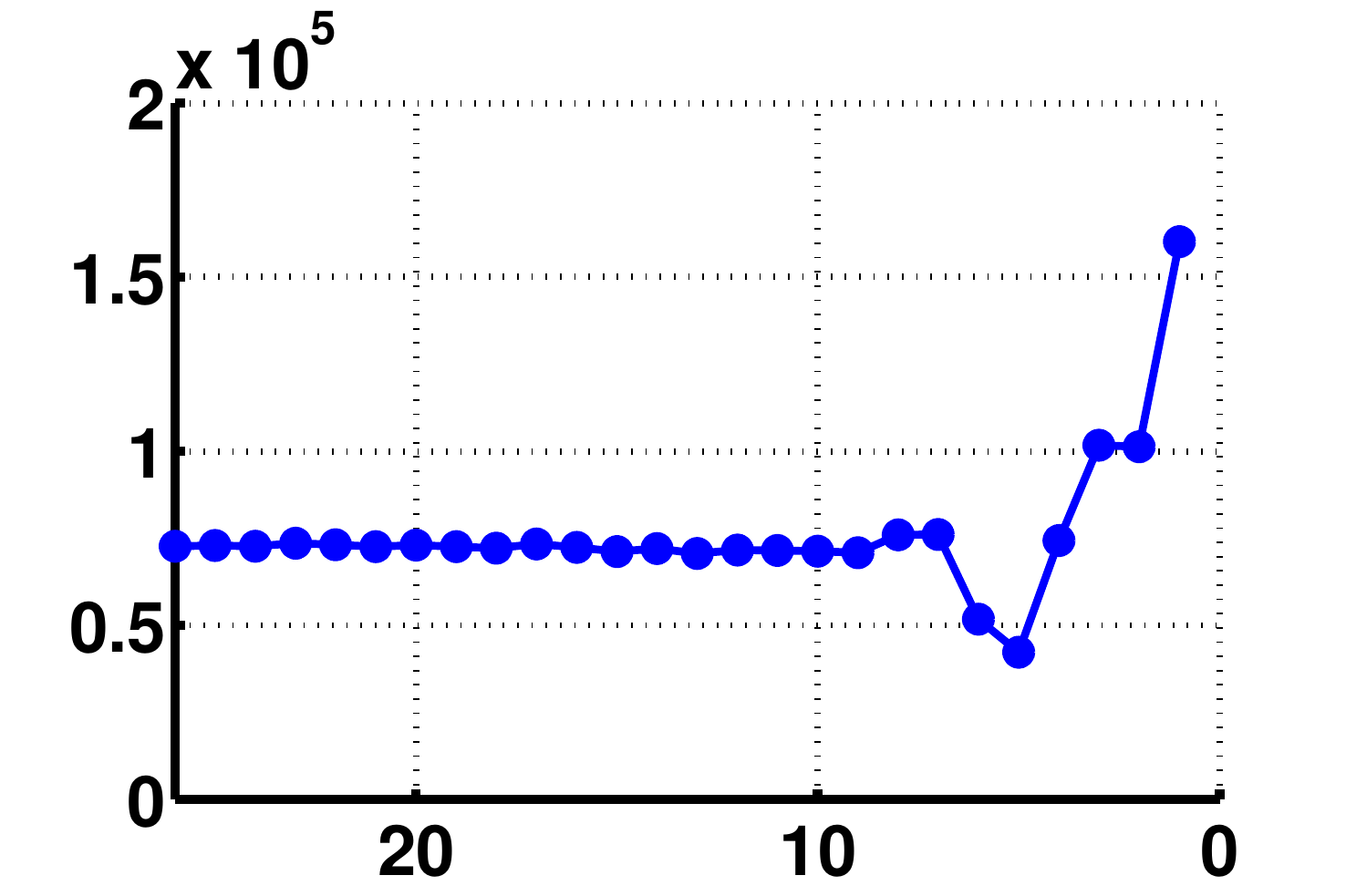}\label{fig:dream_astar_pat}} 
	\subfloat[bzip2-l]{\includegraphics[scale=0.2]{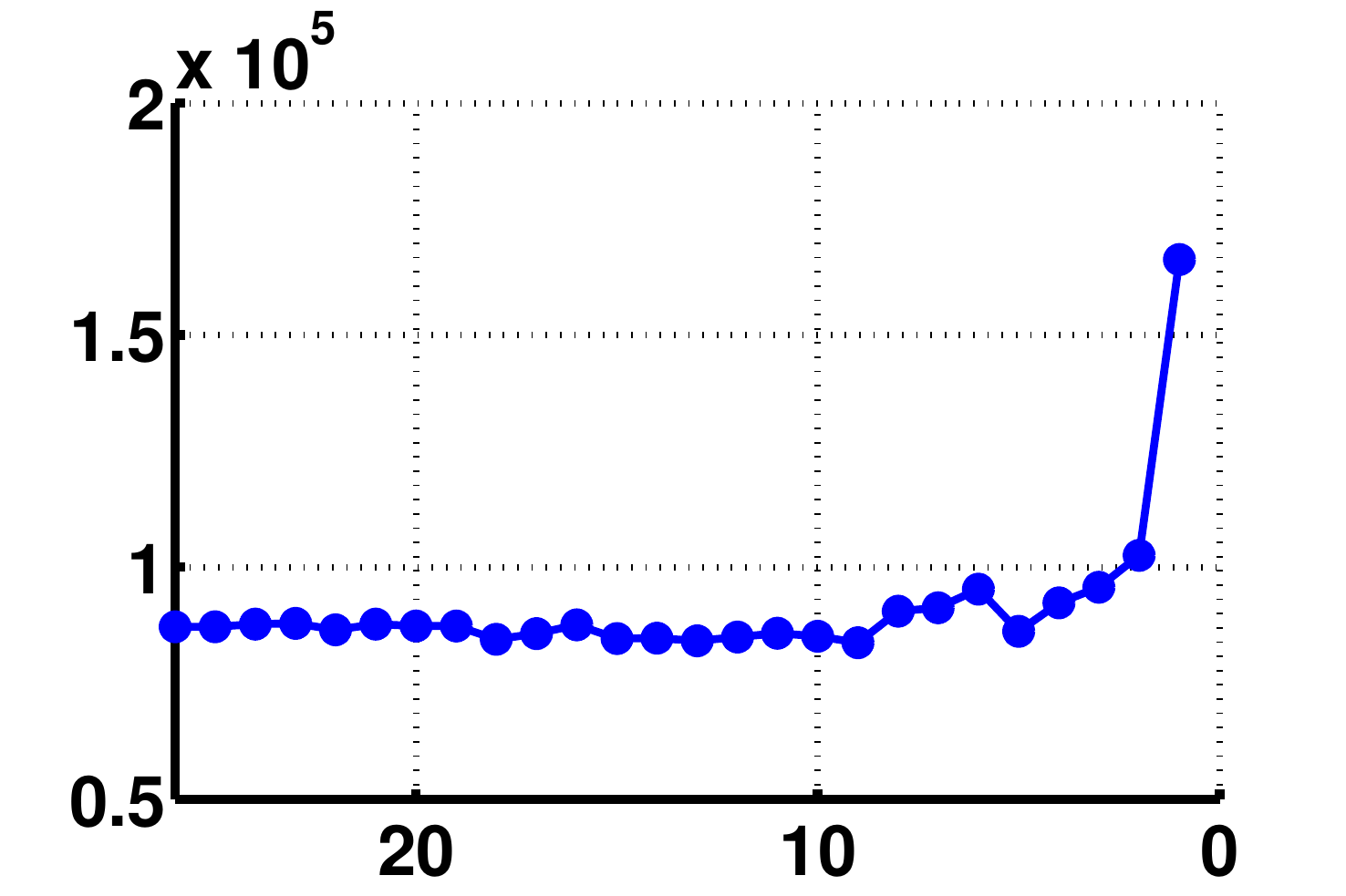}\label{fig:dream_bzip2_I_pat}} 
	\subfloat[bzip2-t]{\includegraphics[scale=0.2]{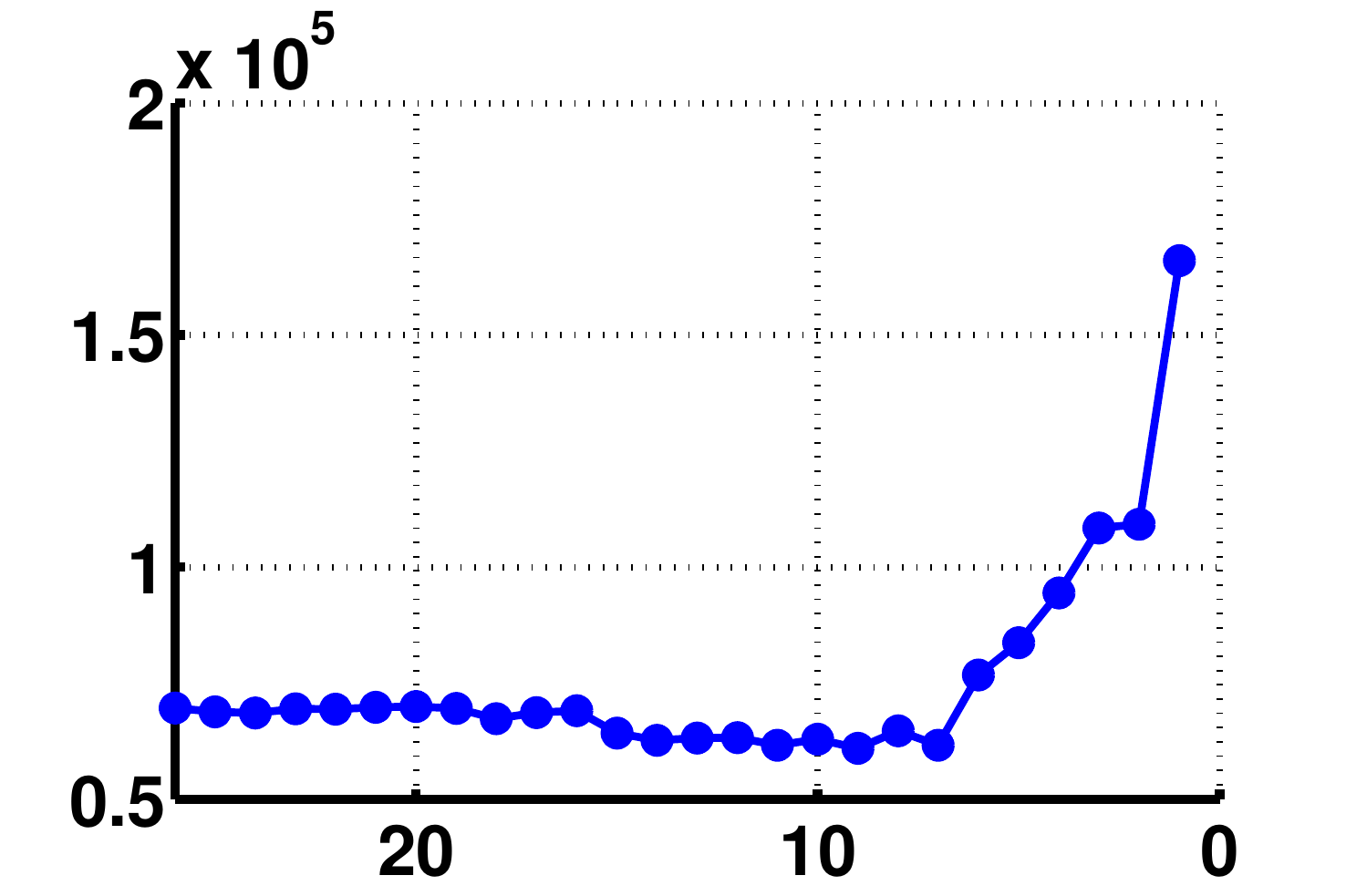}\label{fig:dream_bzip2_t_pat}} 
	\subfloat[cactusADM-b]{\includegraphics[scale=0.2]{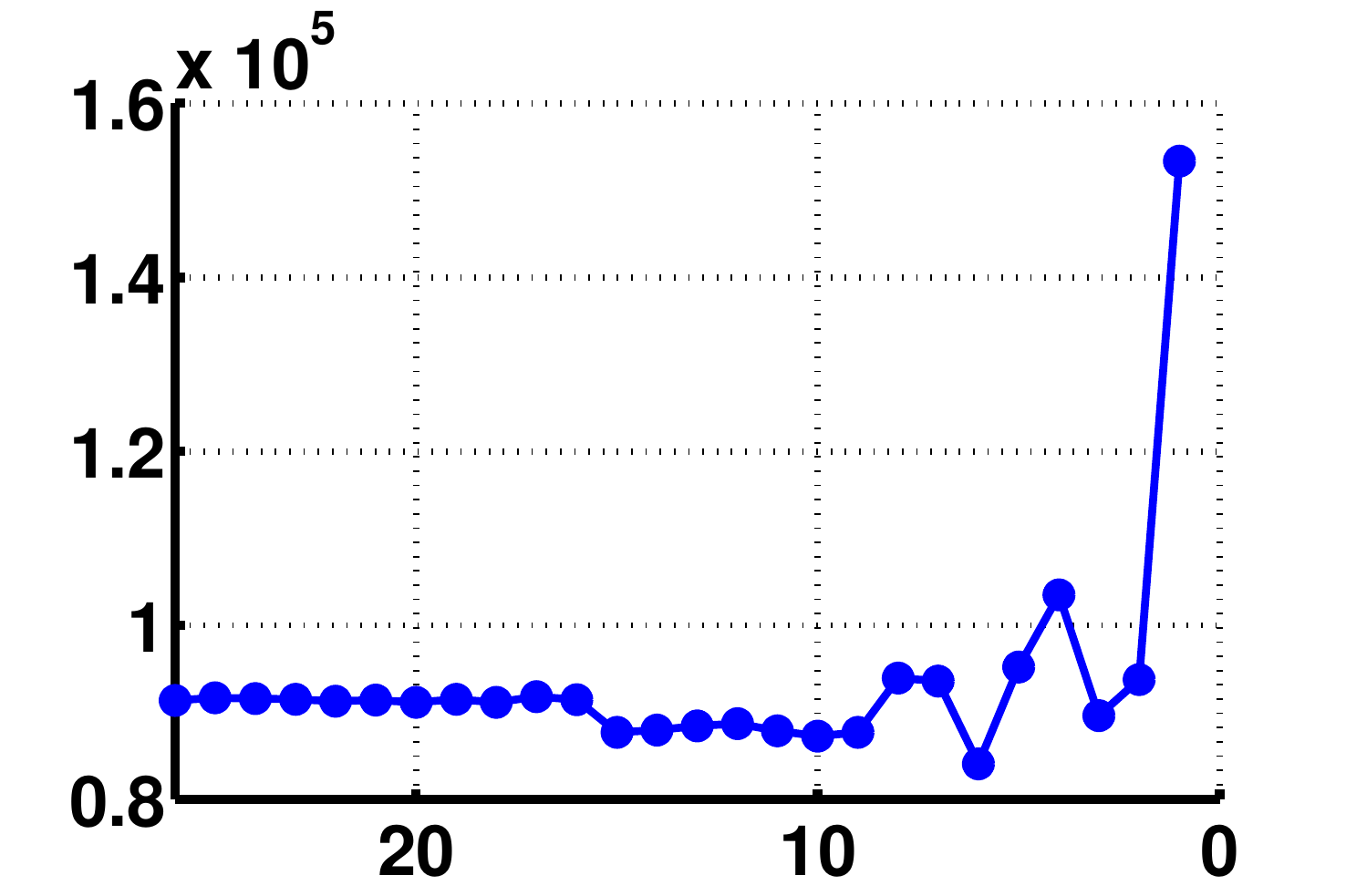}\label{fig:dream_cactusADM_b_pat}}\\ 	
	\subfloat[gcc-1]{\includegraphics[scale=0.2]{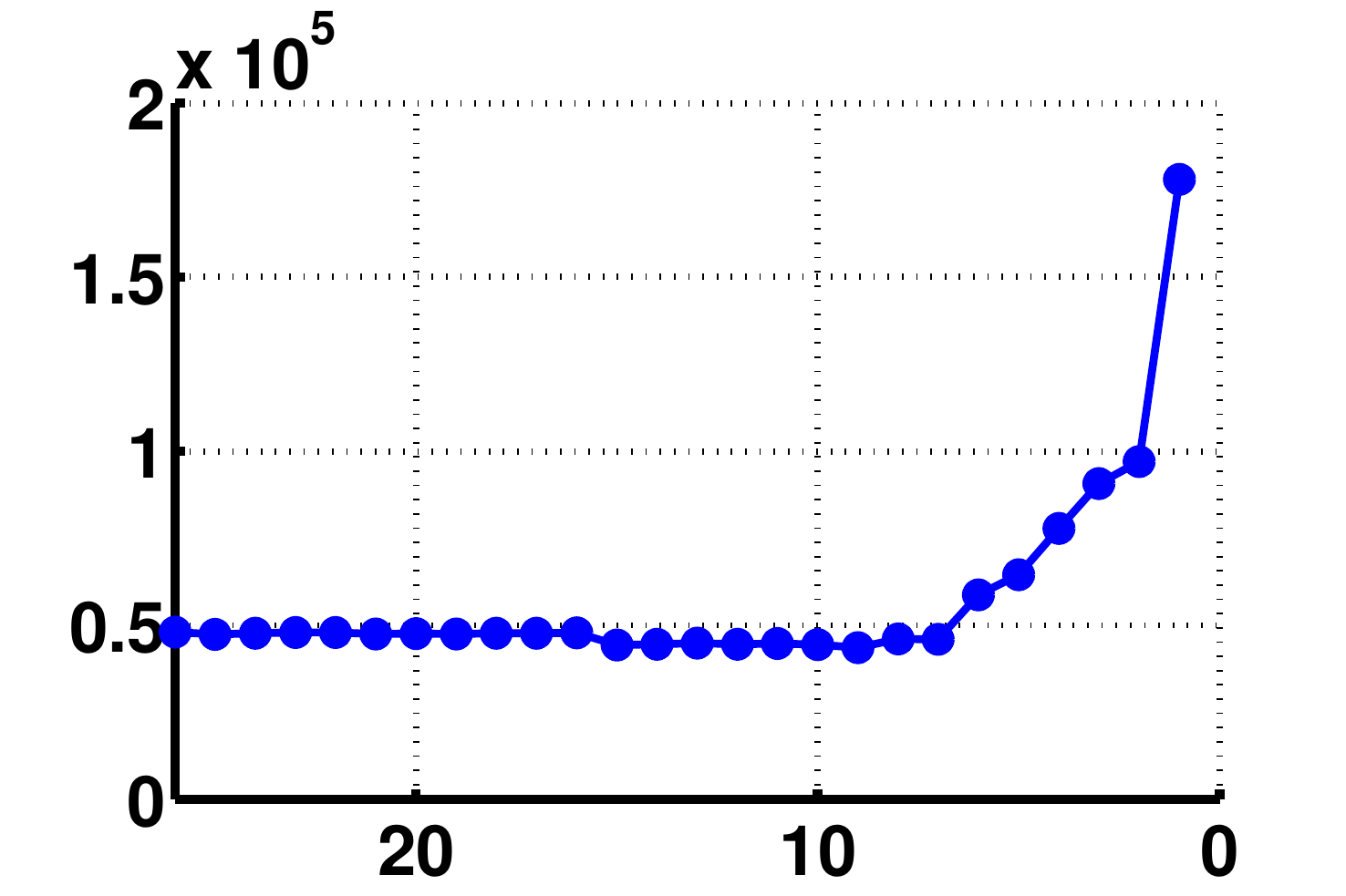}\label{fig:dream_gcc_1_pat}} 	
	\subfloat[gcc-2]{\includegraphics[scale=0.2]{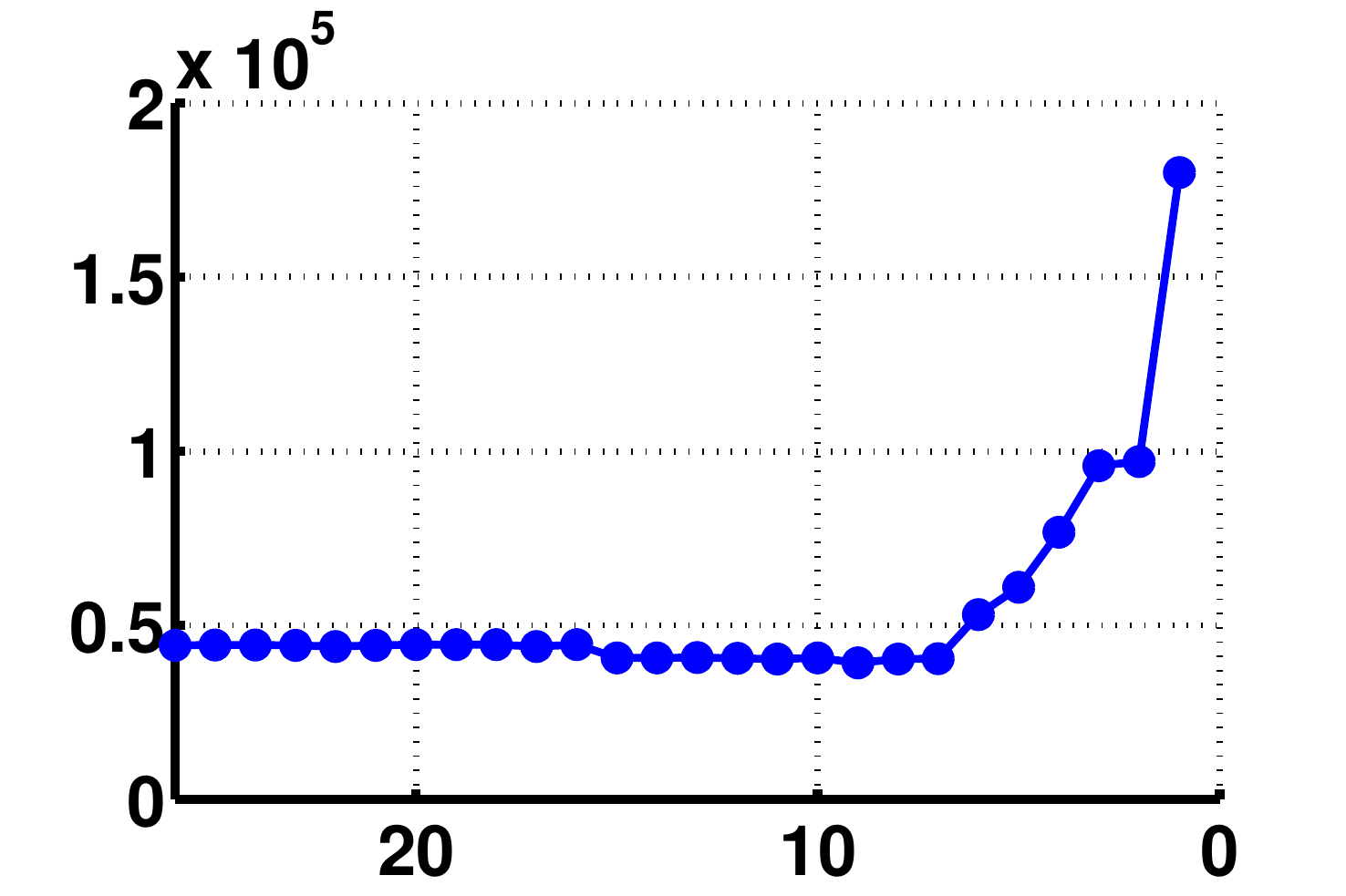}\label{fig:dream_gcc_2_pat}} 
	\subfloat[gcc-c]{\includegraphics[scale=0.2]{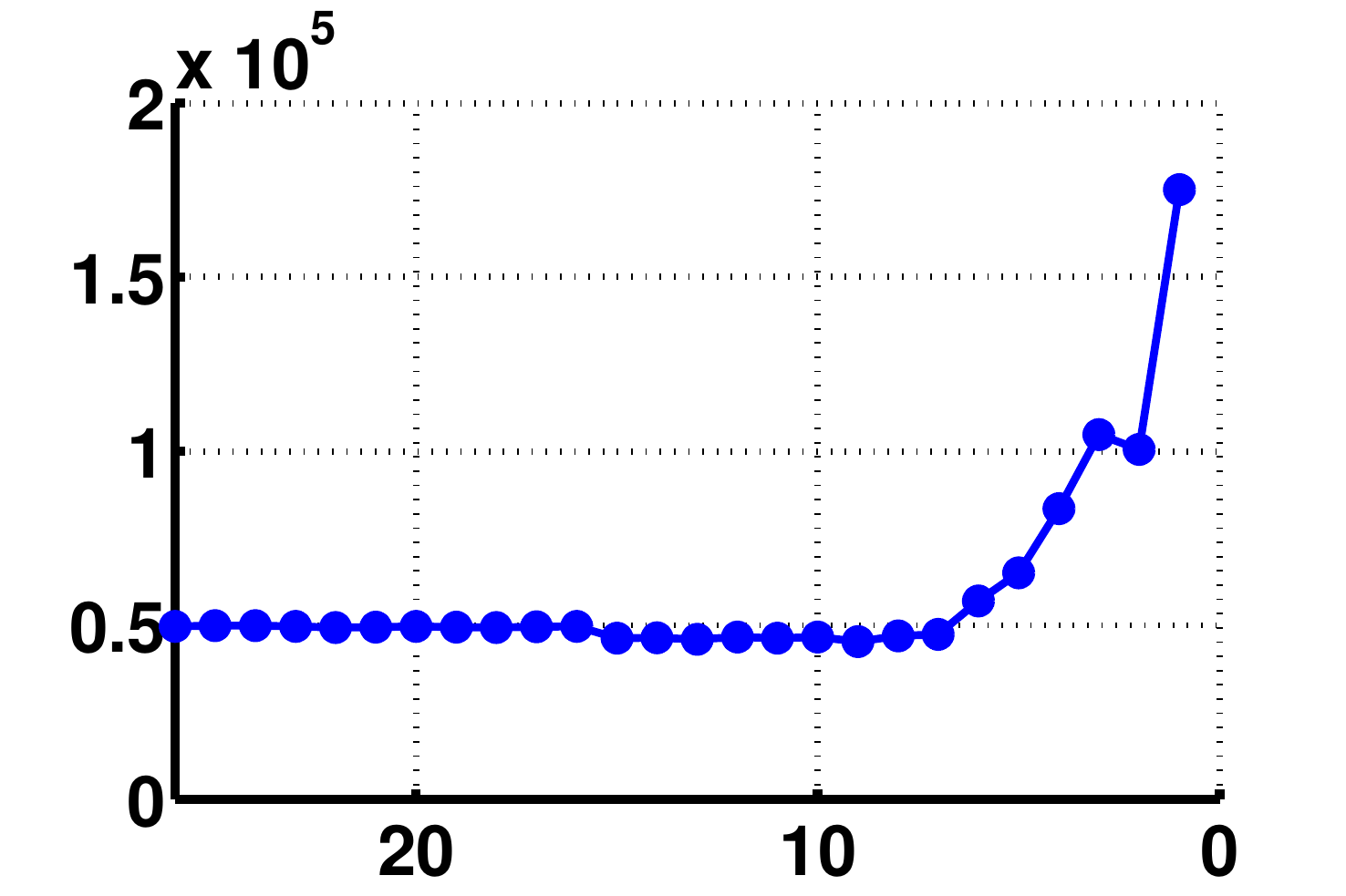}\label{fig:dream_gcc_c_pat}} 
	\subfloat[gcc-cp]{\includegraphics[scale=0.2]{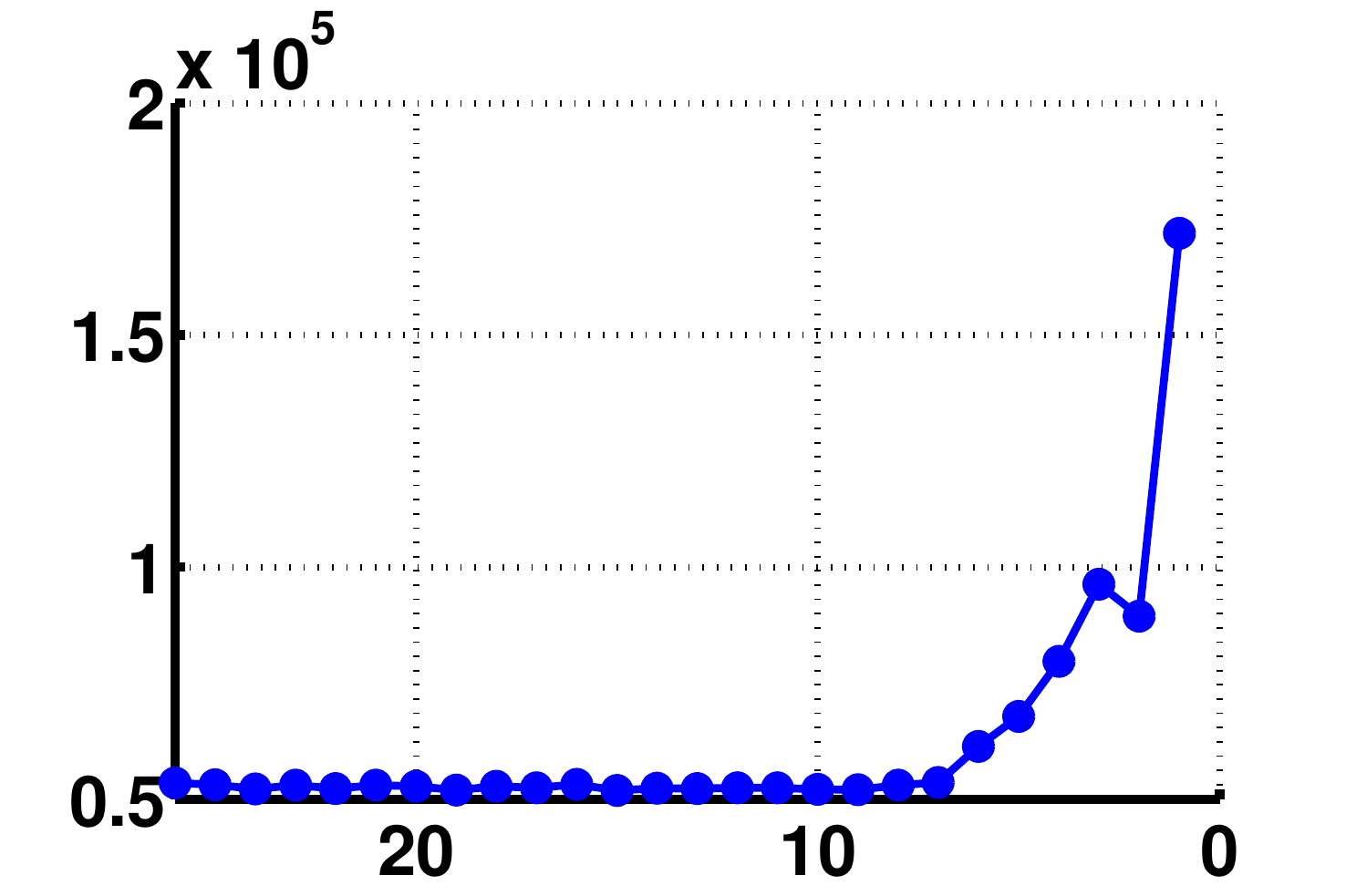}\label{fig:dream_gcc_cp_pat}} 	
	\subfloat[gcc-g]{\includegraphics[scale=0.2]{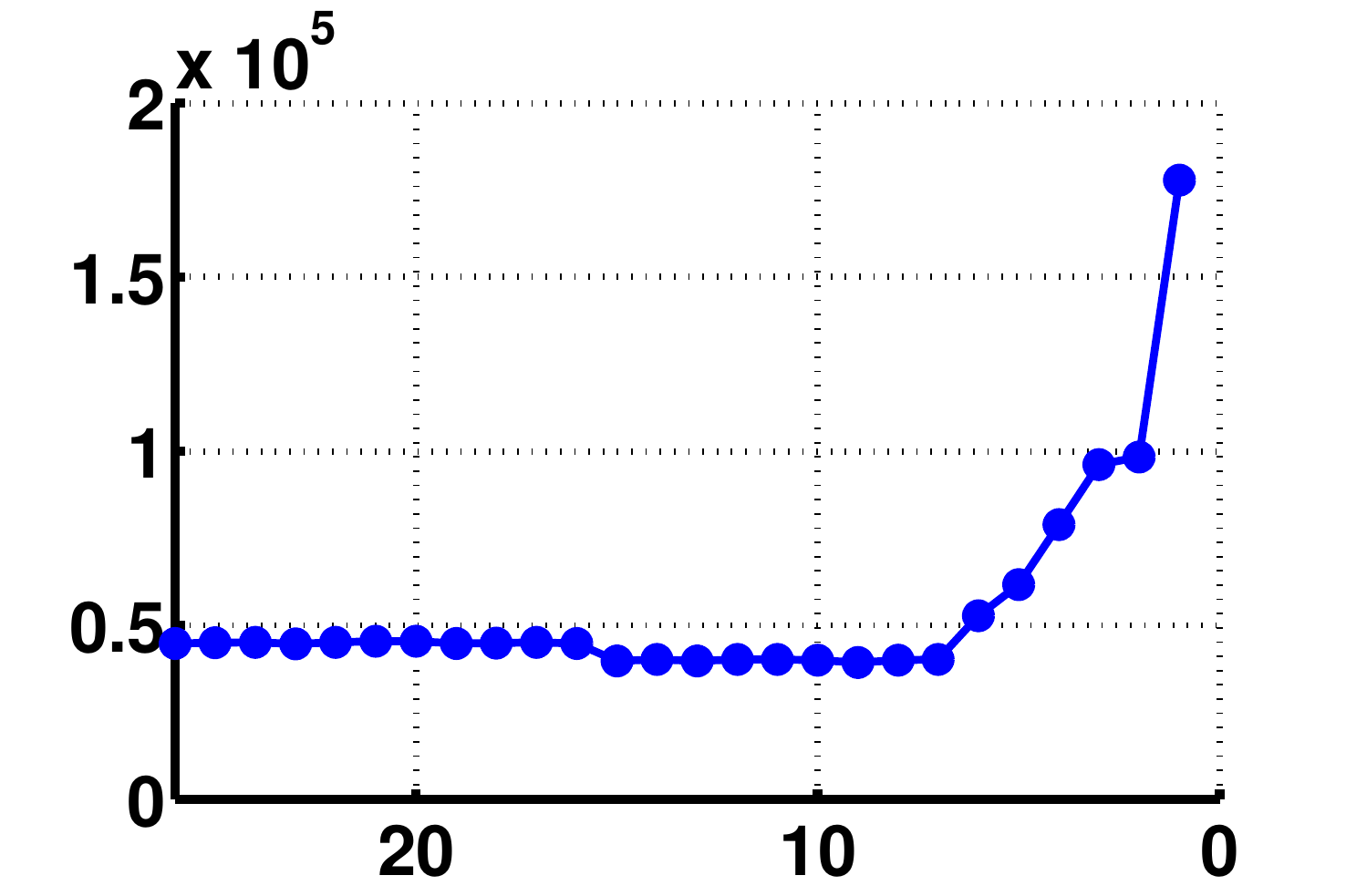}\label{fig:dream_gcc_g_pat}} 	
	\subfloat[gcc-sc]{\includegraphics[scale=0.2]{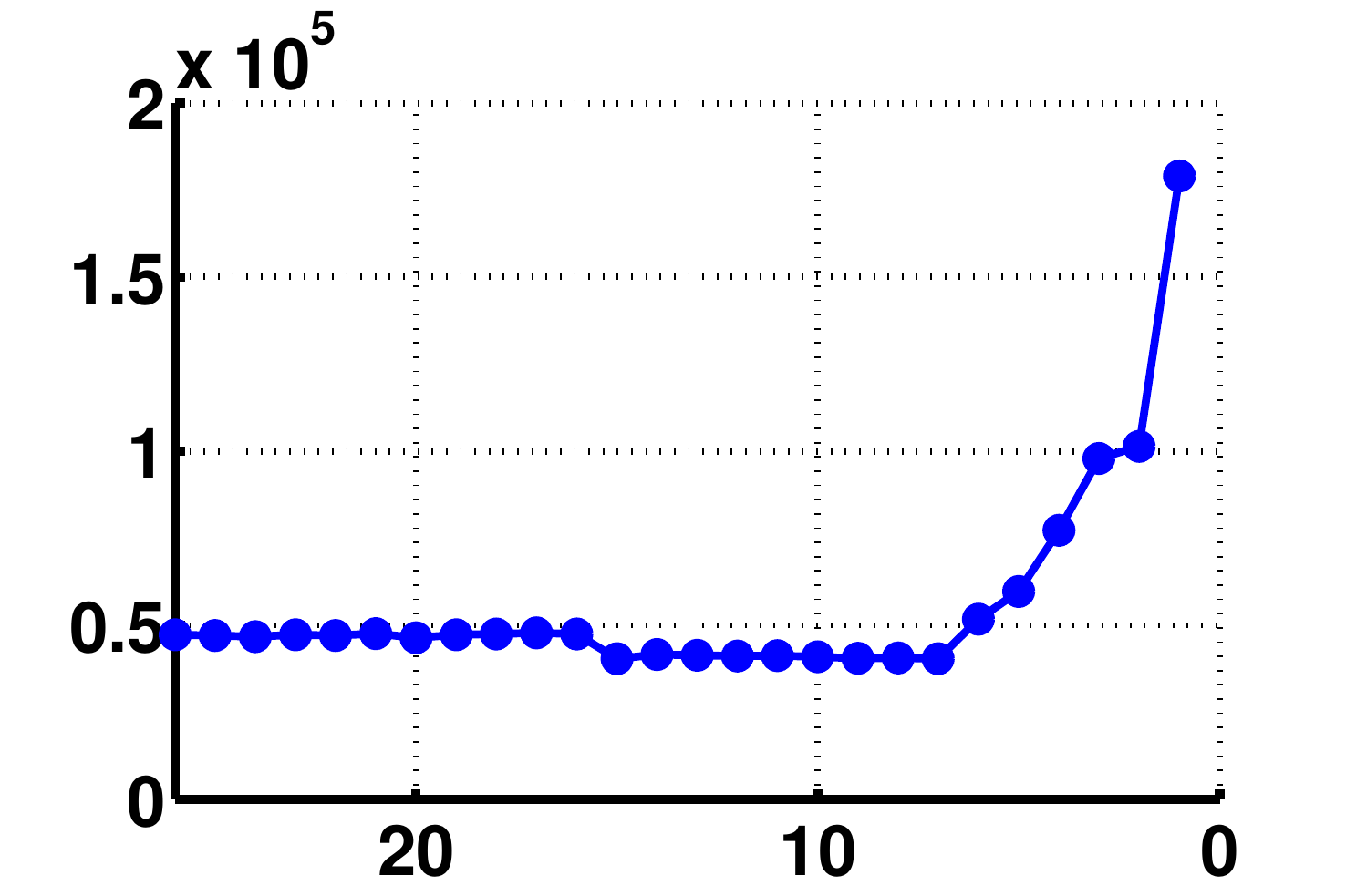}\label{fig:dream_gcc_sc_pat}}\\ 
	\subfloat[GemsFDTD-r]{\includegraphics[scale=0.2]{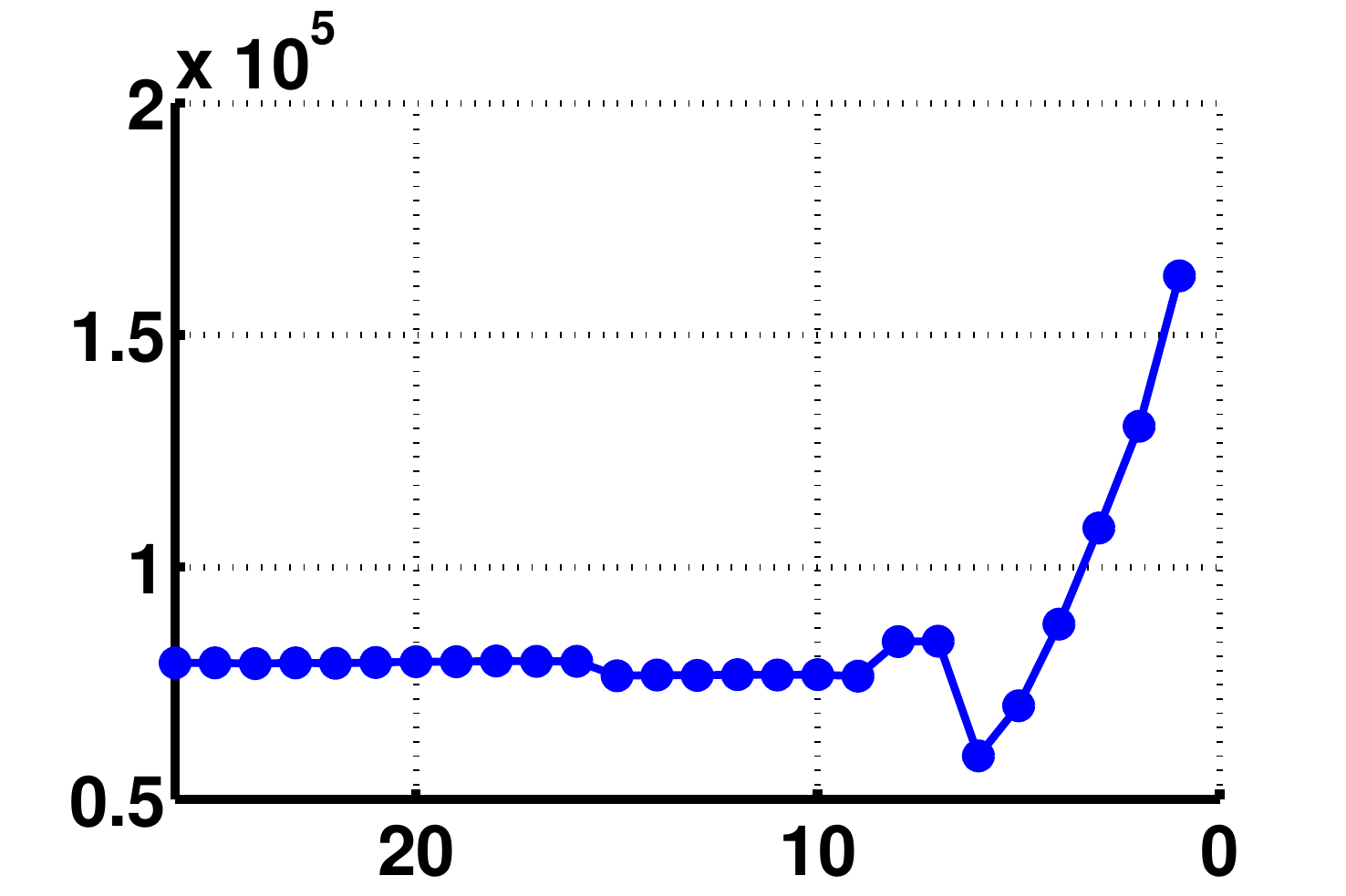}\label{fig:dream_GemsFDTD_r_pat}} 
	\subfloat[leslie3d-l]{\includegraphics[scale=0.2]{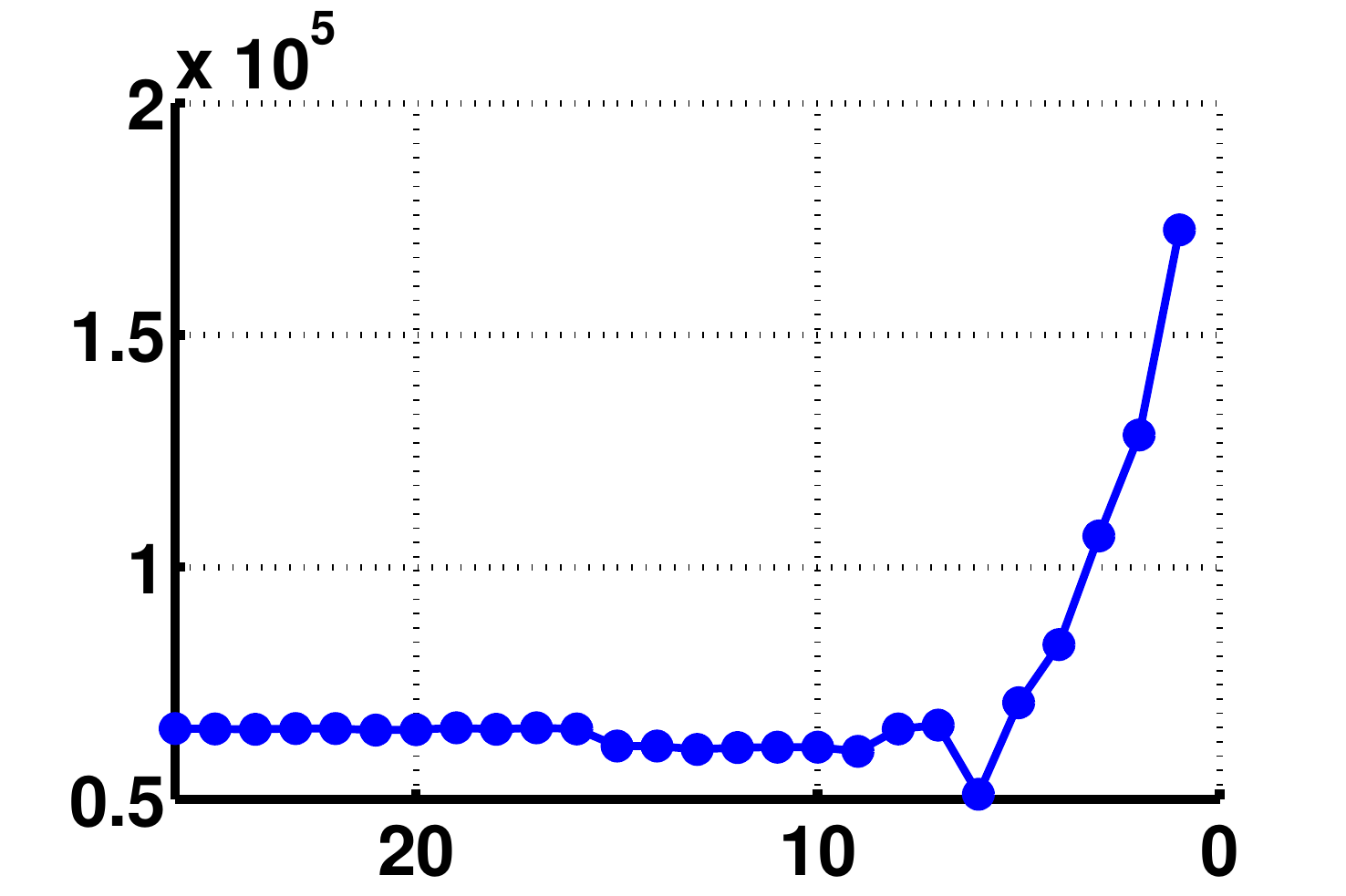}\label{fig:dream_leslie3d_l_pat}}	
	\subfloat[libquantum]{\includegraphics[scale=0.2]{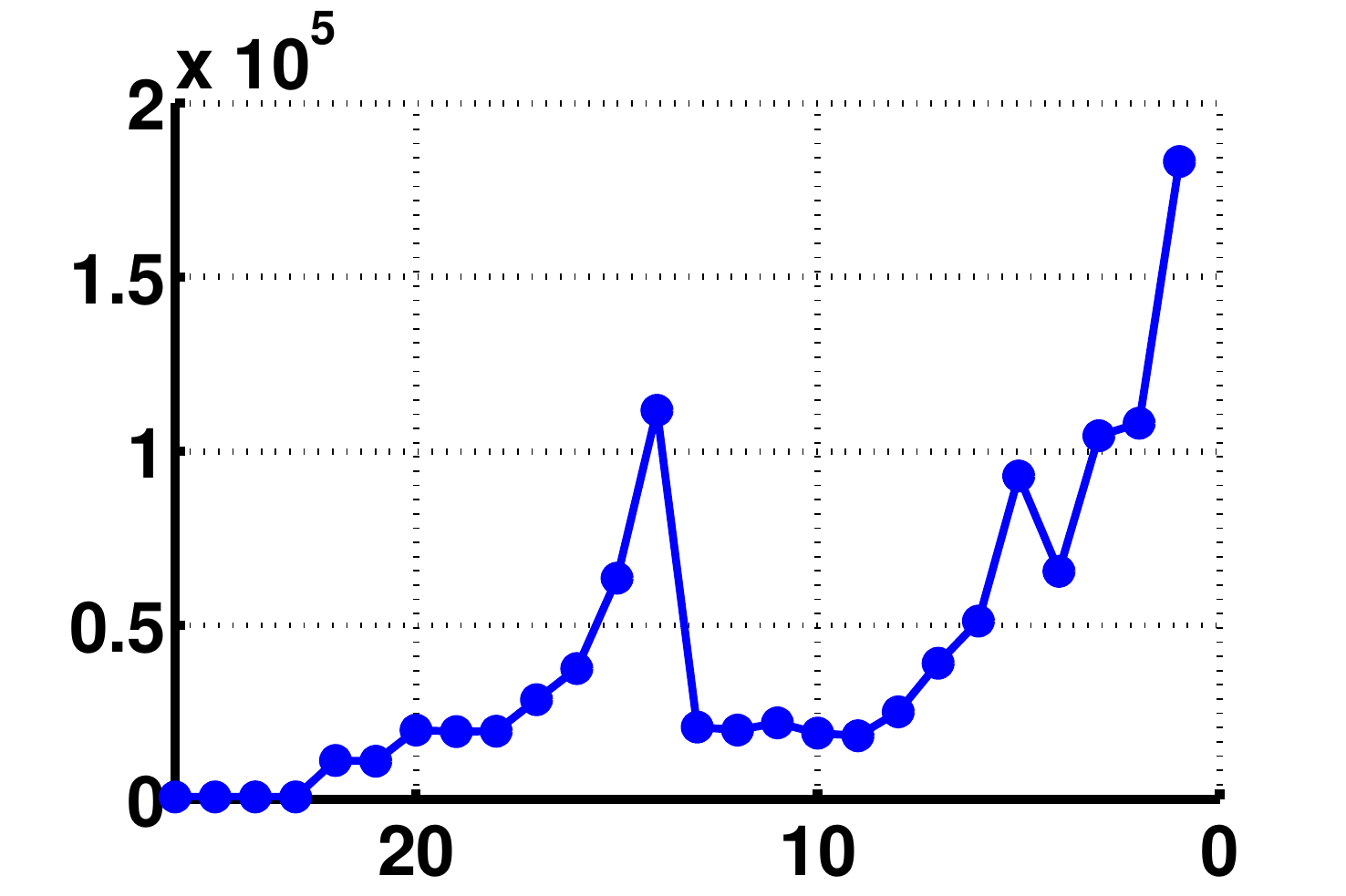}\label{fig:dream_libquantum_pat}} 	
	\subfloat[mcf-r]{\includegraphics[scale=0.2]{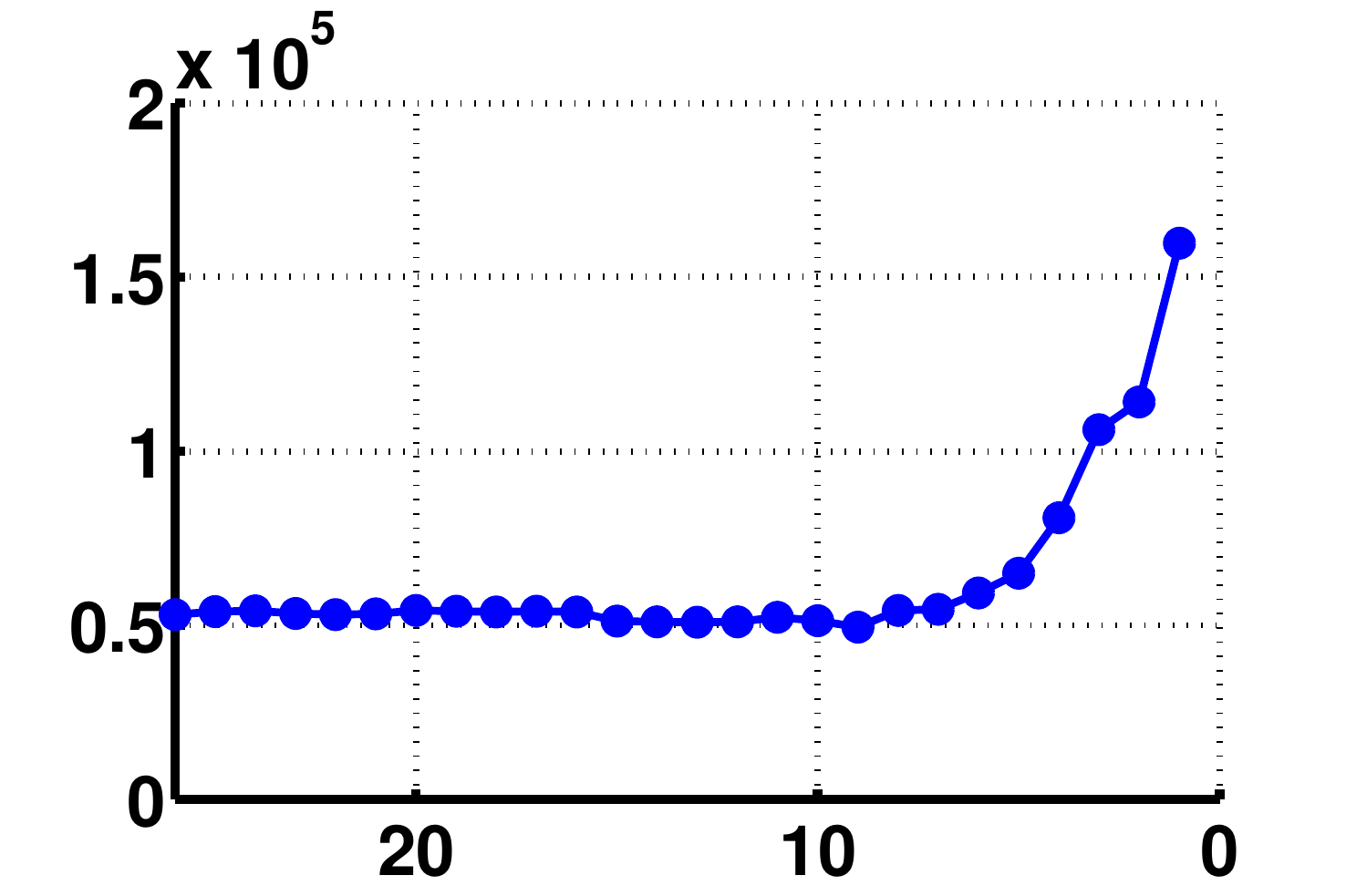}\label{fig:dream_mcf_r_pat}} 
	\subfloat[milc-s]{\includegraphics[scale=0.2]{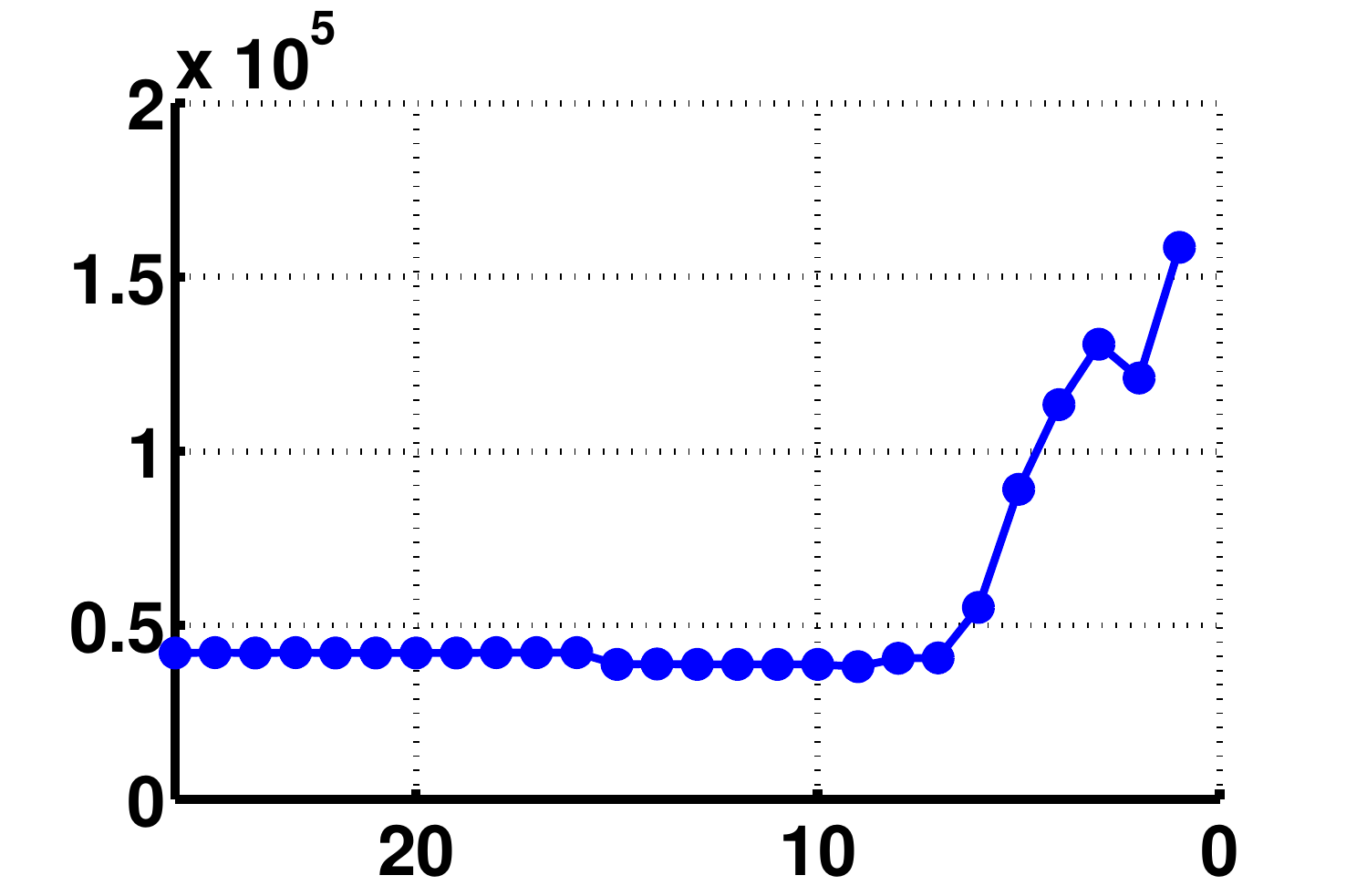}\label{fig:dream_milc_s_pat}} 
	\subfloat[omnetpp-o]{\includegraphics[scale=0.2]{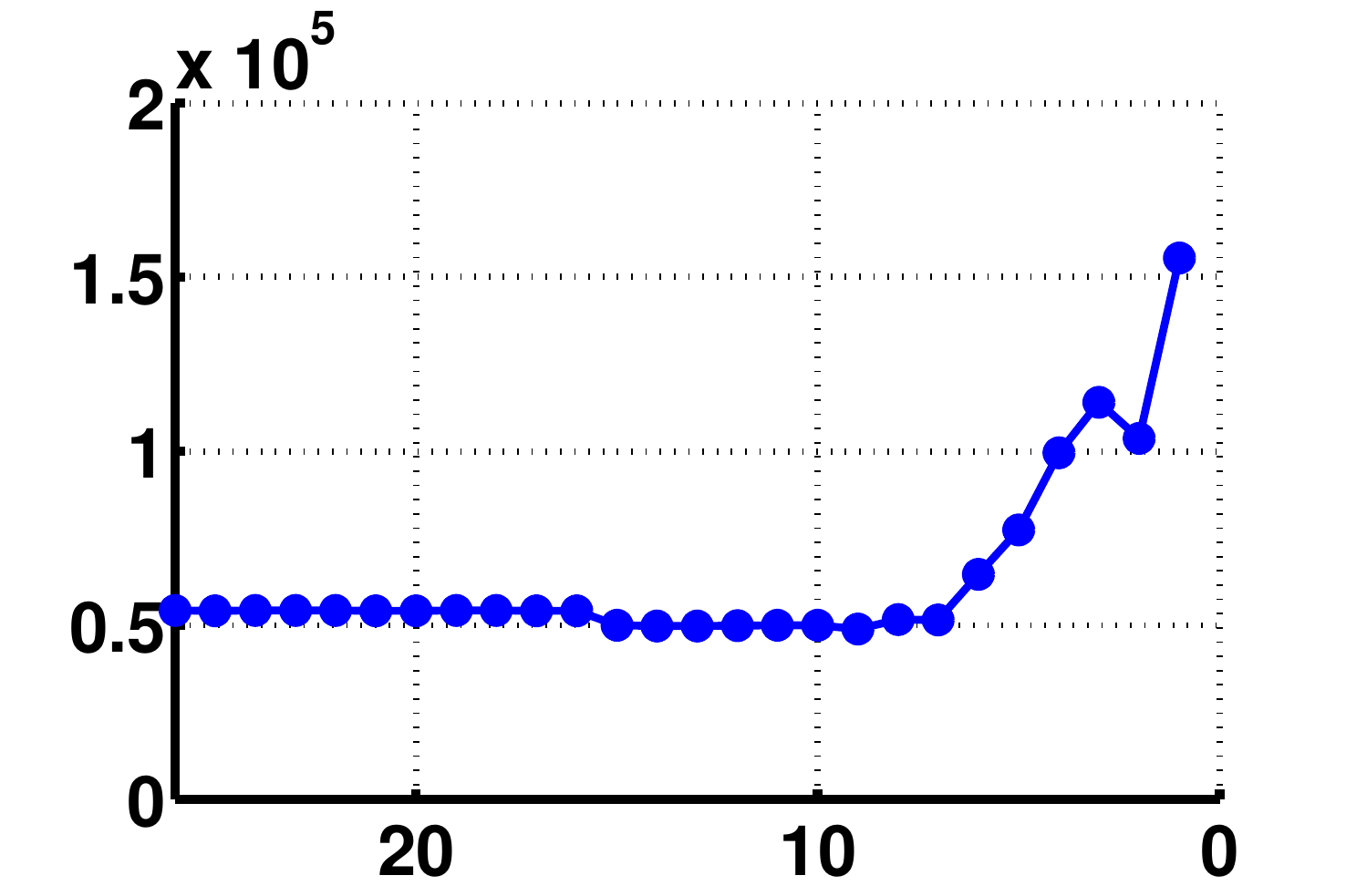}\label{fig:omnetpp_o_pat}}\\ 	
	\subfloat[soplex-r]{\includegraphics[scale=0.2]{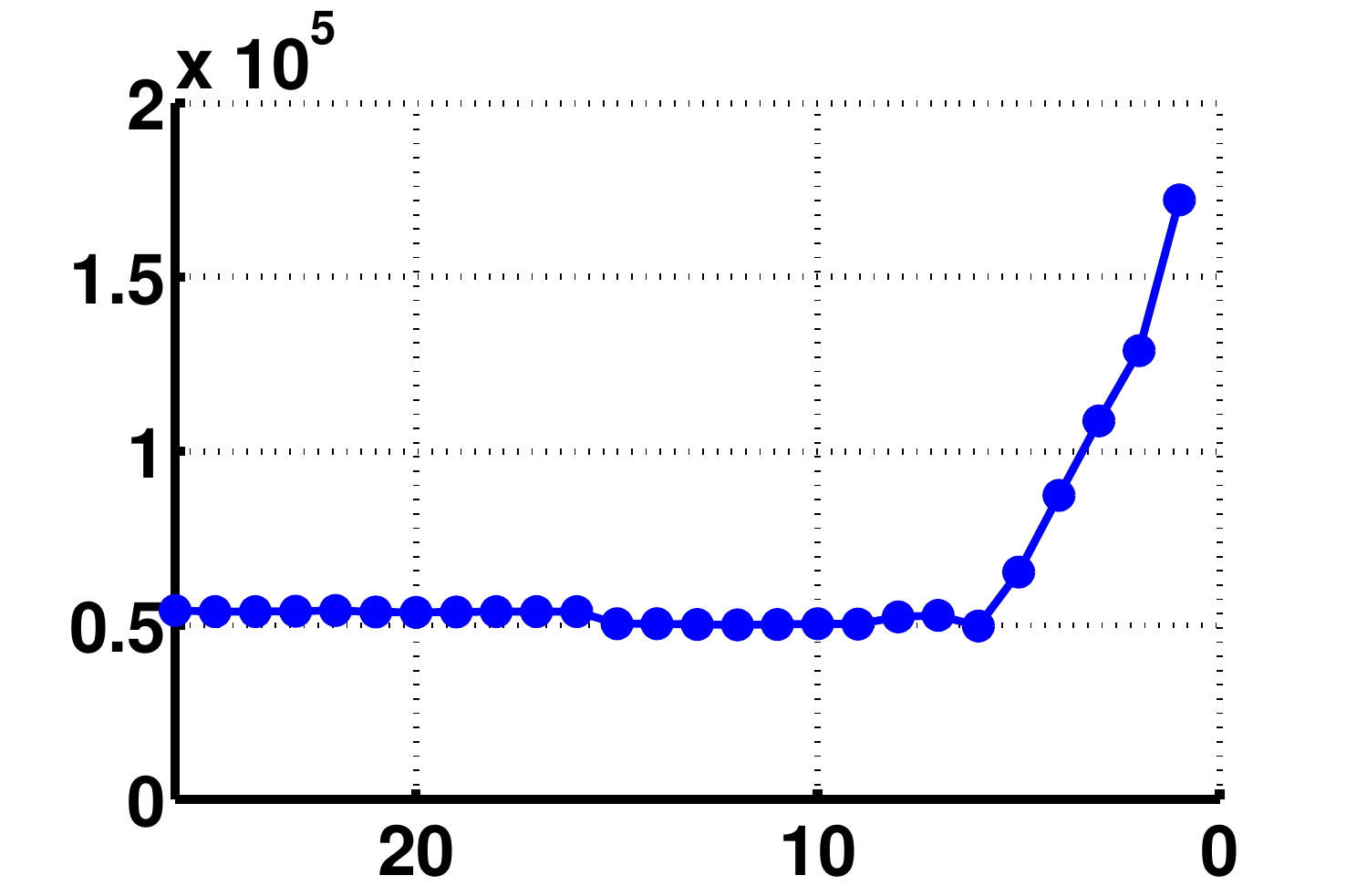}\label{fig:dream_soplex_r_pat}} 	
	\subfloat[sphinx3-a]{\includegraphics[scale=0.2]{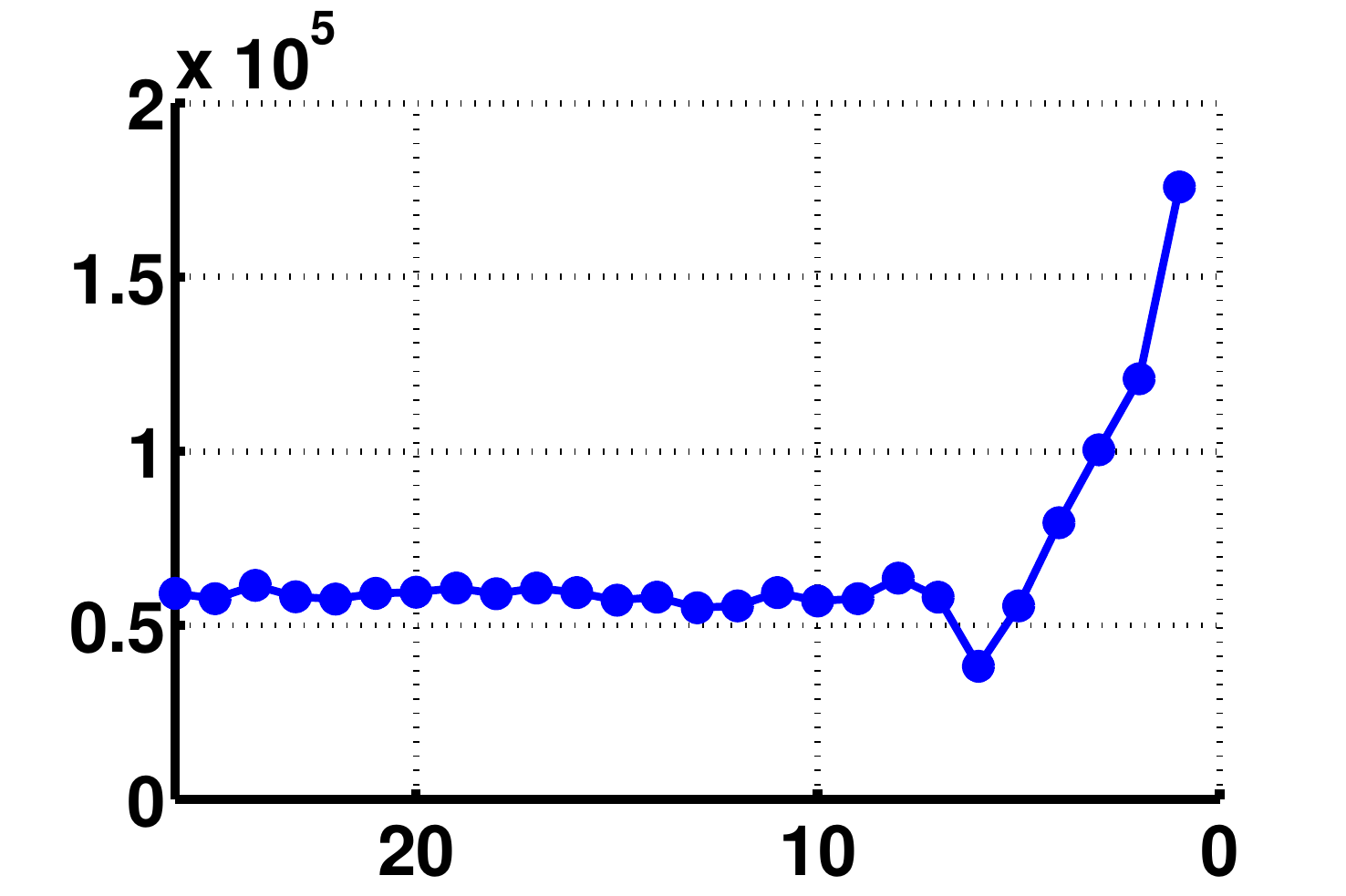}\label{fig:dream_sphinx3_a_pat}} 
	\subfloat[xalancbmk-r]{\includegraphics[scale=0.2]{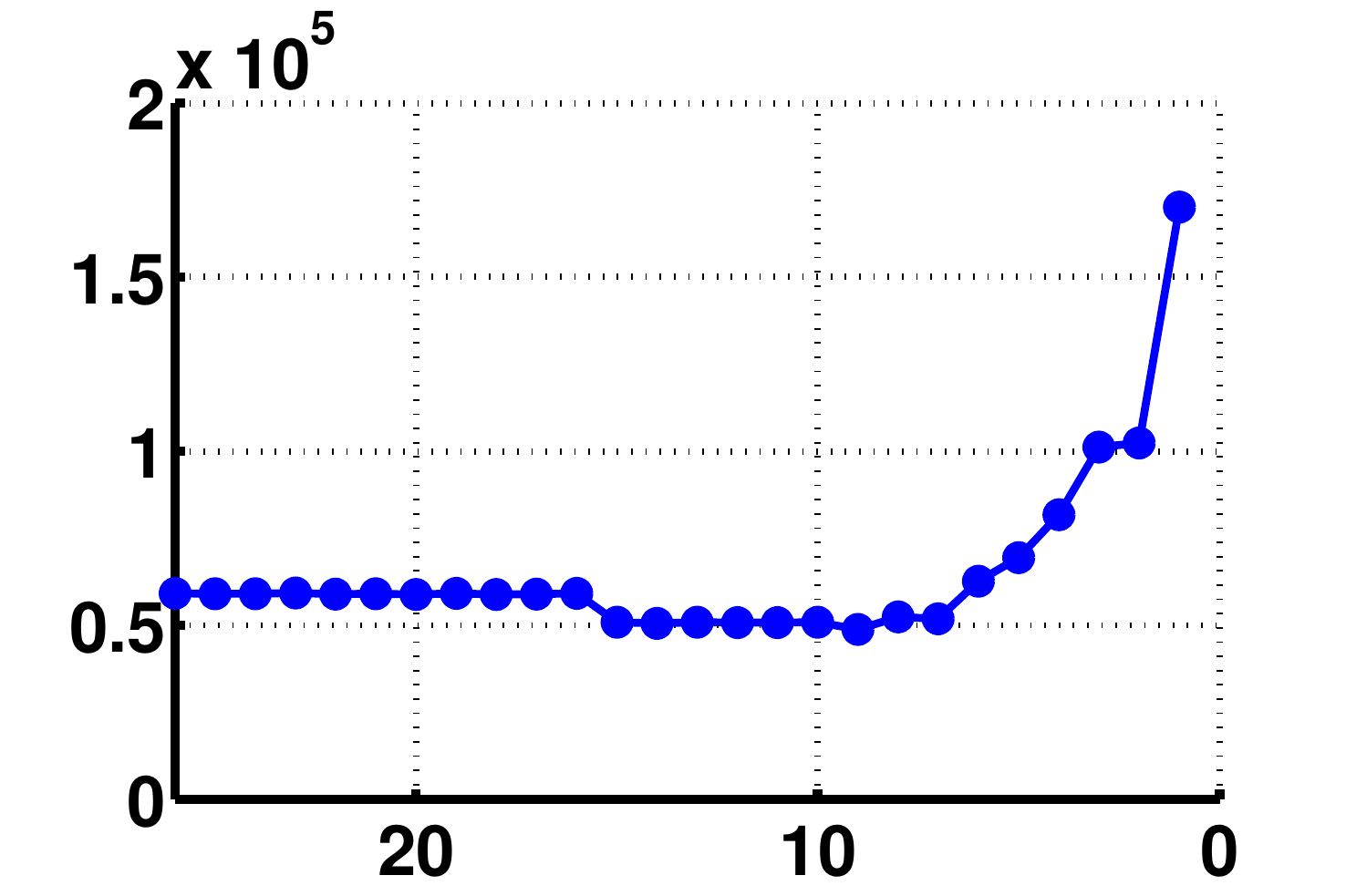}\label{fig:dream_xalancbmk_r_pat}} 
	\subfloat[zeusmp-z]{\includegraphics[scale=0.2]{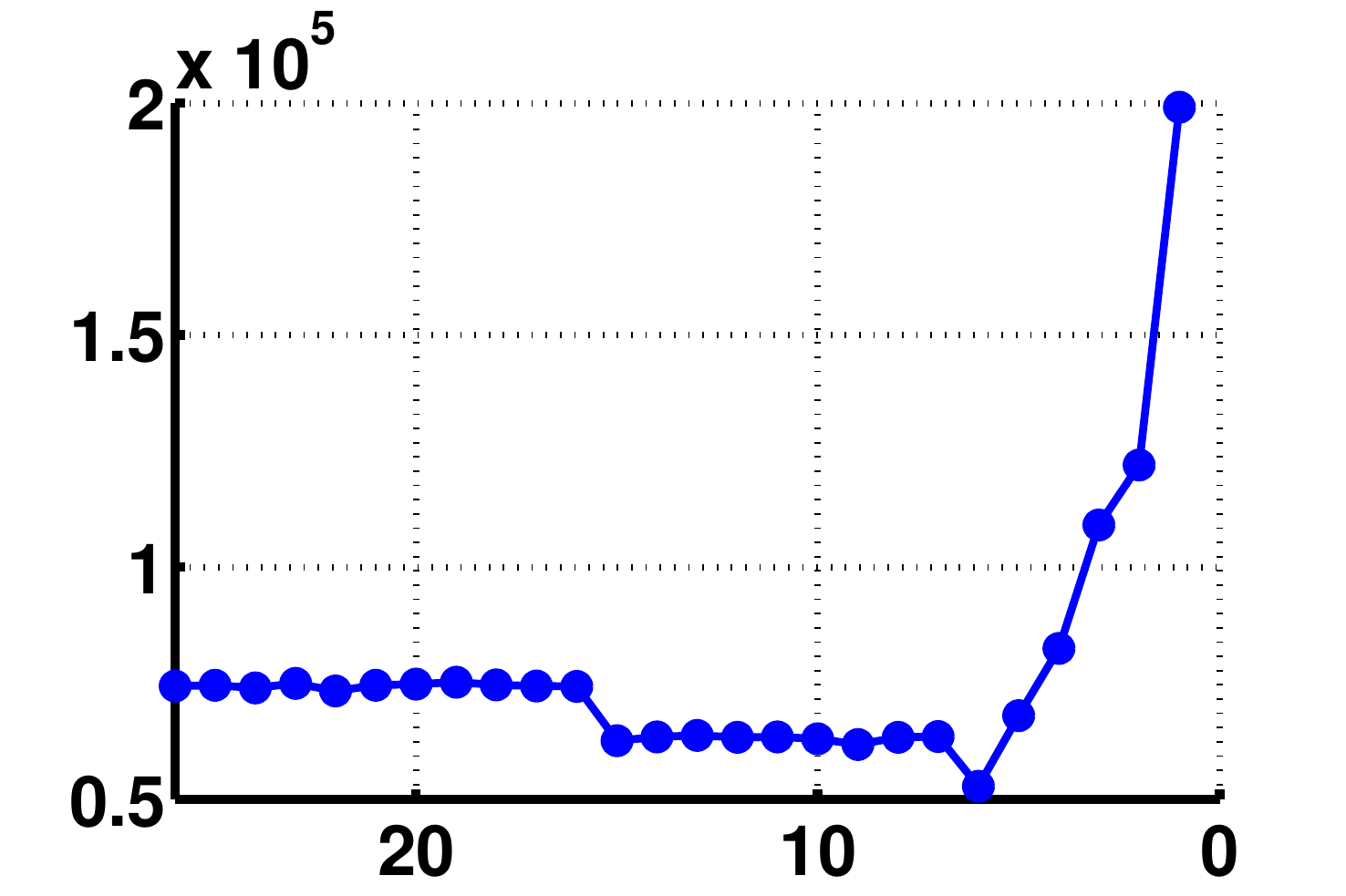}\label{fig:dream_zeusmp_z_pat}}	
	\subfloat[mummer]{\includegraphics[scale=0.2]{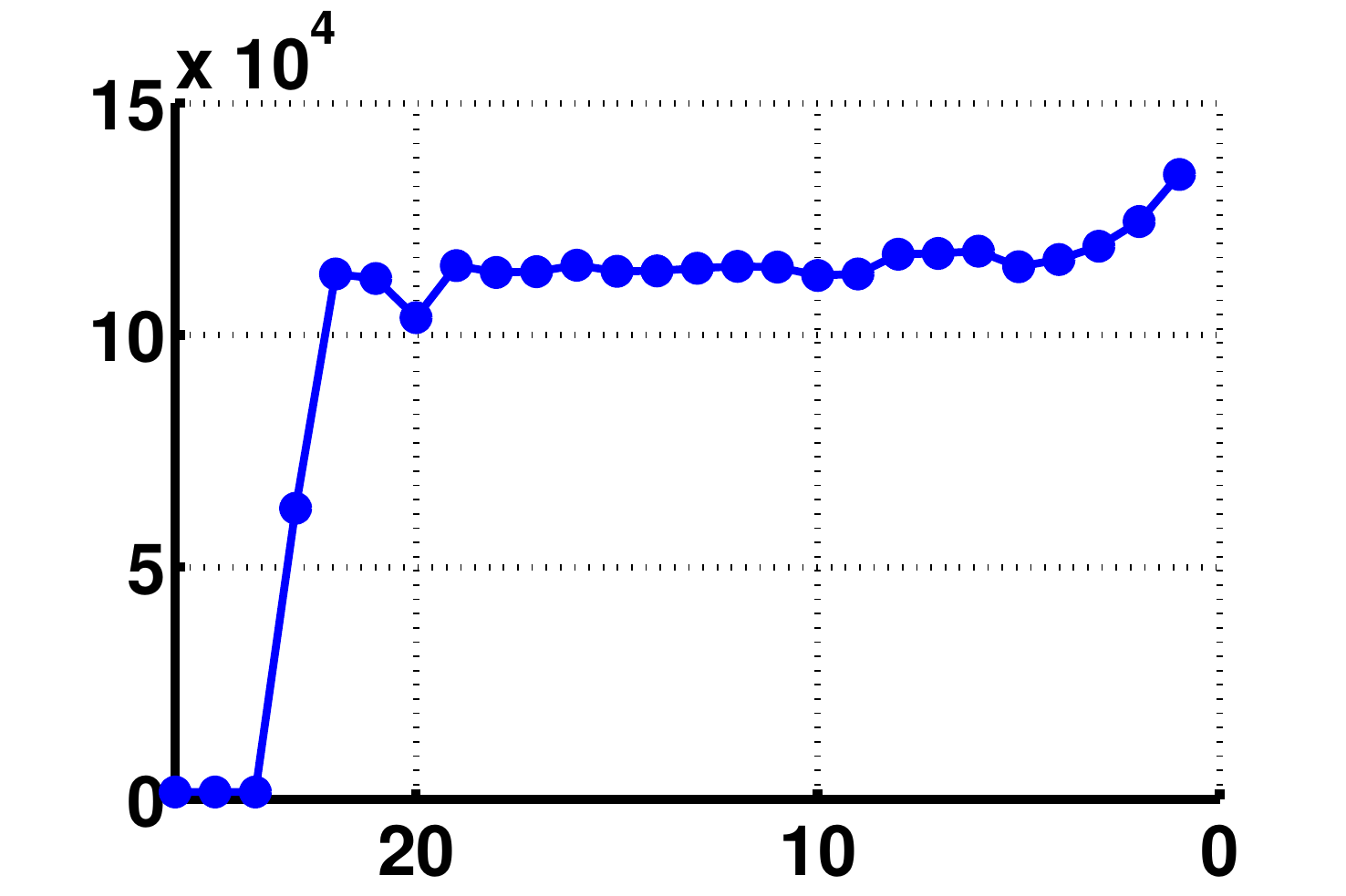}\label{fig:dream_mummer_pat}} 
	\subfloat[tigr]{\includegraphics[scale=0.2]{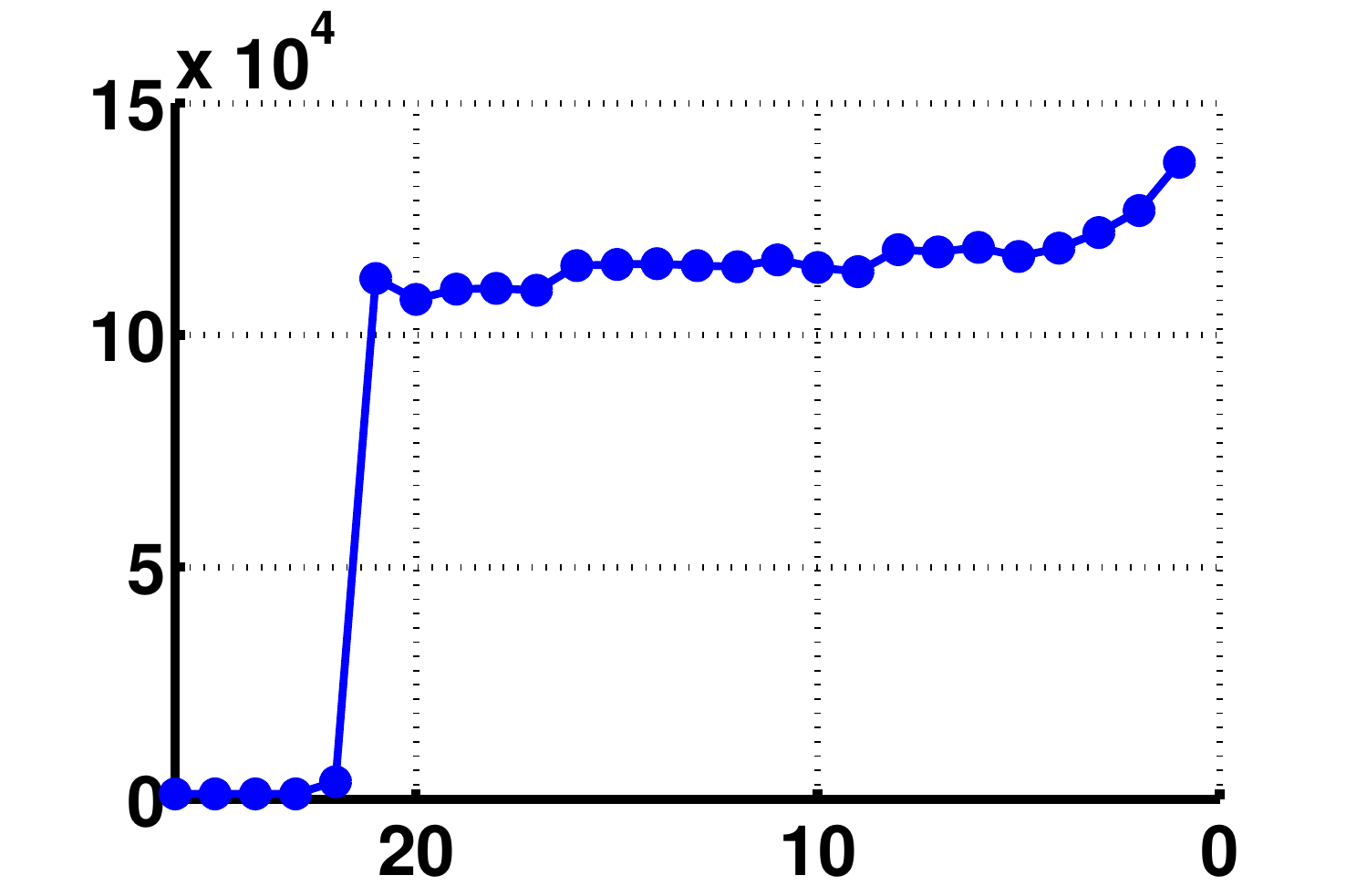}\label{fig:dream_tigr_pat}} 
	
	\caption{Extracted bit-change pattern for all benchmark suites.}	
	\label{fig:dream_Extratced_Pattern}
\end{figure*}

\renewcommand{\thesubfigure}{\alph{subfigure}}

The counter value of the two highlighted bits shows that bit 15 and bit 26 have been changed once and 4 times, respectively, in the last five memory requests. These counters generate a pattern (or signature) that is representative of the current memory access behavior as perceived by the memory controller. Figure \ref{fig:dream_Extratced_Pattern} shows such a signature extracted from these counters for all the benchmarks evaluated in this paper. The X-axis in each plot represents the corresponding counter ID per physical address bits and Y-axis shows the overall bit change rate over the application execution time. There is an exponential growth in the rightmost five bits of almost all the patterns. This is due to spacial locality that implies accessing the sequential physical addresses. Looking at these pattern and the address-mapping schemes presented in Figure \ref{fig:DReAM_Address_Mapping_fig} justifies why the column address bits are typically placed in the bottom of the physical address space. In this way, accessing consecutive cache lines will be mapped to the consecutive columns within the same row (i.e.\ Page Hit).


\subsubsection{Address Mapping Prediction} 
Given the signature for a set of running applications, the next issue is how to generate an optimized address-mapping scheme. The idea is to map the physical address bits with low variation to rows (to reduce the row switching or page conflicts), the physical address bits with medium variation to banks and the physical address bits with highest rate of changes to columns to increase the locality and decrease the page conflicts. Moreover, it is possible to limit DReAM to rearrange only a part of the physical address bits to mitigate the associated cost of the address mapping change in DRAMs that will be discussed later (data migration). For instance, in this paper, DReAM does not rearrange the column-address bits to avoid cache-line-level migration. To produce a new address mapping scheme at run-time, (\lowerromannumeral{1}) the bit-change rate of physical address bit will be monitored for each time-window, (\lowerromannumeral{2}) a new address mapping scheme will be estimated based on each time-window monitoring information, (\lowerromannumeral{3}) the bit-change rate monitored, based on the predefined and new address mapping schemes for each time-window will be compared. If the new address-mapping scheme can improve the bit-change rate in comparison with the baseline address mapping above a desired (and programmable) threshold (for consecutive time-windows defined by `Consistency Threshold') then the new address mapping will be used as the primary address mapping scheme in the system.

\subsubsection{Mathematical Insight}  \label{subsubsec:dream_insight}

Intuitively, DReAM proposes a simple technique to detect an application-specific address mapping scheme based on the physical address bit-change monitoring process. However, the question is to find an analytical proof to show that the application-specific address-mapping scheme predicted using this method can actually improve the performance of the memory system. 

\begin{figure}[!htb]
\centering
\includegraphics[scale=0.09]{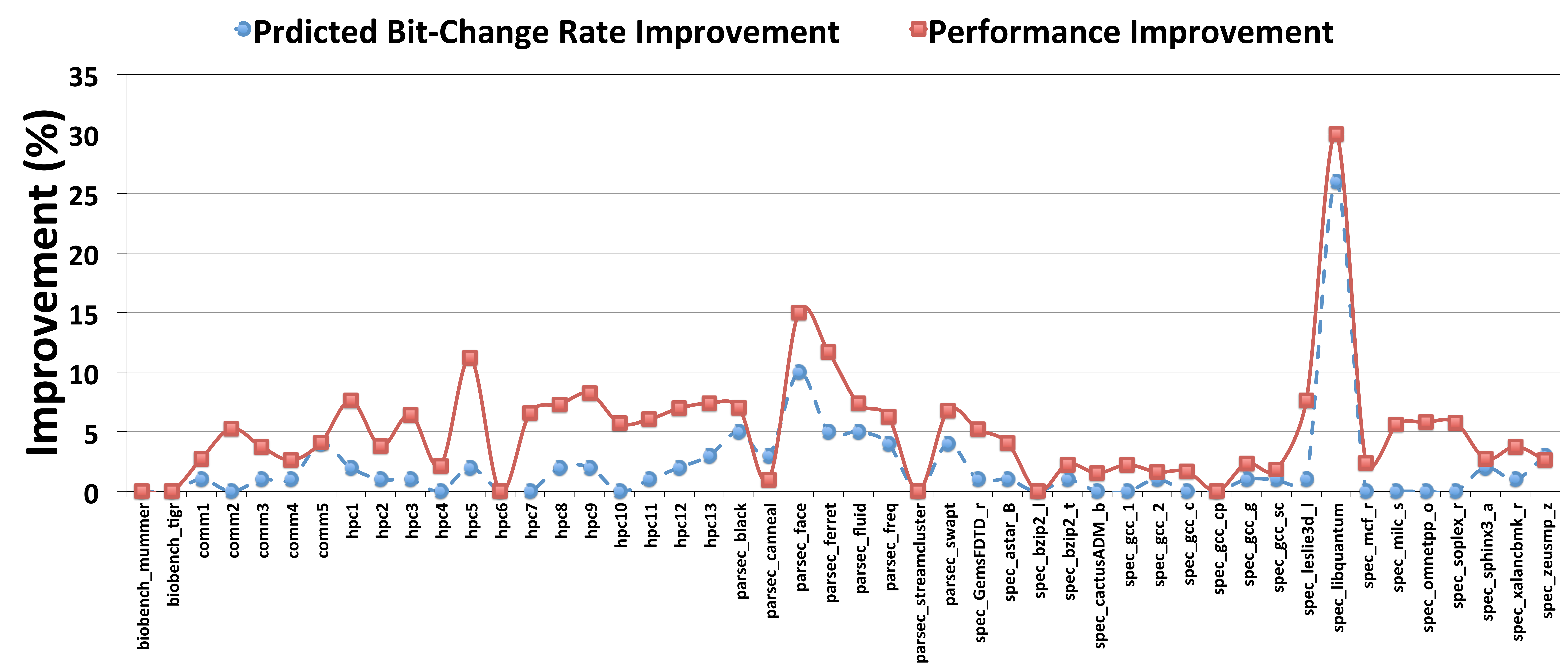}
\vspace{-2mm}
\caption{Bit-change rate improvement vs. the overall performance improvement.}
\label{fig:dream_correlation}
\end{figure}

As discussed, the predicted address-mapping scheme will be exploited only if it can reduce the bit-change rate, in comparison with the baseline address mapping, beyond a certain threshold. This means that DReAM assumes that there is a correlation between the bit-change rate of physical address bits and the performance of DRAMs. To investigate this, the correlation coefficient between the average bit-changed improvement reported by DReAM and the performance improvement of memory system, while using the DReAM address mapping, was investigated. The experimental results shows that there is a strong correlation, 0.89 with a very small P-value (i.e.\ $1.97\times10^{-15}$), between the bit-change rate and the final performance improvement. This justifies why the predicted address mapping scheme proposed by DReAM can improve the performance of DRAMs. Figure \ref{fig:dream_correlation} shows the bit-change rate improvement reported by DReAM and the final performance improvement achieved using the predicted address-mapping scheme by DReAM.

\subsubsection{Mapping-Sensitive Vs. Mapping-Insensitive} 
Looking at the patterns presented in Figure \ref{fig:dream_Extratced_Pattern} and considering the basic principles behind the address mapping prediction explained so far, it is possible to categorise workloads based on their sensitivity to the address mapping. If there is an opportunity to swap a physical address bit with a high change rate (that corresponds to the row address space) with another physical address bit with a smaller change rate (that is not a part of the row address space) then this is called a mapping-sensitive workload. Otherwise, this is categorised as an mapping-insensitive workload. For instance, `stream' (Figure \ref{fig:dream_stream_pat}) is a mapping-insensitive workload since all the bits dedicated to the row address space have a smaller change rate than other bits. On the other hand, `libquantum' (Figure \ref{fig:dream_libquantum_pat}) is a mapping-sensitive workload since there is an opportunity to swap bit 14 with another bit with smaller change rate (let's say bit 10).

\subsection{Data Migration Challenge} \label{subsec:drean_migration_challenge}

Changing the address-mapping scheme of a DRAM, on-the-fly, has a very important obstacle which is the requirement for the Data Migration. Initially, a DRAM places data into memory based on a predefined address mapping scheme. Therefore, changing the address mapping scheme implies that the data previously loaded into the DRAM cannot be accessed using the new address mapping scheme. Thus, before employing the new address mapping, the existing data in DRAMs {\em must} be migrated to a new location based on the new address mapping scheme. This imposes some overhead to the overall performance of memory system. To alleviate this overhead this paper investigate two different scenarios explained as follows.



\subsection{Data Migration Solutions} \label{subsec:drean_migration}


\subsubsection{Scenario 1 - Offline Data Migration} \label{subsubsec:dream_offine_migration} 

This scenario explains the simplest DReAM implementation that imposes a minimal hardware overhead to the overall memory system. In general, this scenario is well suited for application-specific computer architectures, e.g.\ database systems, where a specific application is running on the system over and over. For instance, in a database system, depending on the type of database (e.g. financial, medical etc.), usually only a few specific queries with minor variations are used to search for specific data. Moreover, in the big-data research area running a query over a database might take a few days or weeks. This produces a specific memory access pattern in the system that usually is consistent over a long period of time. In this implementation, the memory access pattern of applications (single or multithread) will be monitored at run-time for a desired period (e.g. it can be a few hours, a few days etc). This period is called Region Of Interest (ROI). Ideally, the ROI should be chosen to be long enough to represent the application access behavior. For instance, if the ROI for a medical database is chosen to be one day, then the memory access pattern of almost all the possible queries that are usually run on the database during the day can be covered by the ROI. In this situation, DReAM will estimate an optimized address-mapping scheme based on the average bit-change rate extracted from the dedicated counters per physical address bits for the entire ROI. This new mapping will be saved on the memory controller and upon rebooting the system user has an option to choose the DReAM address mapping scheme over the baseline from the system BIOS. Thus, whenever user reboots the system the memory controller can employ a new address mapping that is estimated based on DReAM calibration mode. A similar approach has been implemented for Intel-adaptive page policy and a special beta BIOS provided by ASUS that allows the user to choose a desired page closure policy at system start up \cite{website:inteladaptive,dodd2006adaptive}.

In this scenario, there is a penalty for the rebooting process but after that, as far the usual workloads running on the system, the overall performance of memory system will be improved by taking advantage of new address-mapping scheme. This is why this scenario is well suited for the systems with consistent behavior over the time.

\subsubsection{Scenario 2 - Online Data Migration}\label{subsubsec:dream_scenario_2}

This scenario investigates the possibility of performing on-the-fly data migration inside a DRAM device by proposing small modifications to the internal structure of this memory system.



\begin{figure}[!htb]
\centering
\includegraphics[scale=0.31]{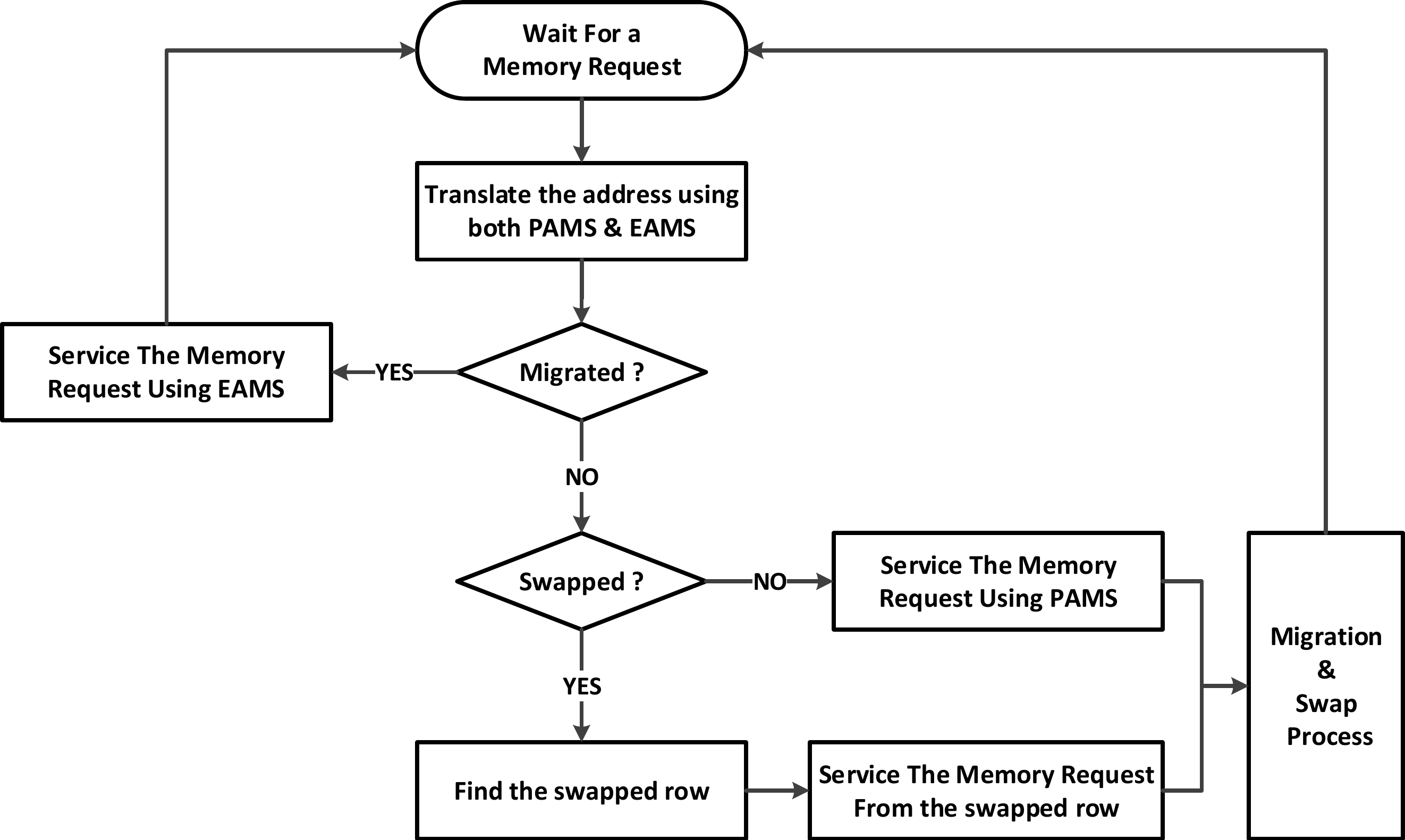}
\caption{DReAM flowchart. }
\label{fig:dream_flowchart}
\end{figure}

{\bf Basic Procedure:} Figure \ref{fig:dream_flowchart} presents the basic flowc\-hart of servicing a memory request while using DReAM considering the second data migration scenario. To minimize the overhead of migration, a row is migrated only when it has been accessed. In practice, this means that the migration occurs gradually on demand. 

On the first access to a row, the requested physical address is translated to the internal structure of the DRAM using both the Predefined Address-Mapping Scheme (PAMS) and the Estimated Address-Mapping Scheme (EAMS). The translated address by PAMS is the source row address and the translated address by EAMS is the destination row address. There are two main functions that might be applied on the requested address in different situations which are Migration and Swap. The requirement for these two functions and what they are will be discussed later on in this section and they are declared here just for initial familiarity to explain the flowchart.

The first step is to determine if the accessed row is in its original location, pointed to by PAMS, or not. Two bits are dedicated to each row in a DRAM bank to keep track of the current status of that row: one bit (Migration-Bit) to determine if the row has been moved to its new location (migrated) and one bit (Swap-Bit) to determine if the row has been swapped (this process will be discussed later). Two tables can be dedicated to accommodate these bits for the entire DRAM module: the Migration Table (MT) and the Swap Table (ST). At this point several situations might happen:

\begin{itemize}[leftmargin=*]
\item If the requested row is in its original location (the migration-bit and swap-bit are 0) then, (\lowerromannumeral{1}) the PAMS will be used to access and service the requested row, (\lowerromannumeral{2}) the requested row will be migrated to the destination location pointed by EAMS, (\lowerromannumeral{3}) if the destination location is occupied by a different row then intuitively the content of destination row also needs to be migrated to a third place. This can produce a chain of unnecessary data migration which is costly. To avoid this, a simple row-swap algorithm is employed which means that in such situations the content of destination row will be swapped by the content of source row (corresponding swap-bit will change to 1).
 
 \item If the requested row has been migrated then the EAMS will be used to access and service the requested row
 
 \item If the requested row has been swapped then, (\lowerromannumeral{1}) the swapped location will be calculated by applying the reverse address-mapping mechanism to the source location, (\lowerromannumeral{2}) step i will be repeated until the swap-bit of the pointed location by reverse address mapping scheme is 0, (\lowerromannumeral{3}) the request will be serviced, (\lowerromannumeral{4}) The requested row will be migrated to the destination location pointed by EAMS, (\lowerromannumeral{5}) a swap will be performed if it is necessary.
 
\end{itemize}

To make all this happens inside DRAM some modification needs to be done to the traditional structure of DRAMs which is explained below.


{\bf Required DRAM Modification:} There are two main requirements for DReAM to perform data migration in a DRAM device:  the capability of bulk data copy inside DRAM and the capability of on-the-fly buffering of the entire row to perform the swap operation. Both of these requirements have been studied individually by previous work to address different issues, using existing subarray level parallelism in DRAMs,\cite{seshadri2013rowclone,kim2012case} which are described in the following.


{\bf Bulk Data Copy in DRAM:} Seshadri \textit{et al}.\ \cite{seshadri2013rowclone} exploits the existing subarrays per bank in DRAMs to copy the entire row from one location to another inside DRAMs. Depending on the location of the source and destination rows, there are three different scenarios that should be considered: (i) copying between two rows within the same subarray (intra-subarray), (ii) copying between two rows in different subarrays in the same bank (inter-subarray), (iii) copying between two rows in different banks (inter-bank).



{\bf Subarray-Level Parallelism:} Kim \textit{et al}.\ \cite{kim2012case} proposed some small modification to DRAMs to be able to exploit existing subarray level parallelism in DRAMs. They discussed three different levels of modification to DRAM to improve the access latency by making subarrays working independently. Part of this work which, is more interesting from the point of view of this paper, is that called MASA. The key idea of MASA is to allow multiple activated subarrays in the same bank. MASA imposes (i) a designated-bit latch to each subarray, (ii)  a new DRAM command, subarray-select (SA-SEL) and (iii) routing of a new global wire. Based on their experimental methodology, they showed that the required extra latches imposes 0.15\% area overhead and consume 72.2~ \si\micro \si\watt \ additional power for each ACTIVATE command. Moreover, they evaluated that there is an extra 0.56 mW of static power in the steady state imposed by multiple activation of subarrays. 

Having explained the above techniques, an overview of the DReAM architecture will be explained in the following sections.

\subsection{DReAM - Overview of Architecture} 

\begin{figure}[!htb]
\centering
\includegraphics[scale=0.4]{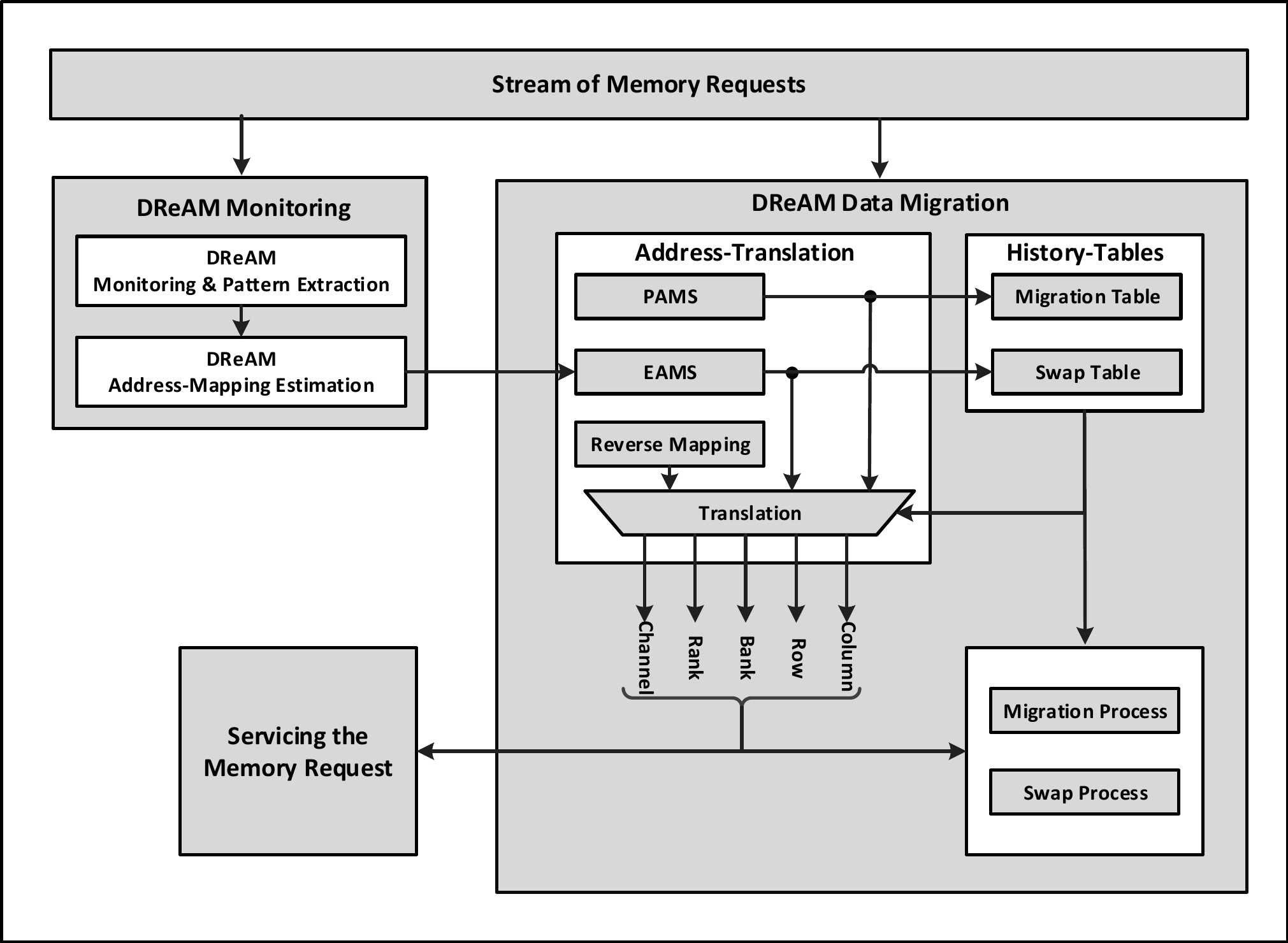}
\caption{DReAM architecture.}
\label{fig:dream_architecture}
\end{figure}

Figure \ref{fig:dream_architecture} presents a high-level overview of the DReAM architecture. DReAM includes two main phases, Address-Mapping Estimation and Online Data Migration.

\subsubsection{Address-Mapping Estimation} 
\begin{figure}[!htb]
\centering
\includegraphics[scale=0.35]{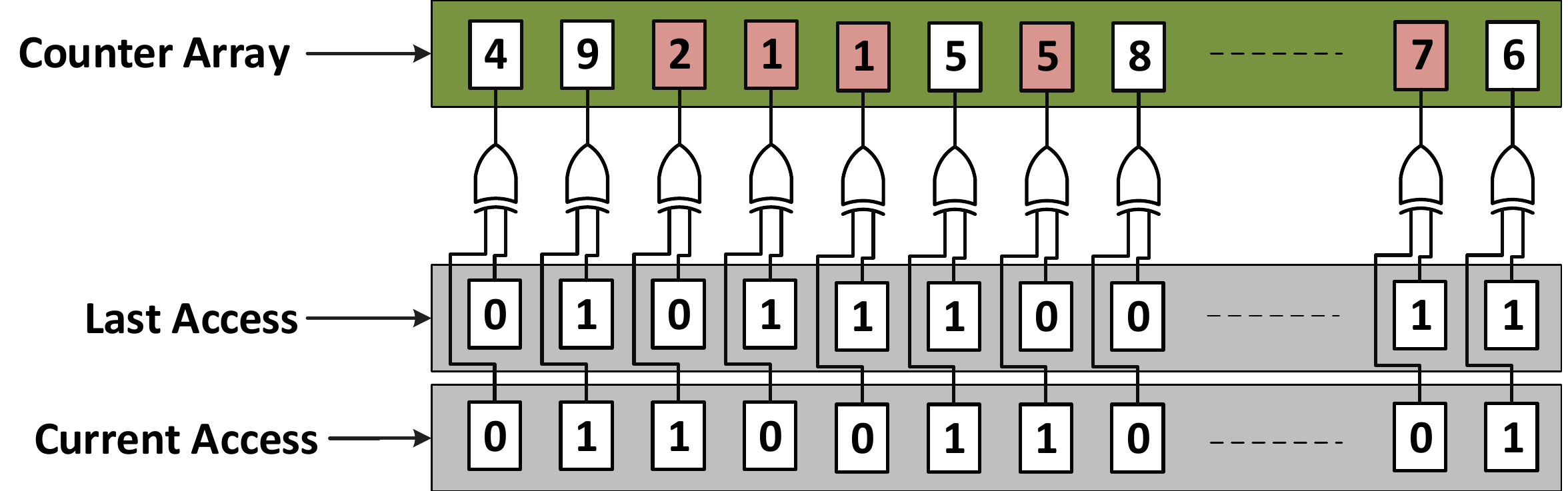}
\caption{DReAM monitoring counter structure.}
\label{fig:dream_monitor}
\end{figure}

Address-Mapping Estimation requires minimal architecture support. Only one counter per physical address bit, a history register to hold the last accessed address and an array of XORs to detect the bit-change between two consecutive memory requests are required to extract the access pattern at run-time. Figure~\ref{fig:dream_monitor} presents a simple overview of such a structure. In this structure each bit of the currently accessed address will be XORed with the corresponding bit of the last accessed address. Then, if there is a difference in the accessed bit the corresponding counter will be incremented. As discussed, this will produce a pattern of physical address bit changes over a period that can be employed to estimate an application-specific address-mapping scheme. 

\subsubsection{Data Migration - Operation} 

\begin{figure}[h!]
	\centering
	\subfloat[Intra-Subarray]{\includegraphics[scale=0.23]{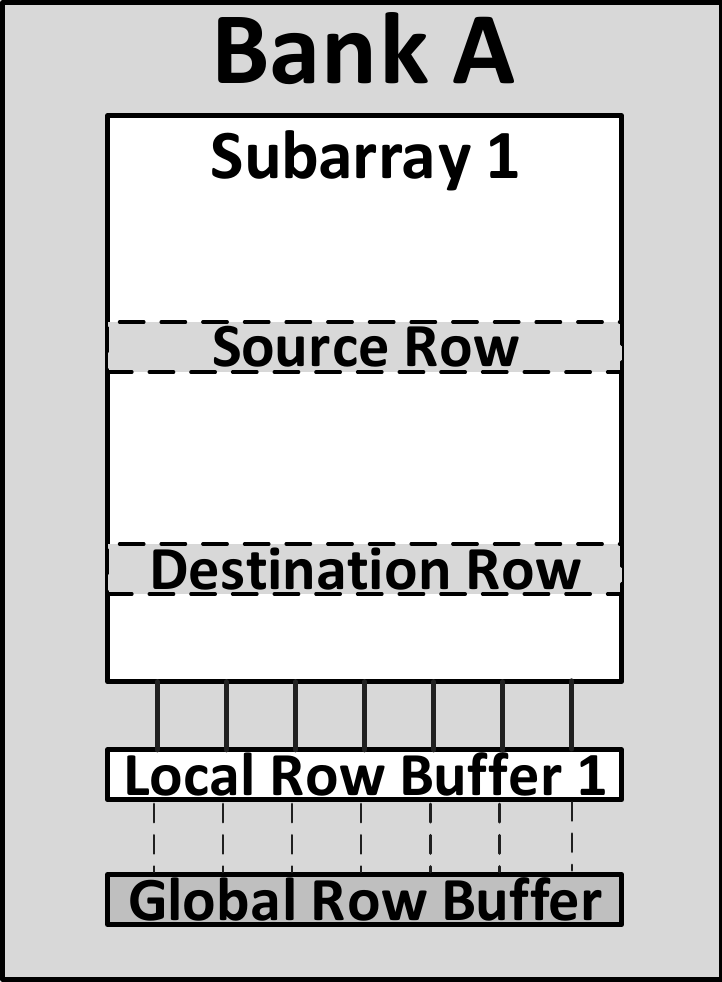}\label{fig:intra_subarray_mig}} 
	\hspace*{5mm}
	\subfloat[Inter-Subarray]{\includegraphics[scale=0.23]{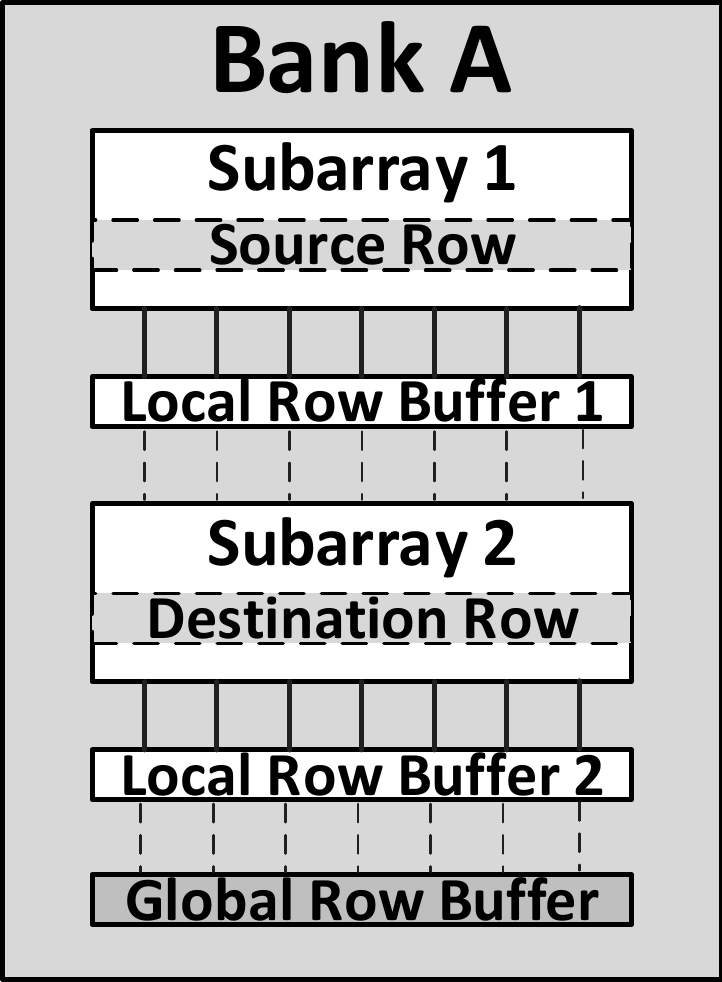}\label{fig:inter_subarray_mig}} 
	\hspace*{5mm}
	\subfloat[Inter-Bank]{\includegraphics[scale=0.23]{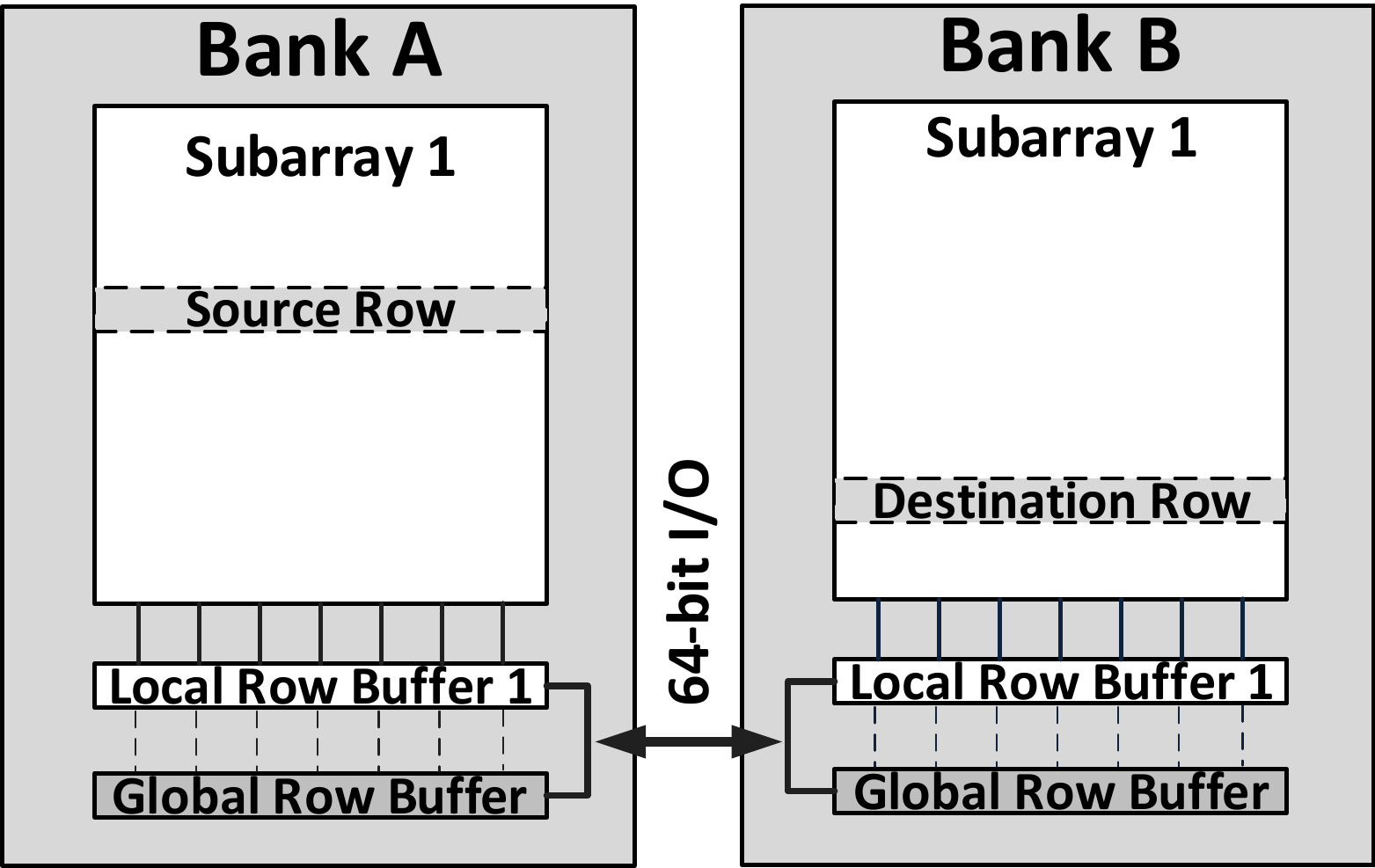}\label{fig:inter_bank_mig}} 		

	\caption{Different data migration scenarios.}	
	\label{fig:DReAM_Data_Migration_scenarios}
\end{figure}

The Data Migration required by DReAM can be described considering the following observations:

First, all the local row buffers (one local row buffer in each subarray) within a bank are connected to the global row buffer using global bitlines  and all the row-buffers (either local or global) within a DRAM device are connected together using a narrow I/O bus (64-bit wide) \cite{itoh2001vlsi,kim2012case}. Second, considering the modification proposed by Kim \textit{et al}.\ \cite{kim2012case} the DRAM module supports MASA. This supports multiple activation of subarrays while only one of them can be connected to the global bitline at a time.
Figure \ref{fig:DReAM_Data_Migration_scenarios} presents the possible scenarios that data migration might happen. To describe the following scenarios it is assumed that the destination row always has been occupied by another row (worst case scenario) and thus a swap process is necessary.  

{\bf Inter and Intra bank Migration:} although Figure \ref{fig:DReAM_Data_Migration_scenarios} presents all different possible scenarios for data migration considering the main purpose of DReAM (i.e.\ reducing page conflicts) the first two scenarios are inefficient. The reason is that in the first two scenarios the data migration happens within the same bank which does not reduce the page conflict occurrence probability. Thus, there is no point of paying extra penalty to perform the data migration for these two scenarios.

{\bf Inter-bank Migration:} in this scenario (Figure \ref{fig:inter_bank_mig}), source and destination rows are in different banks. Therefore, both of source and destination rows can be activated in parallel. Thus, to perform this scenario the memory controller, (\lowerromannumeral{1}) activates both source and destination row and load their contents into their local row-buffer, (\lowerromannumeral{2}) puts bank A into the read mode and puts bank B into the write mode, (\lowerromannumeral{3}) transfers the source row from local row-buffer 1 in bank A to the global row buffer of the bank B using the narrow I/O bus, (\lowerromannumeral{4}) puts bank A into the write mode and puts bank B into the read mode, (\lowerromannumeral{5}) transfers the destination row from local row-buffer 1 in bank B to the global row buffer of the bank A using the narrow I/O bus, (\lowerromannumeral{6}) connects the global bitlines of the global row-buffer in bank A to the source row and the global row-buffer of bank B to the destination row. 

\subsubsection{Data Migration - Timing Overhead} 

The latency overhead imposed by the data migration for each workload is the number of inter-bank migration (Figure \ref{fig:inter_bank_mig}) times cost of transferring a row using the internal narrow I/O (i.e. 64-bit) bus. Considering the transfer rate of 64 bits/clock and a row buffer size of 4 Kbit (per device) then 64 clock cycles are required to transfer a row from one bank to another. Another 64 clock cycles are required in the case that a swap is necessary. Therefore in the worst case scenario, the penalty for each data relocation between two banks is 128 memory clock cycles. Assuming that the CPU clock cycle is 4 times faster than memory clock cycle then the data migration penalty is 512 CPU clock cycles. In a very pessimistic situation it is assumed that the processor will be stalled while the data migration is happening. Therefore the 512 clock cycles times number of required inter-bank data migrations delivers a good estimation of the extra overhead imposed on the overall execution time.

\subsubsection{Rollback Process to Avoid Degradation Loop} 

DReAM predicts an application-specific address mapping scheme based on the monitoring period of the past application access pattern. However, it is not guaranteed that the application access pattern will not change again in the future. Therefore, the predicted address-mapping scheme by DReAM might not be efficient anymore and, as a result, using such an address-mapping scheme might degrade the performance of the DRAMs (i.e.\ Degradation Loop). To work around this issue, DReAM supports `Rollback' procedure. As discussed, DReAM will switch to the predicted address-mapping scheme if the new mapping can improve the bit-change rate in comparison with the baseline, over a predefined threshold, for consecutive time windows. A similar approach will be used to evaluate the efficiency of the predicted address-mapping at run-time. DReAM keeps monitoring the bit-change pattern over the time windows even after a new address-mapping scheme is predicted. If the bit-change improvement of the predicted address mapping scheme no longer outperforms the baseline DReAM will switch back to the predefined address mapping scheme. This triggers the roll back function to return the migrated rows to their original location. In this situation the memory controller can switch between at least two address-mapping scheme based on the application access pattern. A third address mapping scheme can be employed if the rollback process completes which means that all the rows migrated by the previous address-mapping have returned to their original locations.

%% file: EvaluationMethodology.tex
\section{Evaluation Methodology}
\label{sec:methodology}



{\bf Simulator:} USIMM \cite{chatterjee2012usimm} was used as the main simulation platform for these experiments. USIMM was modified to support Permeation-based Page Interleaving \cite{zhang2000permutation} and Minimalist Open-Page scheme plus a full implementation of the DReAM architecture. DReAM was evaluated based on a 4 GB DRAM organized in 1 channel, running single thread applications. To increase the randomness of memory access patterns the size of memory was fixed while running multithread applications. A FR-FCFS scheduling algorithm is used in our experiments. Table~\ref{table:usimm_config_parameter_dream} presents the configuration parameters of USIMM.



\begin{table}[!htb]
\centering
  \begin{tabular}{ c  c  c }
	\hline
	\rowcolor[gray]{.8} \textbf{Model} & \textbf{Description} & \textbf{Value} \\
	\hline
	\multirow{2}{*}{Processor} & Clock Speed & 3.2~GHz \\
	 & ROB size & 32\\
       \hline
	\multirow{7}{*}{Memory System} &  Bus Speed & 800~MHz\\
 	 & Number of Channels & 1\\
	 & Ranks per channel & 1\\
	 & Bank per rank & 8\\
	 & Row per bank & 65,536\\
	 & Cache line size & 64~Byte\\
	\hline
  \end{tabular}
  \caption{USIMM configuration parameters.}
  \label{table:usimm_config_parameter_dream}
\end{table}

\begin{table*}[!htb]
\centering
  \begin{tabular}{| >{\centering\arraybackslash}m{4cm} | >{\centering\arraybackslash}m{4cm} |  >{\centering\arraybackslash}m{4cm} | >{\centering\arraybackslash}m{4cm} |}
	\hline
	\multicolumn{2}{|c|}{\textbf{SPEC}} & \textbf{PARSEC} & \textbf{COMMERCIAL} \\	
	\hline
	(a) GemsFDTD\_r & (k) astar\_B & (u) canneal & (D1) comm1\\	
	\hline
	(b) bzip2\_l & (l) bzip2\_t & (v) streamcluster & (D2) comm2\\	
	\hline
	(c) cactusADM\_b & (m) gcc\_1 & (w) blackschols & (D3) comm3\\	
	\hline
	(d) gcc\_2 & (n) gcc\_c & (x) facesim & (D4) comm4\\	
	\hline
	(e) gcc\_cp & (o) gcc\_g & (y) ferret & (D5) comm5\\	
	\hline
	(f) gcc\_sc & (p) mcf\_r & (z) fluidanimate & \textbf{BIOBENCH}\\	
	\hline
	(g) milc\_s & (q) omnetpp\_o & (A) freqmine & (E) mummer\\	
	\hline
	(h) soplex\_r & (r) sphinx3\_a & (B) swaption &  (F) tigr\\
	\hline
	(i) xalancbmk\_r & (s) zeusmp\_z & \textbf{HPC} & \cellcolor[gray]{0} \\
	\hline
	(j) libquantum & (t) leslie & (C) hpc1 - hpc13 & \cellcolor[gray]{0} \\
	\hline
  \end{tabular}
  \caption{Evaluated workloads and benchmark suites. }
  \label{table:dream_workloads}
\end{table*}

{\bf Address Mapping Schemes:}  The memory access pattern, and as a result the number of page conflicts in DRAMs, can be affected by the predefined memory address mapping scheme. The experiments consider three different address mappings presented in Figure~\ref{fig:DReAM_Address_Mapping_fig}. The experimental results presented in Section \ref{subsec:DReAMMotivaiton} (Figure \ref{fig:motivation_address_mapping}) show that the Permutation-based Page interleaving policy (Mapping 2) performs best for most of the workloads. Therefore, this address mapping scheme is employed as a fair baseline to compare with the DReAM scheme. 

{\bf Workloads:} the workloads include a wide range of memory intensive applications (i.e.\ 48 workloads) from different  benchmark suites (PARSEC \cite{bienia2008parsec}, SPEC \cite{dixit1991spec}, BIOBENCH \cite{albayraktaroglu2005biobench}, HPC and COMMERCIAL) and representative regions of interest for each application. Table~\ref{table:dream_workloads} lists the workloads and their corresponding benchmark suites. An identifier is assigned to each application to facilitate the naming of multithread workloads constructed from these applications. To increase the variety of memory access patterns, USIMM was set up for multithread applications to evaluate 20 randomly selected workload mixes; a combination of 4-thread and 8-thread applications. Table \ref{table:dream_multicore_workloads} lists these multithread workloads employing the identifier of single thread workloads presented in Table \ref{table:dream_workloads}.

\begin{table}[!htb]
\centering
  \begin{tabular}{| m{2.7cm} | m{5.3cm} | }
	\hline
	\multicolumn{2}{|c|}{\textbf{Multithread Workloads}} \\		
	\hline
	M1:C13-C5-x-t & M11:C9-C13-C5-w-x-t-j-q\\
	\hline
	M2:C9-w-j-q & M12:C8-C3-w-x-y-a-t-j\\
	\hline
	M3:w-x-y-t & M13:C8-C5-x-y-a-t-p-q\\
	\hline
	M4:C8-C5-t-p & M14:C9-C12-C13-C9-C12-C12-p-q\\
	\hline
	M5:t-t-p-g & M15:C13-x-t-g-p-t-p-g\\
	\hline
	M6:C8-w-p-q & M16:C8-C3-C5-w-C5-C5-p-q\\
	\hline
	M7:C3-C5-C5-C5 & M17:C9-w-y-w-w-a-t-g\\
	\hline
	M8:C9-w-y-w & M18:C13-C3-x-C13-a-a-p-g\\
	\hline
	M9:C12-C13-a-a & M19:C12-C13-y-a-a-a-g-q\\
	\hline
	M10:x-t-j-q & M20:x-y-p-a-x-a-p-q\\
	\hline
  \end{tabular}
  \caption{Randomly selected multithread workloads. }
  \label{table:dream_multicore_workloads}
\end{table}

%% file: ResultsAndDiscussion.tex
\vspace{-5mm}
\section{Results and Discussions}
\label{sec:dream_results}

\subsection{Performance Analysis} 

\begin{figure*}[!htb]

\centering
\includegraphics[scale=0.25]{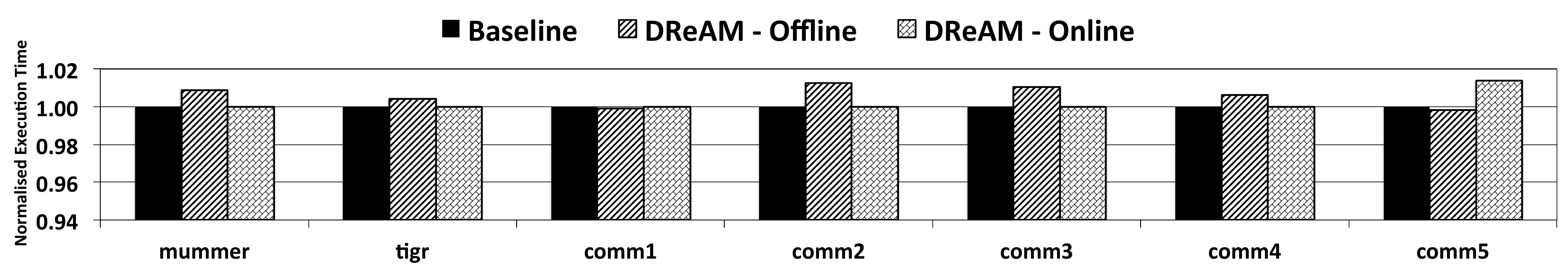}
\vspace{-2mm}
\caption{Results for BIOBENCH and COMMERCIAL benchmark suites.}
\label{fig:dream_performance_improvement_Comm}

\centering
\includegraphics[scale=0.25]{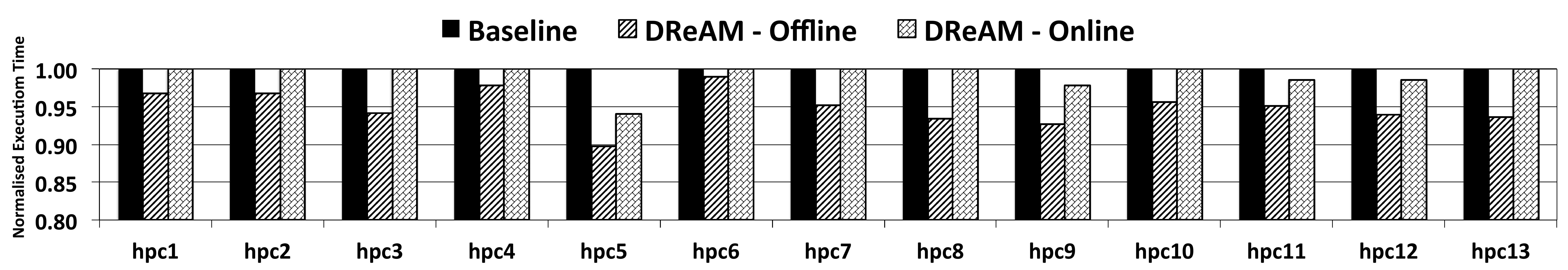}
\vspace{-2mm}
\caption{Results for HPC benchmarks.}
\label{fig:dream_performance_improvement_HPC}

\centering
\includegraphics[scale=0.25]{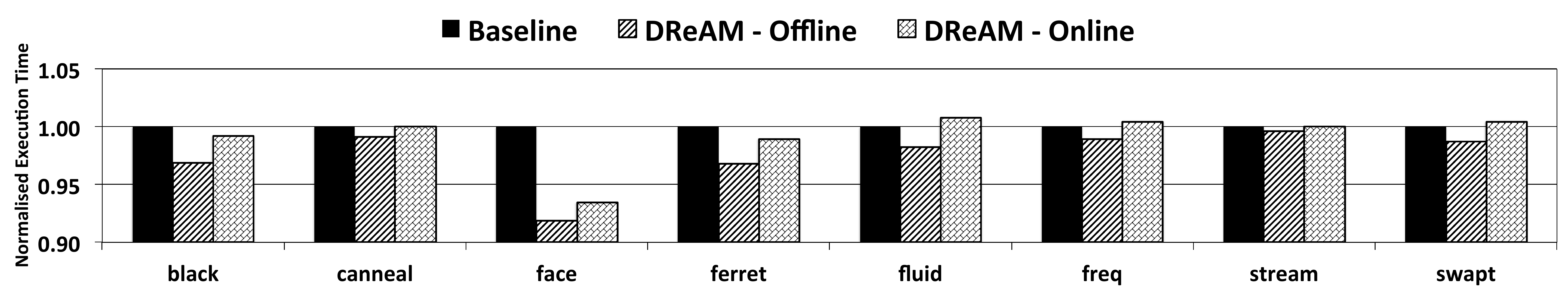}
\caption{Results for the PARSEC benchmark suite.}
\label{fig:dream_performance_improvement_PARSEC}

\centering
\includegraphics[scale=0.25]{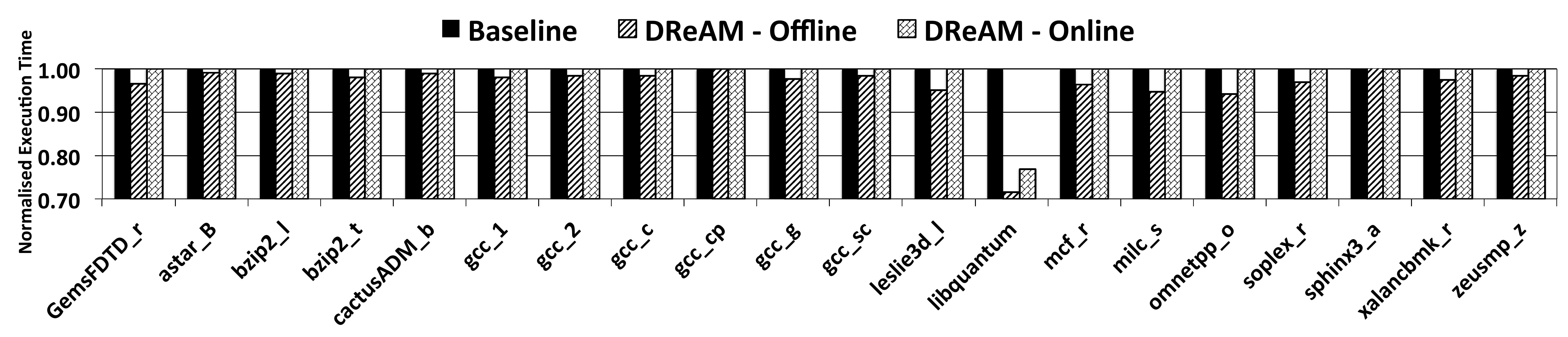}
\vspace{-3mm}
\caption{Results for the SPEC benchmark suite.}
\label{fig:dream_performance_improvement_SPEC}

\centering
\includegraphics[scale=0.25]{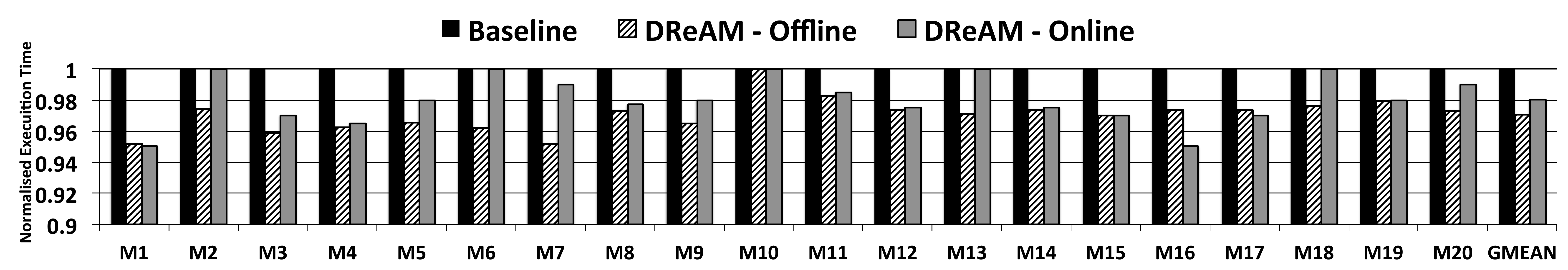}
\vspace{-2mm}
\caption{Results for multithreaded benchmarks.}
\label{fig:performance_improvement_multithread}

\end{figure*}

In this section the performance of DReAM is investigated. Before jumping to the result graphs, the following summary might be helpful: (\lowerromannumeral{1}) The performance numbers presented in this section are normalized to the baseline (Permutation-based address mapping) which delivers the best average execution time among three address-mapping schemes presented in Figure \ref{fig:DReAM_Address_Mapping_fig}. (\lowerromannumeral{2}) As discussed, the offline mapping is desired only in the case of applications with a consistent behavior and will be achieved after a long calibration period. Therefore, the rebooting cost will be negligible considering the long-period running application. Thus, in the results presented in Figures \ref{fig:dream_performance_improvement_Comm} to \ref{fig:performance_improvement_multithread} the cost of rebooting is ignored in the case of DReAM-Offline and only the efficiency of the address mapping detected by DReAM is investigated, in comparison with the baseline mapping.

Figure \ref{fig:dream_performance_improvement_Comm} presents the execution time for BIOBENCH and PARSEC benchmarks. This result suggests that the baseline address-mapping scheme is good enough for the workloads presented in these benchmarks and DReAM does not have margin to predict a better address mapping scheme. Therefore, there is no bit-change rate improvement when using DReAM in comparison with the baseline. The small degradation by DReAM-Offline (i.e.\ around 1\%) manifested in Figure \ref{fig:dream_performance_improvement_Comm} is due to slightly different access patterns caused by reordering the baseline address bits. This can be counted as noise. On the other hand, DReAM-Online mitigates this issue by on-the-fly checking the bit-change improvement, between two consecutive time windows, against a predefined threshold. For instance in these experiments DReAM-Online employs the new address mapping only if it can improve the bit change rate by more than 7\%. Thus, although DReAM cannot predict a better address mapping scheme than the baseline it does not degrade the performance for most of the cases. A similar behavior can be observed in Figures \ref{fig:dream_performance_improvement_HPC}-\ref{fig:performance_improvement_multithread}.

Overall, DReAM-Offline outperforms the permutation based address-mapping scheme (the best evaluated baseline) by 5\%, on average, and up to 28\% across all the workloads. In the case of DReAM-Online, 12 workloads satisfy the DReAM's threshold at run-time (i.e.\ improve the bit change rate by more than 7\%) and for these workloads DReAM-Online outperforms the baseline by 4.5\%, on average, and up to 23\%. Figure \ref{fig:performance_improvement_multithread} depicts the execution time for the randomly selected multithread workloads presented in Table \ref{table:dream_multicore_workloads}. These results show that DReAM can still predict a better address mapping scheme than the baseline even in the case of multithread workloads which produce a highly random memory access pattern. Looking into the results from a different angle suggests that DReAM outperforms the best evaluated baseline address mapping on average by 9\% and 2\% for mapping-sensitive and mapping-insensitive workloads respectively.

Considering the results presented in Figure \ref{fig:dream_performance_improvement_SPEC}, {\em libquantum} achieves a significant performance improvement taking advantage of DReAM. To understand this outcome, it is useful to check the extracted pattern for this workload presented earlier in Figure \ref{fig:dream_libquantum_pat}. This figure shows that there is a high change rate for bit 14. This bit is mapped to the rows address space increasing the possibility of accessing different rows within the same bank (i.e.\ Page Conflict) and so imposes a significant performance overhead. DReAM simply assigns this bit to the another address space (e.g. bank or column address space) by replacing it with a bit with a minimal change rate. In this situation, the excessive change rate of this bit increases the possibility of interleaving the accesses to different banks which improves the level of parallelism (or access locality, if using column address space) in the system and as a result improves the performance significantly. A similar argument explain the significant performance for the other workloads, such as `face'.

\subsection{Data Relocation Analysis} 

As discussed, the data relocation required by DReAM is composed of two main phases: Migration and Rollback. In the following some statistical analysis of migrations and rolls back required by DReAM will be discussed.


{\bf Migration vs. Rollback:} The experimental results (presented in Figure \ref{fig:dream_performance_improvement_Comm} to Figure \ref{fig:dream_performance_improvement_SPEC}) show that 12 standard workloads undergo dynamic data relocation. Out of these 12 workloads only two workloads require data rollback which are `ferret', with 10\% of data relocation spent on data rollback, and `libquantum', with 39\% of data relocation spent on data rollback.


{\bf Inter Bank vs.\ Intra Bank Data Relocation:} The results show that 87.5\% of data relocation happens between banks (Inter-bank relocation) and 12.5\% happens within banks (Intra-bank relocation). As mentioned, DReAM does not perform the Intra-bank scenarios to reduce the cost of data relocation.


\subsection{Storage Overhead and Scalability} \label{subsec:dream_storage}


{\bf Address Mapping Prediction:} As discussed, there is only one counter and one XOR gate per physical address bit plus one history buffer to keep track of the last access address is required to extract the monitoring pattern. Thus, assuming a sampling window of 250K memory requests, 18-bit counters times the number of physical address bits are the main storage overhead for the first phase of DReAM. Our experimental results show that this number is no more than 60 bytes for 512~GB DRAM.

\begin{figure}[!htb]
\centering
\includegraphics[scale=0.13]{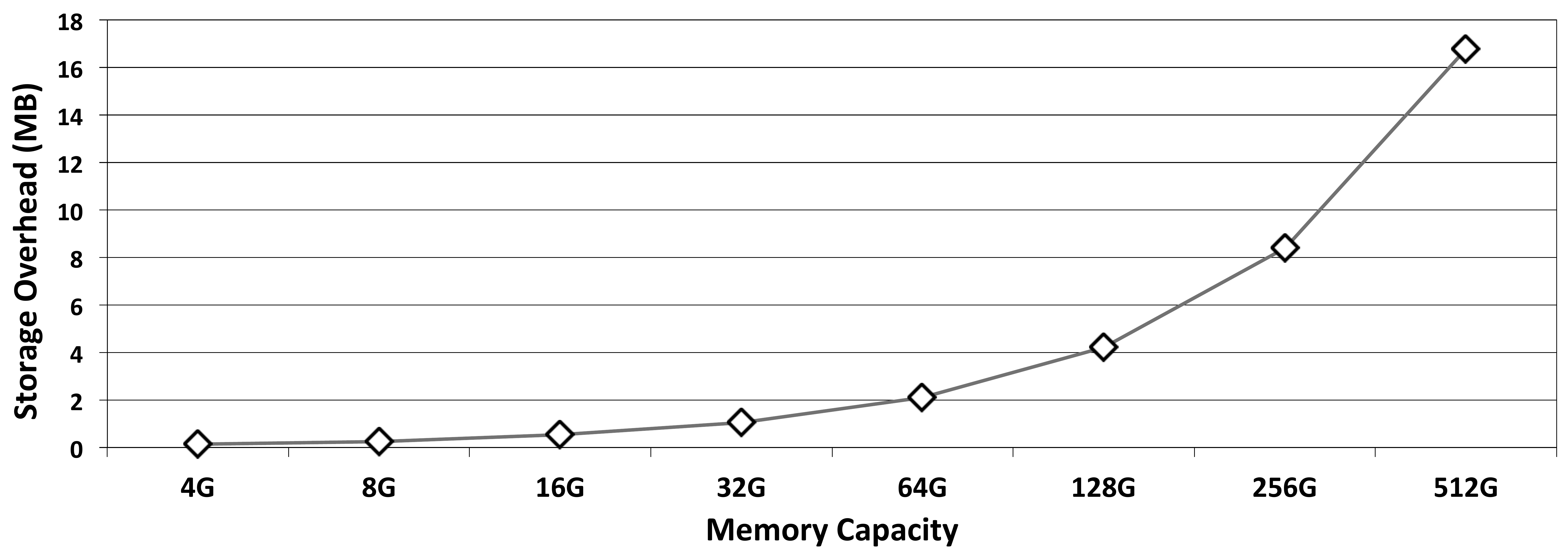}
\vspace{-2mm}
\caption{DReAM data migration overhead.}
\label{fig:migration_cost}
\end{figure}

{\bf Data Migration:} the online data migration requires to keep track of migrated and swapped pages. Therefore the required  MT and ST impose extra storage overhead to the overall memory system. Figure \ref{fig:migration_cost} depicts the overall storage overhead imposed by online data migration. This result shows that DReAM imposes a storage overhead of $3\times 10^{-5} ~\%$ to the overall DRAM size. Depending on the implementation choice the MT and ST can be implemented as a part of the memory controller or as  Metadata inside the DRAM.

%% file: RelatedWork.tex
\section{Related Work}

The shortfalls of DRAMs with respect to page conflicts are widely recognized in the area of memory system design. Prior work proposed a wide range of different techniques such as memory interleaving schemes, scheduling algorithms and some architectural modifications to the current structure of DRAMs to mitigate this issue. For instance, Zhang \textit{et al}.\ \cite{zhang2000permutation} proposed a page interleaving scheme to reduce page conflicts and exploit data locality. Hsu \textit{et al}. \cite{hsu1993performance} proposed another memory interleaving scheme to address the same issue. There are many other interesting works in the area of developing new scheduling algorithms\cite{ebrahimi2011parallel,ipek2008self,kim2010atlas,kim2011thread,mutlu2007stall,mutlu2008parallelism,nesbit2006fair} that prioritize servicing certain memory requests to reduce page conflicts and improve the memory performance. Some other types of work in this area are those that propose either a new architecture for DRAMs or a small modification to the traditional structure of these memory systems. For instance, Sudan \textit{et al}.\ \cite{sudan2010micro} proposes a technique to recognise the highly accessed data in DRAM and place them in the same row to improve the data locality. Kim \textit{et al}.\ \cite{kim2012case} proposed a technique to exploit the existing subarray level parallelism in DRAMs to improve the bank conflicts. PARDIS by Bojnordi \textit{et al}. \cite{bojnordi2012pardis} is a programmable memory controller that can be configured using a specific instruction set architecture (ISA). Although the focus of this work was not on developing optimized address-mapping scheme they configured PARDIS by the application-specific address mapping heuristic achieved by offline profiling analysis and presented a good performance improvement in the memory system.

%% file: Conclusion.tex
\section{Conclusions}
This paper has introduced DReAM which is a novel hardware technique based on approximating the entropy of each memory address bit for a set of memory requests. DReAM presents three main contributions: first, a low-cost pattern recognition technique is developed to extract the memory access pattern at run-time. Then, a methodology is proposed to estimate an optimized address-mapping based on the detected access pattern. Finally, a technique is proposed for the on-the-fly migration of data within DRAMs to reduce page conflicts. An extensive performance evaluation was carried out with 48 different workloads from 5 benchmark suites and 20 multithreaded applications. In summary, DReAM-Offline outperforms the permutation-based address mapping scheme (the state-of-the-art mapping) by 5\%, on average, and up to 28\% across all workloads. In the case of DReAM-Online, 12 workloads satisfy DReAM's threshold at run-time (i.e.\ improve the bit change rate over 7\%) and for these workloads DReAM-Online outperforms the baseline by 4.5\%, on average, and up to 23\%. Categorising workloads to mapping sensitive and insensitive, DReAM outperforms the best evaluated baseline address mapping on average by 9\% and 2\% for the first and second category respectively. Overall, DReAM is the first on-the-fly mechanism capable of generating workload specific address-mappings without requiring running applications to be stopped. 



